\pdfoutput=1
\documentclass[12pt,a4paper]{article}

\usepackage{lineno}  
\usepackage{graphicx}  
\usepackage{xspace} 
\usepackage{color}
\usepackage{colortbl}
\usepackage{amsmath} 
\usepackage{ifthen} 
\usepackage{multirow}

\newboolean{pdflatex}
\setboolean{pdflatex}{true} 
\newboolean{articletitles}
\setboolean{articletitles}{true} 

\newboolean{uprightparticles}
\setboolean{uprightparticles}{false} 

\usepackage{amssymb}
\usepackage{amsfonts}
\usepackage{upgreek} 

\usepackage{hyperref}    

\usepackage[all]{hypcap} 
\usepackage{afterpage}

\graphicspath{{fig/};{figures/}}

\def\lhcb {\mbox{LHCb}\xspace}
\def\ux85 {\mbox{UX85}\xspace}

\ifthenelse{\boolean{uprightparticles}}%
{

 \def\Ppsi        {\ensuremath{\uppsi}\xspace}

 \def\PDelta      {\ensuremath{\Delta}\xspace}
 \def\PXi      {\ensuremath{\Xi}\xspace}
 \def\PLambda      {\ensuremath{\Lambda}\xspace}
 \def\PSigma      {\ensuremath{\Sigma}\xspace}
 \def\POmega      {\ensuremath{\Omega}\xspace}
 \def\PUpsilon      {\ensuremath{\Upsilon}\xspace}

 \def\PB      {\ensuremath{\mathrm{B}}\xspace}
 
 \def\PD      {\ensuremath{\mathrm{D}}\xspace}

 \def\PJ      {\ensuremath{\mathrm{J}}\xspace}
 \def\PK      {\ensuremath{\mathrm{K}}\xspace}

 \def\Pb      {\ensuremath{\mathrm{b}}\xspace}
 \def\Pc      {\ensuremath{\mathrm{c}}\xspace}

 \def\Pi      {\ensuremath{\mathrm{i}}\xspace}

 \def\Ps      {\ensuremath{\mathrm{s}}\xspace}

}
{

 \def\Ppsi        {\ensuremath{\psi}\xspace}
 
 \mathchardef\PDelta="7101
 \mathchardef\PXi="7104
 \mathchardef\PLambda="7103
 \mathchardef\PSigma="7106
 \mathchardef\POmega="710A
 \mathchardef\PUpsilon="7107
 
 \def\PB      {\ensuremath{B}\xspace}
 
 \def\PD      {\ensuremath{D}\xspace}

 \def\PJ      {\ensuremath{J}\xspace}
 \def\PK      {\ensuremath{K}\xspace}

 \def\Pb      {\ensuremath{b}\xspace}
 \def\Pc      {\ensuremath{c}\xspace}

 \def\Pi      {\ensuremath{i}\xspace}

 \def\Ps      {\ensuremath{s}\xspace}

}

\def\squark    {\ensuremath{\Ps}\xspace}

\def\cquark    {\ensuremath{\Pc}\xspace}

\def\bquark    {\ensuremath{\Pb}\xspace}

\def\kaon  {\ensuremath{\PK}\xspace}
  \def\Kbar  {\kern 0.2em\overline{\kern -0.2em \PK}{}\xspace}

\def\Kz    {\ensuremath{\kaon^0}\xspace}
\def\Kzb   {\ensuremath{\Kbar^0}\xspace}
\def\KzKzb {\ensuremath{\Kz \kern -0.16em \Kzb}\xspace}
\def\Kp    {\ensuremath{\kaon^+}\xspace}
\def\Km    {\ensuremath{\kaon^-}\xspace}

\def\KpKm  {\ensuremath{\Kp \kern -0.16em \Km}\xspace}

  \def\Dbar    {\kern 0.2em\overline{\kern -0.2em \PD}{}\xspace}
\def\D       {\ensuremath{\PD}\xspace}

\def\Dz      {\ensuremath{\D^0}\xspace}
\def\Dzb     {\ensuremath{\Dbar^0}\xspace}
\def\DzDzb   {\ensuremath{\Dz {\kern -0.16em \Dzb}}\xspace}
\def\Dp      {\ensuremath{\D^+}\xspace}
\def\Dm      {\ensuremath{\D^-}\xspace}

\def\DpDm    {\ensuremath{\Dp {\kern -0.16em \Dm}}\xspace}

\def\B       {\ensuremath{\PB}\xspace}
  \def\Bbar    {\kern 0.18em\overline{\kern -0.18em \PB}{}\xspace}
\def\Bb      {\ensuremath{\Bbar}\xspace}

\def\Bzb     {\ensuremath{\Bbar^0}\xspace}

\def\Bs      {\ensuremath{\B^0_\squark}\xspace}
\def\Bsb     {\ensuremath{\Bbar^0_\squark}\xspace}
\def\Bdb     {\ensuremath{\Bbar^0}\xspace}

\def\jpsi     {\ensuremath{{\PJ\mskip -3mu/\mskip -2mu\Ppsi\mskip 2mu}}\xspace}

  \def\Y#1S{\ensuremath{\PUpsilon{(#1S)}}\xspace}

\def\Lbar {\ensuremath{\kern 0.1em\overline{\kern -0.1em\Lambda\kern -0.05em}\kern 0.05em{}}\xspace}

\def\to                 {\ensuremath{\rightarrow}\xspace}

\def\CP                {\ensuremath{C\!P}\xspace}

\newcommand{\DGs}{\ensuremath{\Delta\Gamma_{\squark}}\xspace}

\newcommand{\Gs}{\ensuremath{\Gamma_{\squark}}\xspace}

\def\AT#1     {\ensuremath{A_{\mathrm{T}}^{#1}}\xspace}           

\def\C#1      {\ensuremath{\mathcal{C}_{#1}}\xspace}                       
\def\Cp#1     {\ensuremath{\mathcal{C}_{#1}^{'}}\xspace}                    
\def\Ceff#1   {\ensuremath{\mathcal{C}_{#1}^{\mathrm{(eff)}}}\xspace}        
\def\Cpeff#1  {\ensuremath{\mathcal{C}_{#1}^{'\mathrm{(eff)}}}\xspace}       
\def\Ope#1    {\ensuremath{\mathcal{O}_{#1}}\xspace}                       
\def\Opep#1   {\ensuremath{\mathcal{O}_{#1}^{'}}\xspace}                    

\newcommand{\ket}[1]{\ensuremath{|#1\rangle}}              

\newcommand{\tev}{\ensuremath{\mathrm{\,Te\kern -0.1em V}}\xspace}
\newcommand{\gev}{\ensuremath{\mathrm{\,Ge\kern -0.1em V}}\xspace}
\newcommand{\mev}{\ensuremath{\mathrm{\,Me\kern -0.1em V}}\xspace}
\newcommand{\kev}{\ensuremath{\mathrm{\,ke\kern -0.1em V}}\xspace}
\newcommand{\ev}{\ensuremath{\mathrm{\,e\kern -0.1em V}}\xspace}
\newcommand{\gevc}{\ensuremath{{\mathrm{\,Ge\kern -0.1em V\!/}c}}\xspace}
\newcommand{\mevc}{\ensuremath{{\mathrm{\,Me\kern -0.1em V\!/}c}}\xspace}
\newcommand{\gevcc}{\ensuremath{{\mathrm{\,Ge\kern -0.1em V\!/}c^2}}\xspace}
\newcommand{\gevgevcccc}{\ensuremath{{\mathrm{\,Ge\kern -0.1em V^2\!/}c^4}}\xspace}
\newcommand{\mevcc}{\ensuremath{{\mathrm{\,Me\kern -0.1em V\!/}c^2}}\xspace}

\def\m    {\ensuremath{\rm \,m}\xspace}

\def\mm   {\ensuremath{\rm \,mm}\xspace}

\def\mum  {\ensuremath{\,\upmu\rm m}\xspace}

\def\gsim{{~\raise.15em\hbox{$>$}\kern-.85em
          \lower.35em\hbox{$\sim$}~}\xspace}
\def\lsim{{~\raise.15em\hbox{$<$}\kern-.85em
          \lower.35em\hbox{$\sim$}~}\xspace}

\newcommand{\Real}{\ensuremath{\mathcal{R}e}\xspace}

\def\pt         {\mbox{$p_{\rm T}$}\xspace}

\def\photos     {\mbox{\textsc{Photos}}\xspace}
\def\evtgen     {\mbox{\textsc{EvtGen}}\xspace}
\def\pythia     {\mbox{\textsc{Pythia}}\xspace}

\def\geant      {\mbox{\textsc{Geant4}}\xspace}

\def\gauss      {\mbox{\textsc{Gauss}}\xspace}

\def\tell1  {TELL1\xspace}
\def\ukl1   {UKL1\xspace}

\def \m {m_{hh}}
\def \angmu {\theta_{\jpsi}}
\def \angpi {\theta_{hh}}

\def \Bq {B_s^{0}}
\def \Bqb {\overline{B}{}_s^{0}}
\def \dv {{\rm d}}

\def \ch {\cosh \frac{\DGs t}{2}}
\def \sh {\sinh \frac{\DGs t}{2}}
\def \Ab {\overline{A}}

\def\Lz {\ensuremath{\PLambda}\xspace}
\def\Lbar {\ensuremath{\kern 0.1em\overline{\kern -0.1em\PLambda}}\xspace}

\usepackage{cite}
\usepackage{mciteplus}
\usepackage{epsfig,accents}
\newcommand*{\fancybar}{\scalebox{.4}{(}\raisebox{-1.7pt}{--}\scalebox{.4}{)}}
\newcommand*{\brabar}[1]{\accentset{\fancybar}{#1}}

\begin{document}
\renewcommand{\thefootnote}{\fnsymbol{footnote}}
\setcounter{footnote}{1}
\begin{titlepage}

\belowpdfbookmark{Title page}{title}

\pagenumbering{roman}
\vspace*{-1.5cm}
\centerline{\large EUROPEAN ORGANIZATION FOR NUCLEAR RESEARCH (CERN)}
\vspace*{1.5cm}
\hspace*{-5mm}\begin{tabular*}{16cm}{lc@{\extracolsep{\fill}}r}
\vspace*{-12mm}\mbox{\!\!\!\includegraphics[width=.12\textwidth]{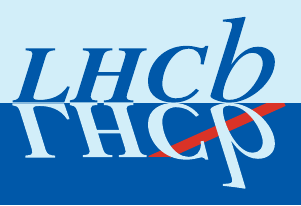}}& & \\
 & & CERN-PH-EP-2013-024\\
 & & LHCb-PAPER-2013-069\\  
 & & \today \\ 
 & & \\
\end{tabular*}

\vspace*{2.0cm}

{\bf\boldmath\Large
\begin{center}
Measurement of resonant  and \CP components in $\Bsb\rightarrow J/\psi \pi^+\pi^-$ decays
\end{center}
}

\vspace*{1.0cm}
\begin{center}
\normalsize {
The LHCb collaboration\footnote{Authors are listed on the following pages.}
}
\end{center}


\begin{abstract}
  \noindent
The resonant structure of the decay $\Bsb\to\jpsi\pi^+\pi^-$ is studied using data corresponding to 3~fb$^{-1}$ of integrated luminosity from $pp$ collisions by the LHC and collected by the LHCb detector. Five interfering $\pi^+\pi^-$ states are required to describe the decay: $f_0(980),~f_0(1500),~f_0(1790),~f_2(1270)$, and $f_2^{\prime}(1525)$. An alternative model including these states and a non-resonant $\jpsi \pi^+\pi^-$ component also provides a good description of the data. Based on the different transversity components measured for the spin-2 intermediate states, the final state is found to be compatible with being entirely \CP-odd. The \CP-even part is found to be $<2.3$\% at 95\% confidence level. The $f_0(500)$ state is not observed, allowing a limit to be set on the absolute value of the mixing angle with the $f_0(980)$ of $<7.7^{\circ}$ at 90\% confidence level, consistent with a tetraquark interpretation of the $f_0(980)$ substructure.

\end{abstract}

\vspace*{2.0cm}
\vspace{\fill}


\vspace*{1.0cm}
\begin{center}
\hspace*{6mm}Submitted to Phys. Rev. D\\
\end{center}
\vspace{\fill}

{\footnotesize
\centerline{\copyright~CERN on behalf of the \lhcb collaboration, license \href{http://creativecommons.org/licenses/by/3.0/}{CC-BY-3.0}.}}
\vspace*{2mm}






\end{titlepage}

\newpage
\setcounter{page}{2}
\mbox{~}

\centerline{\large\bf LHCb collaboration}
\begin{flushleft}
\small
R.~Aaij$^{41}$,
B.~Adeva$^{37}$,
M.~Adinolfi$^{46}$,
A.~Affolder$^{52}$,
Z.~Ajaltouni$^{5}$,
J.~Albrecht$^{9}$,
F.~Alessio$^{38}$,
M.~Alexander$^{51}$,
S.~Ali$^{41}$,
G.~Alkhazov$^{30}$
% ...

{\footnotesize \it
$ ^{1}$Centro Brasileiro de Pesquisas F\'{i}sicas (CBPF), Rio de Janeiro, Brazil\\
% ...
$ ^{5}$Clermont Universit\'{e}, Universit\'{e} Blaise Pascal, CNRS/IN2P3, LPC, Clermont-Ferrand, France\\
% ...
$ ^{9}$Fakult\"{a}t Physik, Technische Universit\"{a}t Dortmund, Dortmund, Germany\\
% ...
$ ^{30}$Petersburg Nuclear Physics Institute (PNPI), Gatchina, Russia\\
% ...
$ ^{37}$Universidad de Santiago de Compostela, Santiago de Compostela, Spain\\
$ ^{38}$European Organization for Nuclear Research (CERN), Geneva, Switzerland\\
% ...
$ ^{41}$Nikhef National Institute for Subatomic Physics, Amsterdam, The Netherlands\\
% ...
$ ^{46}$H.H. Wills Physics Laboratory, University of Bristol, Bristol, United Kingdom\\
% ...
$ ^{51}$School of Physics and Astronomy, University of Glasgow, Glasgow, United Kingdom\\
$ ^{52}$Oliver Lodge Laboratory, University of Liverpool, Liverpool, United Kingdom\\
% ...
}
\end{flushleft}


\cleardoublepage

\renewcommand{\thefootnote}{\arabic{footnote}}
\setcounter{footnote}{0}



\pagestyle{plain} 
\setcounter{page}{1}
\pagenumbering{arabic}



%
\clearpage

\renewcommand{\thefootnote}{\arabic{footnote}}
\setcounter{footnote}{0}










\section{Introduction}
\label{sec:Introduction}
\begin{figure}[h]
\vskip -.4cm
\begin{center}
\includegraphics[width=3.2in]{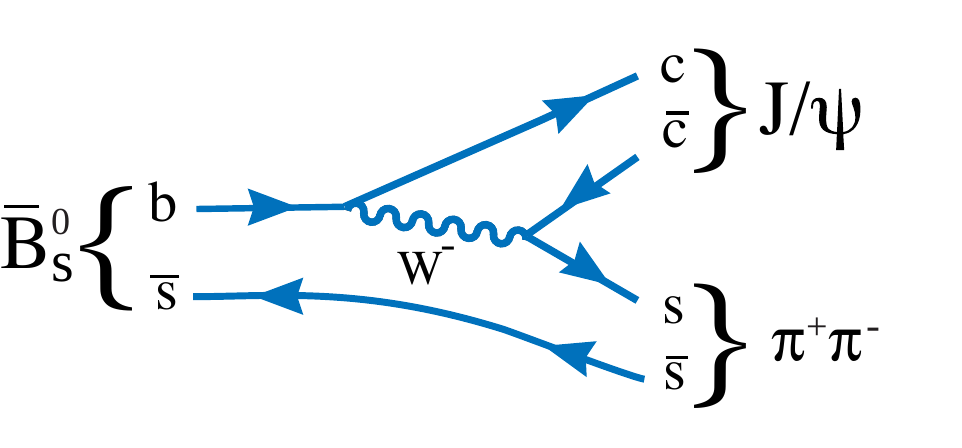}
\end{center}\label{feyn1}
\vskip -.5cm
\caption{Leading order diagram for $\Bsb$ decays into $J/\psi \pi^+\pi^-$.}
\end{figure}

\CP violation studies in the $\Bsb \to \jpsi \pi^+\pi^-$ decay mode complement studies using $\Bsb \to \jpsi \phi$ and improve the final accuracy in the \CP-violating phase, $\phi_s$, measurement \cite{Aaij:2013oba}.  While the \CP content was previously shown to be more than 97.7\% \CP-odd at 95\% confidence level (CL), it is important to determine the size of any \CP-even components as these could ultimately affect the uncertainty on the final result for $\phi_s$. Since the $\pi^+\pi^-$ system can form light scalar mesons, such as  the $f_0(500)$ and $f_0(980)$, we can investigate if these states have a quark-antiquark or tetraquark structure, and  determine the mixing angle between these states \cite{Stone:2013eaa}.  The tree-level Feynman diagram for the process is shown in Fig.~\ref{feyn1}.

We have previously studied the resonance structure in $\Bsb \to \jpsi \pi^+\pi^-$ decays using  data corresponding to an integrated luminosity of 1~fb$^{-1}$  \cite{LHCb:2012ae}.\footnote{ Charged conjugated modes are also used when appropriate.} In this paper we use 3~fb$^{-1}$ of luminosity, and also change the analysis technique substantially. 
Here the $\pi^+\pi^-$ mass, and all three decay angular distributions are used to determine the resonant and non-resonant components. Previously the angle between the decay planes of $\jpsi\to\mu+\mu^-$ and $\pi^+\pi^-$ in the $\Bsb$ rest frame, $\chi$, was integrated over. This simplified the analysis, but sacrificed some precision and also prohibited us from measuring separately the helicity $+1$ and $-1$ components of any $\pi^+\pi^-$ resonance, knowledge of which would permit us to evaluate the  \CP composition of resonances with spin greater than or equal to 1.
Since one of the particles in the final state, the $J/\psi$, has spin-1  its three decay amplitudes must be considered, while the  $\pi^+\pi^-$ system is described as the coherent sum of  resonant and possibly non-resonant amplitudes.


\section{Amplitude formula for $\Bsb \to \jpsi h^+ h^-$}
\begin{figure}[!t]
  \centering
\includegraphics[width=0.9\textwidth]{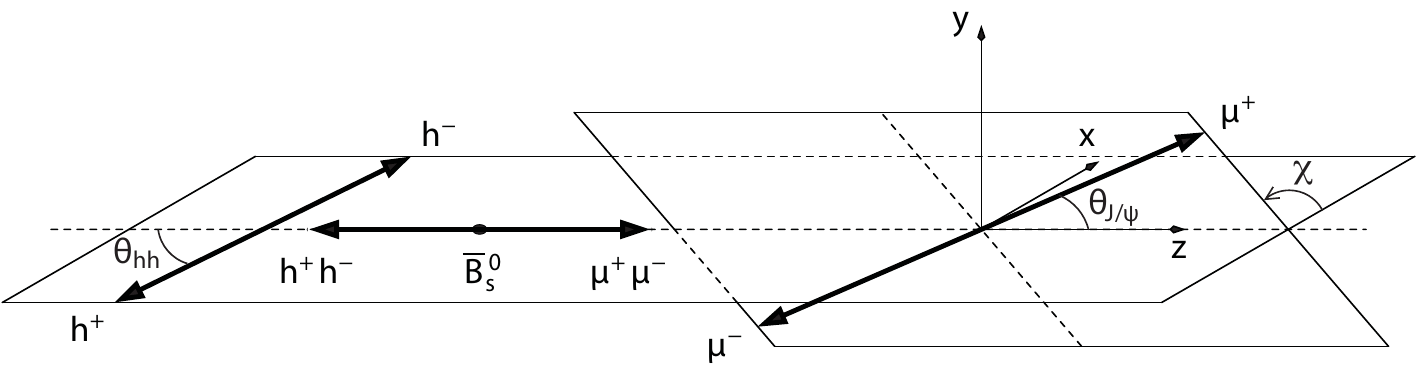}
  \caption{Definition of helicity angles. For details see text.}
  \label{fig:helicityAngles}
\end{figure}

The decay of $\Bsb\to \jpsi h^+ h^-$, where $h$ denotes a pseudoscalar meson, followed by $J/\psi\rightarrow \mu^+\mu^-$ can be described by four variables.
We take the invariant mass of $h^+h^-$ ($m_{hh}$) and three helicity angles defined as
(i) $\angmu$, the angle between the $\mu^+$ direction in the $\jpsi$ rest frame with respect to the $\jpsi$ direction in the $\Bsb$ rest frame;  (ii) $\angpi$, the angle between the $h^+$ direction in the $h^+h^-$ rest frame with respect to the $h^+h^-$ direction in the $\Bsb$ rest frame,
and (iii) $\chi$, the angle between the $\jpsi$ and $h^+h^-$ decay planes in the $\Bsb$ rest frame. Figure~\ref{fig:helicityAngles} shows these angles pictorially\footnote{These definitions are the same for $\Bq$ and $\Bqb$, namely,  $\mu^+$ and $h^+$  are used to define the angles in both cases.}. In this paper $hh$ is equivalent to $\pi^+\pi^-$.

From the time-dependent decay rate of $\brabar{B}_s^0\to \jpsi h^+h^-$ derived in Ref.~\cite{Zhang:2012zk}, the time-integrated and flavor-averaged decay rate is proportional to the function
\begin{align}\label{Eq:SPDF}
S(m_{hh}, \angpi, \angmu, \chi)=&|A(m_{hh}, \angpi, \angmu, \chi)|^2+|\Ab(m_{hh}, \angpi, \angmu, \chi)|^2\nonumber\\
&-2{\cal D}\,\Real\left(\frac{q}{p}A^*(m_{hh}, \angpi, \angmu, \chi)\Ab(m_{hh}, \angpi, \angmu, \chi)\right),
\end{align}
where $\brabar{A}$, the amplitude of $\brabar{B}_s^0\to \jpsi h^+h^-$ at proper time $t=0$, is a function of $m_{hh},\angmu,\angpi,\chi$, and is summed over all resonant (and possibly non-resonant) components; $q$ and $p$ are complex parameters that describe the relation between mass and flavor eigenstates \cite{Bigi:2000yz}. The interference term arises because we must sum the \Bsb and \Bs amplitudes before squaring. Even when integrating over proper time, the terms proportional to $\sinh\left(\DGs t/2\right)$ do not vanish because of the finite $\DGs$ in the \Bs system, where $\DGs$ is the width difference between the light and the heavy mass eigenstates. The factor $\cal D$  is
\begin{equation}
{\cal D} = \frac{\int_{0}^{\infty}{\varepsilon}(t) e^{-\Gs t}\sh \dv t}{\int_{0}^{\infty}{\varepsilon}(t)e^{-\Gs t}\ch \dv t},
\end{equation}
where $\Gs$ is the average $\Bs$ decay width, and ${\varepsilon}(t)$ is the detection efficiency as a function of $t$. For a uniform efficiency, ${\cal D}= \DGs/(2\Gs)$ and is $(6.2\pm0.9)$\%~\cite{PDG}.

The amplitude, $A_R(m_{hh})$, is used to describe the mass line-shape of the resonance $R$, that in most cases is a Breit-Wigner function. It is combined with the $\Bb$  resonance decay properties to form the expression
\begin{equation}\label{eq:DP}
{\cal A}_R(\m)=\sqrt{2J_R+1} \sqrt{P_R P_B}  F_B^{(L_B)} F_R^{(L_R)} A_R(m_{hh})\left(\frac{P_B}{m_B}\right)^{L_B}\left(\frac{P_R}{m_{hh}}\right)^{L_R}.
\end{equation}
Here $P_B$ is the \jpsi momentum in the $\overline{B}^0_s$ rest frame, $P_R$ is the momentum of either of the two hadrons in the dihadron rest frame, $m_{B}$ is the $\overline{B}^0_s$ mass, $J_R$ is the spin of $R$, $L_B$ is the orbital angular momentum between the $J/\psi$ and $h^+h^-$ system, and $L_R$ the orbital angular momentum in the $h^+h^-$ decay, and thus is the same as the spin of the $h^+h^-$ resonance. $F_B^{(L_B)}$ and $F_R^{(L_R)}$ are the Blatt-Weisskopf barrier factors for the $\overline{B}^0_s$ and $R$ resonance, respectively~\cite{LHCb:2012ae}. The factor $\sqrt{P_RP_B}$ results from converting the phase space of  the natural Dalitz-plot variables $m^2_{hh}$ and  $m^2_{\jpsi h^+}$ to that of $m_{hh}$ and $\cos\angpi$ \cite{Dalitz:1953cp}. We must sum over all final states, $R$, so for each $\jpsi$ helicity,  denoted by $\lambda$, equal to $0$, $+1$, and $-1$ we have

\begin{equation}
\label{eq:heart}
{\cal \brabar{H}}_{\lambda}(m_{hh},\angpi) = \sum_R  \brabar{\bf h}_\lambda^R {\cal A}_R(\m) d_{-\lambda,0}^{J_R}(\angpi),
\end{equation}
where $\brabar{\bf h}_\lambda^R$ are the complex coefficients for each helicity amplitude and the Wigner $d$-functions are listed in Ref.~\cite{PDG}.

The decay rates, $|\brabar{A}(m_{hh},\angpi,\angmu,\chi)|^2$, and the interference term, $\hspace{-10mm}A^*(m_{hh},\angpi,\angmu,\chi)\Ab(m_{hh},\angpi,\angmu,\chi)$, can be written as functions of ${\cal \brabar{H}}_{\lambda}(m_{hh},\angpi)$, $\angmu$ and $\chi$. These relationships are given in Ref.~\cite{Zhang:2012zk}.
In order to use the \CP relations, it is convenient to replace the helicity complex coefficients $\brabar{\textbf{h}}_\lambda^R$ by the complex transversity coefficients $\brabar{\textbf{a}}_{\tau}^R$ using the relations
\begin{eqnarray}
\brabar{\textbf{h}}_0^R &=& \brabar{\textbf{a}}_{0}^R,\nonumber\\
\brabar{\textbf{h}}_+^R&=& \frac{1}{\sqrt{2}}(\brabar{\textbf{a}}_\parallel^R+\brabar{\textbf{a}}_\perp^R),\nonumber\\
\brabar{\textbf{h}}_-^R&=& \frac{1}{\sqrt{2}}(\brabar{\textbf{a}}_\parallel^R-\brabar{\textbf{a}}_\perp^R).
\end{eqnarray}
Here $\brabar{\textbf{a}}_{0}^R$ corresponds to longitudinal polarization of the $\jpsi$ meson, and the other two coefficients correspond to polarizations of the $\jpsi$ meson and $h^+h^-$ system transverse to the decay axis: $\brabar{\textbf{a}}_\parallel^R$ for parallel polarization of the $\jpsi$ and $h^+h^-$, and $\brabar{\textbf{a}}_\perp^R$ for  perpendicular polarization.

Assuming no direct \CP violation, as this has not been observed in $\Bsb\to\jpsi\phi$ decays \cite{Aaij:2013oba}, the relation between the $\overline{B}^0_s$ and $B^0_s$ variables is  $\bar{\textbf{a}}^R_\tau=\eta^R_\tau\textbf{a}^R_\tau$, where $\eta^R_\tau$ is \CP eigenvalue of the $\tau$  transversity component for the intermediate state $R$, where $\tau$ denotes $0$, $\parallel$, or $\perp$ component. 
The final state \CP parities for S, P,  and D-waves are given in Table~\ref{CPPart}.

\begin{table}[b]
\centering
\caption{\CP parity for different spin resonances. Note that spin-0 only has the transversity component $0$.}\label{CPPart}
\begin{tabular}{c|ccc}\hline
Spin& $\eta_0$& $\eta_\parallel$& $\eta_\perp$\\\hline
0 & $-1$ & ~-- &~--\\
1 & ~~1 & ~~1 & $-1$\\
2 & $-1$ & $-1$ & ~~1\\ \hline
\end{tabular}
\end{table}

In this analysis a fit determines the amplitude strength $a_\tau^R$ and the phase $\phi_\tau^R$ of the amplitude
 \begin{equation}\label{eq:amp}
\textbf{a}^R_\tau=a_\tau^R e^{i\phi_\tau^R}
\end{equation}
for each resonance $R$ and each transversity $\tau$. For the $\tau=\perp$ amplitude, the $L_B$ value of a spin-1 (or -2) resonance is 1 (or 2); the other transversity components have two possible $L_B$ values of 0 and 2 (or 1 and 3) for spin-1 (or -2) resonances. In this analysis the lower one is used. It is verified that our results are insensitive to the $L_B$ choices.

\section{Data sample and detector}
The data sample corresponds to an integrated luminosity of $3\,{\rm fb}^{-1}$ collected with the \lhcb detector~\cite{LHCb-det} using $pp$ collisions. One-third of the data was acquired at a center-of-mass energy of 7\tev, and the remainder at 8\tev. The detector is a single-arm forward
spectrometer covering the \mbox{pseudorapidity} range $2<\eta <5$,
designed for the study of particles containing \bquark or \cquark
quarks. The detector includes a high-precision tracking system
consisting of a silicon-strip vertex detector surrounding the $pp$
interaction region, a large-area silicon-strip detector located
upstream of a dipole magnet with a bending power of about
$4{\rm\,Tm}$, and three stations of silicon-strip detectors and straw
drift tubes~\cite{LHCb-DP-2013-003} placed downstream.
The combined tracking system provides a momentum\footnote{We work in units where $c = 1$.} measurement with
relative uncertainty that varies from 0.4\% at 5\gev to 0.6\% at 100\gev,
and impact parameter (IP) resolution of 20\mum for
tracks with large transverse momentum ($\pt$). Different types of charged hadrons are distinguished by information
from two ring-imaging Cherenkov detectors (RICH)~\cite{LHCb-DP-2012-003}. Photon, electron and
hadron candidates are identified by a calorimeter system consisting of
scintillating-pad and preshower detectors, an electromagnetic
calorimeter and a hadronic calorimeter. Muons are identified by a
system composed of alternating layers of iron and multiwire
proportional chambers~\cite{LHCb-DP-2012-002}.

The trigger consists of a hardware stage, based
on information from the calorimeter and muon systems, followed by a
software stage that applies a full event reconstruction~\cite{Aaij:2012me}. Events selected for this analysis are triggered by a $J/\psi\to\mu^+\mu^-$ decay, where the
$J/\psi$ is required at the software level to be consistent with coming from the decay of a $\Bsb$ meson by use either of IP requirements or detachment of the $J/\psi$ from the primary vertex (PV). In the simulation, $pp$ collisions are generated using
\pythia~\cite{Sjostrand:2006za,*Sjostrand:2007gs}
 with a specific \lhcb
configuration~\cite{LHCb-PROC-2010-056}.  Decays of hadronic particles
are described by \evtgen~\cite{Lange:2001uf}, in which final state
radiation is generated using \photos~\cite{Golonka:2005pn}. The
interaction of the generated particles with the detector and its
response are implemented using the \geant
toolkit~\cite{Allison:2006ve, *Agostinelli:2002hh} as described in
Ref.~\cite{LHCb-PROC-2011-006}.

\section{Event selection}\label{sec:mfit}
Preselection criteria are implemented to preserve a large fraction of the signal events, and are identical to those used  in Ref.~\cite{Aaij:2013zpt}. A $\Bsb \to \jpsi \pi^+\pi^-$ candidate is reconstructed by combining a $\jpsi \to \mu^+\mu^-$ candidate with two pions of opposite charge. To ensure good track reconstruction, each of the four particles in the $\Bsb$ candidate is required to have the track fit $\chi^2$/ndf to be less than 4, where ndf is the number of degrees of freedom of the fit. The $\jpsi\to \mu^+\mu^-$ candidate is formed by two identified muons of opposite charge, having $\pt$ greater than 500\,\mev, and with a geometrical fit vertex $\chi^2$ less than 16. Only candidates with dimuon invariant mass between $-48$\,\mev and $+43$\,\mev from  the  observed $\jpsi$ mass peak are selected, and are then constrained to the $\jpsi$ mass~\cite{PDG} for subsequent use.

Pion candidates are required to each have $\pt$ greater than 250\,\mev, and the sum, $\pt(\pi^+)+\pt(\pi^-)$ larger than 900\,\mev. Both pions must have $\chi^2_{\rm IP}$ greater than 9 to reject particles produced from the PV. The $\chi^2_{\rm IP}$ is
computed as the difference between the $\chi^2$ of the PV
reconstructed with and without the considered track. Both pions
must also come from a common vertex with $\chi^2{\rm /ndf}<16$, and form a vertex with the $\jpsi$ with a $\chi^2$/ndf less than 10 (here ndf equals five). Pion candidates are identified using the RICH and muon systems. The particle
identification makes use of the logarithm of the likelihood
ratio comparing two particle hypotheses (DLL). For pion
selection we require DLL$(\pi-K)>-10$ and DLL$(\pi-\mu)>-10$.

The $\Bsb$ candidate must have a flight distance of more than 1.5\,\mm. The angle between the combined momentum vector of the decay products and the vector formed from the positions of the PV and the decay vertex (pointing angle) is required to be less than $2.5^\circ$.

\begin{figure}[hbtp]
\begin{center}
\includegraphics[width=0.6\textwidth]{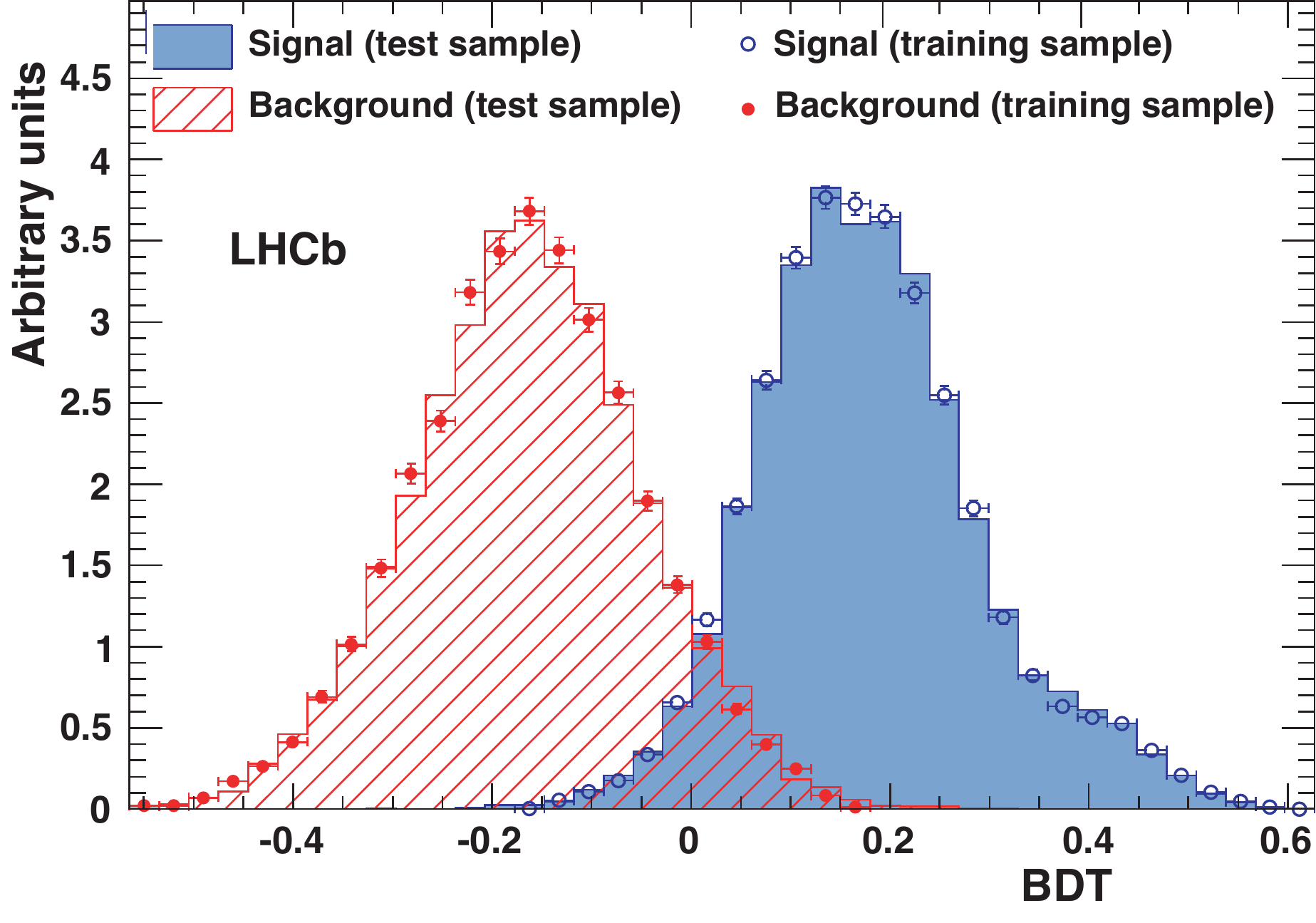}
\vskip -0.5cm
\caption{Distributions of the BDT classifier for
both training and test samples of $\jpsi\pi^+\pi^-$ signal and background
events. The signal samples are from simulation and the
background samples are from data.}
\end{center}\label{bdt}
\end{figure}

Events satisfying this preselection are then further filtered using a multivariate analyzer based on a boosted
decision tree (BDT) technique~\cite{Hocker:2007ht}. The BDT uses eight variables that
are chosen to provide separation between signal and background.
These are the minimum of DLL($\mu-\pi$) of the $\mu^+$ and $\mu^-$, $\pt(\pi^+)+\pt(\pi^-)$, the minimum of $\chi^2_{\rm IP}$ of the $\pi^+$ and $\pi^-$, and the $\Bsb$ properties of vertex $\chi^2$, pointing angle, flight distance, $\pt$ and $\chi^2_{\rm IP}$. The BDT is trained on a simulated sample of $\Bsb\to \jpsi \pi^+\pi^-$ signal events and a background data sample from the sideband $5566<m(J/\psi \pi^+\pi^-)< 5616$\,\mev. Then the BDT is tested on independent samples.
The distributions of BDT classifier for signal and background samples are shown in Fig.~\ref{bdt}.  By maximizing the signal significance we set the requirement that the classifier is greater than zero, which has a signal efficiency of 95\% and rejects 90\% of the  background.

\begin{figure}[t]
\begin{center}
\includegraphics[width=0.8\textwidth]{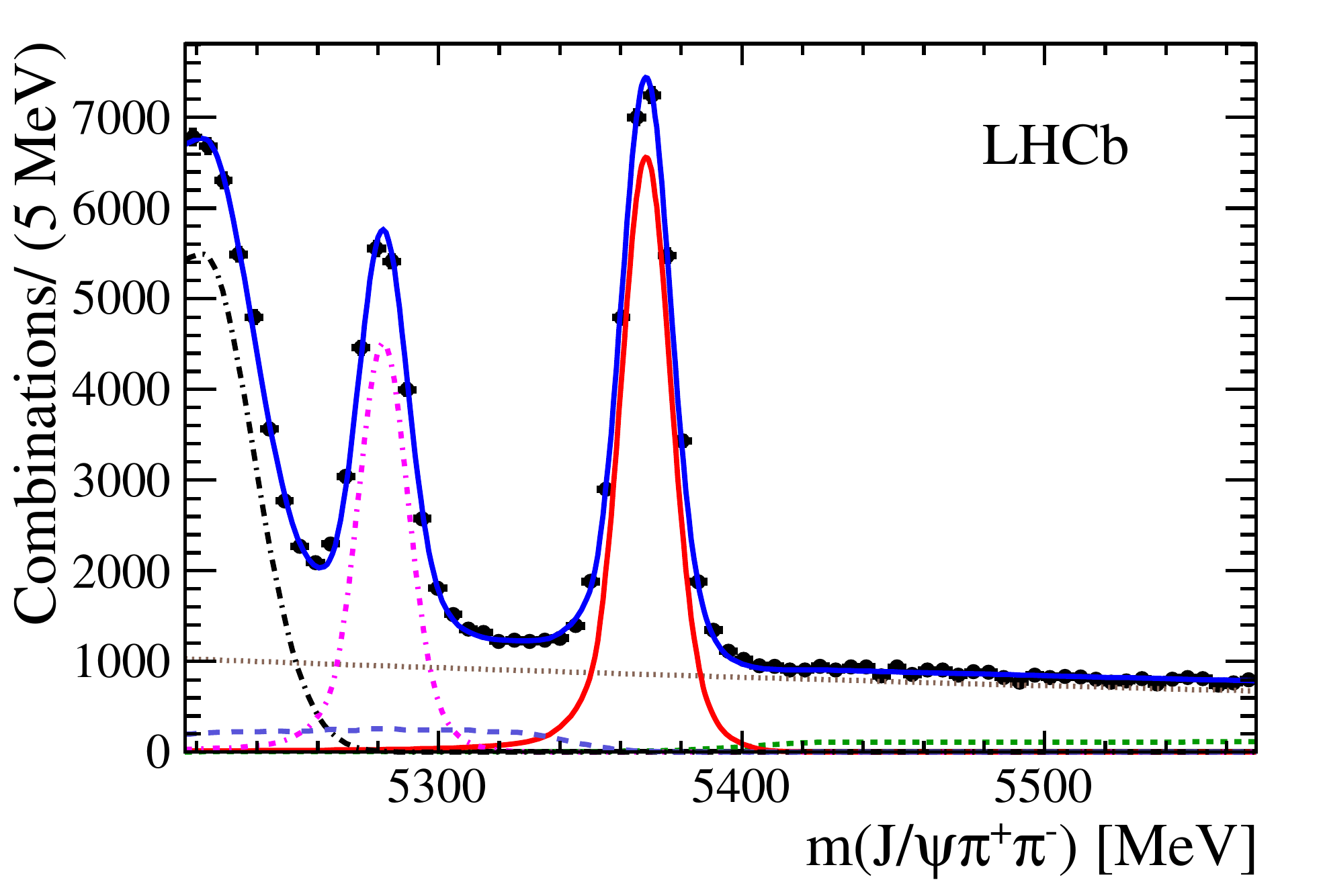}
\end{center}\label{fitmass}
\vskip -1cm
\caption{Invariant mass of $J/\psi \pi^+\pi^-$ combinations. The data have been fitted with double Crystal Ball signal and several background functions. The (red) solid curve shows the $\Bsb$ signal, the (brown) dotted line shows the combinatorial background, the (green) short-dashed line shows the $B^-$ background, the (purple) dot-dashed curve is $\Bzb\rightarrow J/\psi \pi^+\pi^-$, the (light blue) long-dashed line is the sum of $\Bsb\rightarrow J/\psi\eta'$, $\Bsb\rightarrow J/\psi\phi$ with  $\phi\to\pi^+\pi^-\pi^0$ backgrounds and the $\Lz_b^0 \to \jpsi K^- p$ reflection, the (black) dot-long dashed curve is the $\Bdb\rightarrow J/\psi K^- \pi^+$ reflection and the (blue) solid curve is the total.}
\end{figure}

The invariant mass of the selected $J/\psi\pi^+\pi^-$ combinations is shown in Fig.~\ref{fitmass}. There is a large peak at the $\Bsb$ mass and a smaller one at the $\Bzb$ mass on top of a background. A double Crystal Ball function with common means models the radiative tails and is used to fit each of the  signals. The known $\Bsb - \Bzb$ mass difference \cite{PDG} is used to constrain the difference in mean values.   Other components in the fit model take into account contributions from  $B^-\rightarrow J/\psi K^-(\pi^-)$, $\Bsb\rightarrow J/\psi\eta'$ with $\eta'\rightarrow \rho^0 \gamma$,
$\Bsb\rightarrow J/\psi\phi $ with $\phi\rightarrow \pi^+\pi^-\pi^0$ backgrounds and $\Bdb\rightarrow J/\psi K^- \pi^+$ and $\Lz_b^0\to \jpsi K^- p$ reflections, where the $K^-$ in the former, and both $K^-$ and $p$ in the latter, are misidentified as pions. The shape of the $\Bdb\rightarrow J/\psi \pi^+\pi^-$ signal is taken to be the same as that of the \Bsb. The combinatorial background shape is taken from like-sign combinations that are the sum of $\pi^+\pi^+$ and $\pi^-\pi^-$ candidates, and was found to be well described by an exponential function in previous studies \cite{LHCb:2012ae,Stone:2009hd}.
The shapes of the other components are taken from simulation with their yields allowed to vary. The $\Lz_b^0 \to \jpsi K^- p$ reflection yield in the fit region is constrained to the expected number $2145\pm201$, which is obtained from study of the events in the control region of $5066<m(J/\psi \pi^+\pi^-)< 5141$\,\mev. The mass fit gives $27396\pm 207$ signal and $7075\pm101$ background candidates,  leading to the signal fraction $f_{\rm sig}=(79.5\pm0.2)\%$, within $\pm20$ MeV of the $\Bsb$ mass peak.  The effective r.m.s. mass resolution is 9.9~MeV.

\section{Probability density function construction}\label{Formalism}
The correlated distributions of four variables $\m$,  $\cos \angpi$, $\cos \angmu$,  and $\chi$ are fitted using the candidates within $\pm20$ MeV of the $\Bsb$ mass peak. To improve the resolution of these variables we perform a kinematic fit constraining the $\Bsb$ and $J/\psi$ masses to their world average mass values \cite{PDG}, and recompute the final state momenta.

The overall PDF given by the sum of signal, $S$, and background functions is
\begin{eqnarray}\label{eq:pdf}
F(\m ,\angpi, \angmu, \chi)&=&\frac{f_{\rm sig}}{{\cal{N}}_{\rm sig}}\varepsilon(\m ,\angpi, \angmu, \chi) S(\m ,\angpi, \angmu, \chi)\nonumber\\&+&(1-f_{\rm sig}) B(\m ,\angpi, \angmu, \chi),
\end{eqnarray}
where  $\varepsilon$ is the detection efficiency, and $B$ is the background PDF discussed later in Sec.~\ref{sec:bkg}. 
The normalization factor for signal is given by
\begin{eqnarray}
{\cal{N}}_{\rm sig}&=&\int \! \varepsilon(m_{hh}, \angpi, \angmu, \chi) S(m_{hh}, \angpi, \angmu, \chi) \,
\dv \,\m \, \dv\cos\angpi \, \dv\cos\angmu \,\dv\chi.
\end{eqnarray}
The signal function $S$ is defined in Eq.~(\ref{Eq:SPDF}), where ${\cal D} = (8.7\pm1.5)\%$, taking into account the acceptance~\cite{LHCb:2012ad}, and choosing a phase convention $q/p=e^{-i\phi_s}$. The phase $\phi_s$ is fixed to the standard model value
of $-0.04$ radians~\cite{Charles:2011va}. Our results are found to be insensitive to the value of $\phi_s$ used within the 95\% CL limits set by the LHCb measurement  \cite{Aaij:2013oba}.

\subsection{Data distributions of the Dalitz-plot}
\begin{figure}[!t]
\begin{center}
\includegraphics[width=4.5 in]{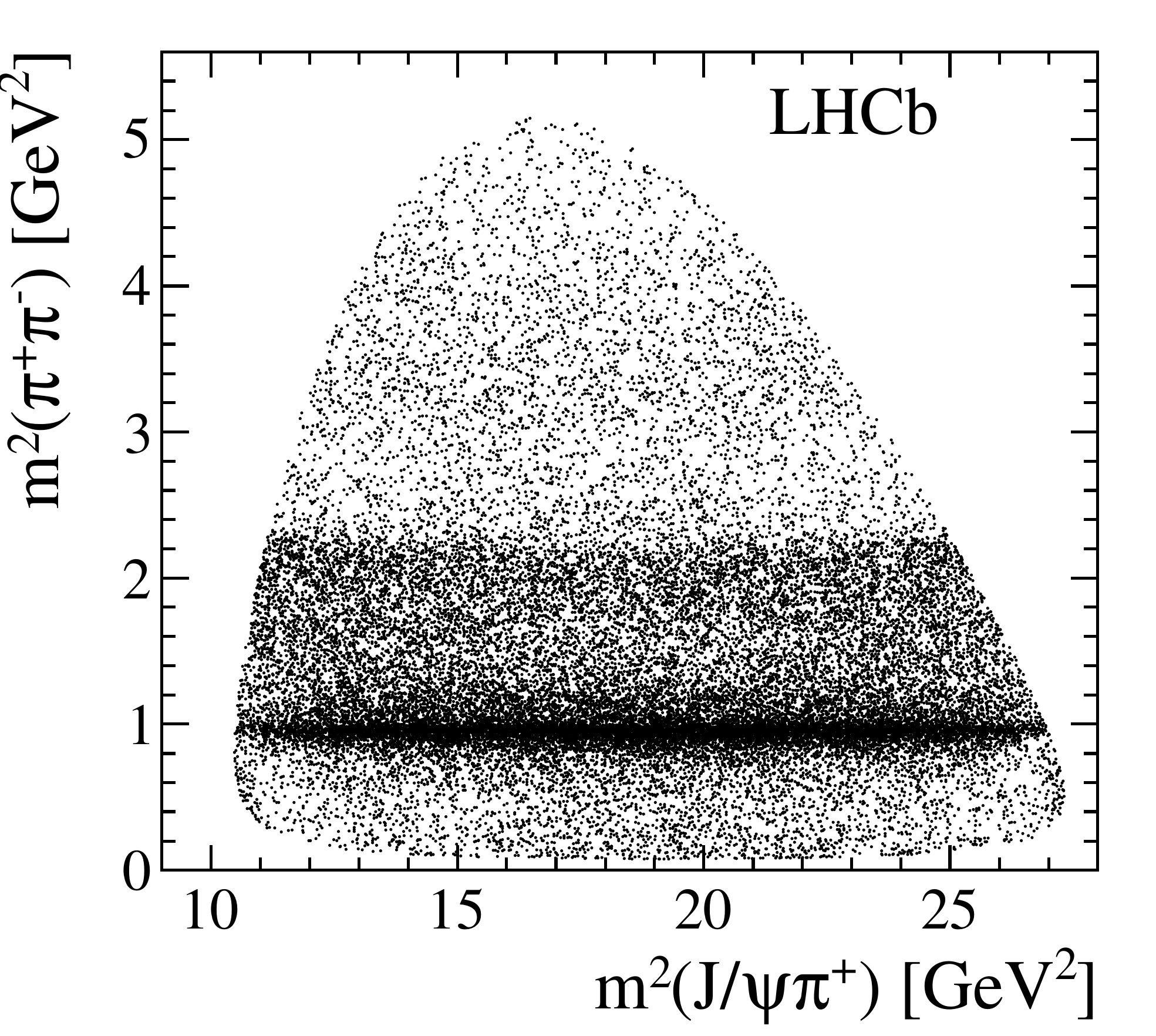}
\caption{Distribution of $m^2(\pi^+\pi^-)$ versus $m^2(J/\psi\pi^+)$ for all events within $\pm20$ MeV of the $\Bsb$ mass peak.}
\end{center}
\label{dalitz-1}
\end{figure}

\begin{figure}[b!]
\begin{center}
\includegraphics[width=0.5\textwidth]{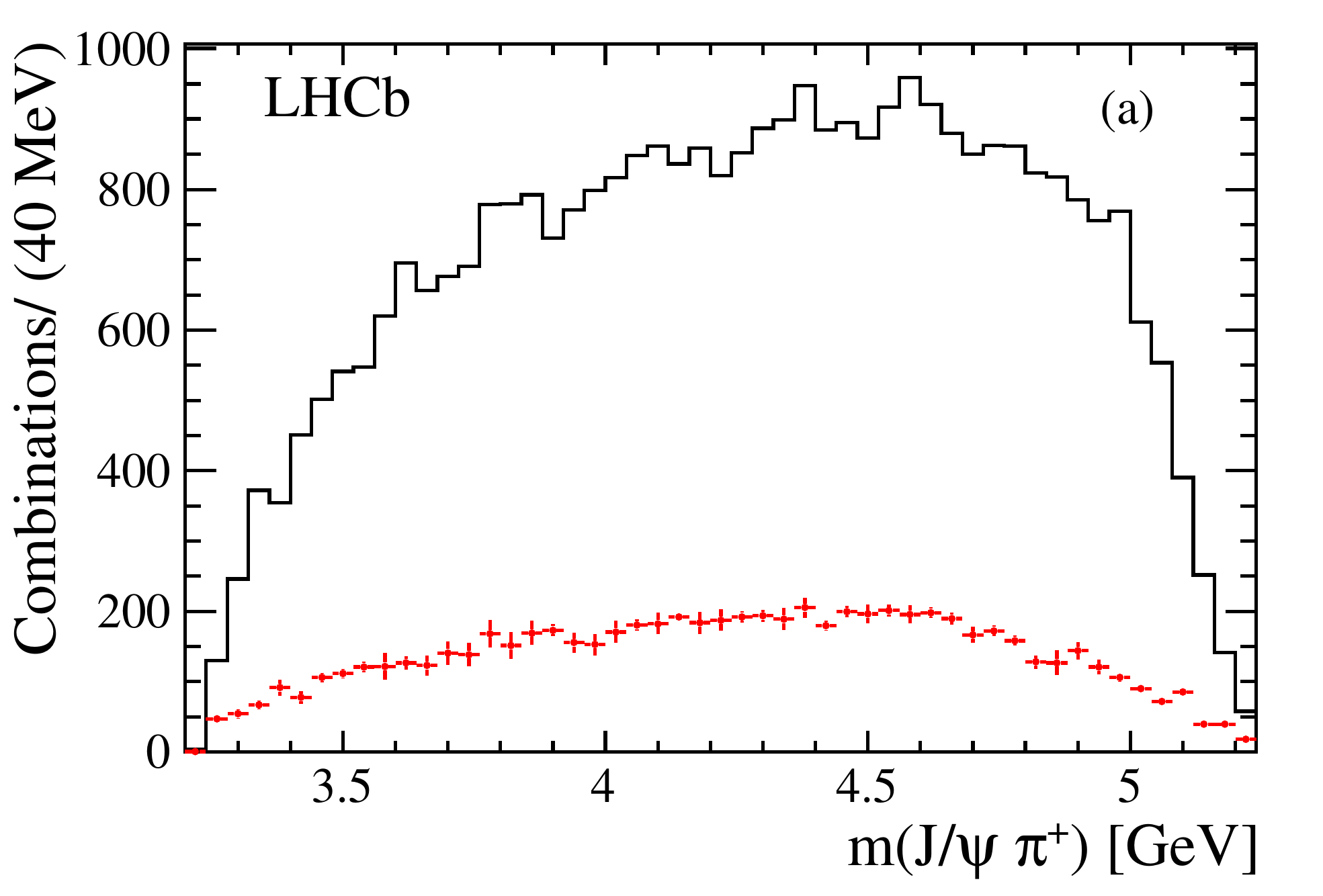}%
\includegraphics[width=0.5\textwidth]{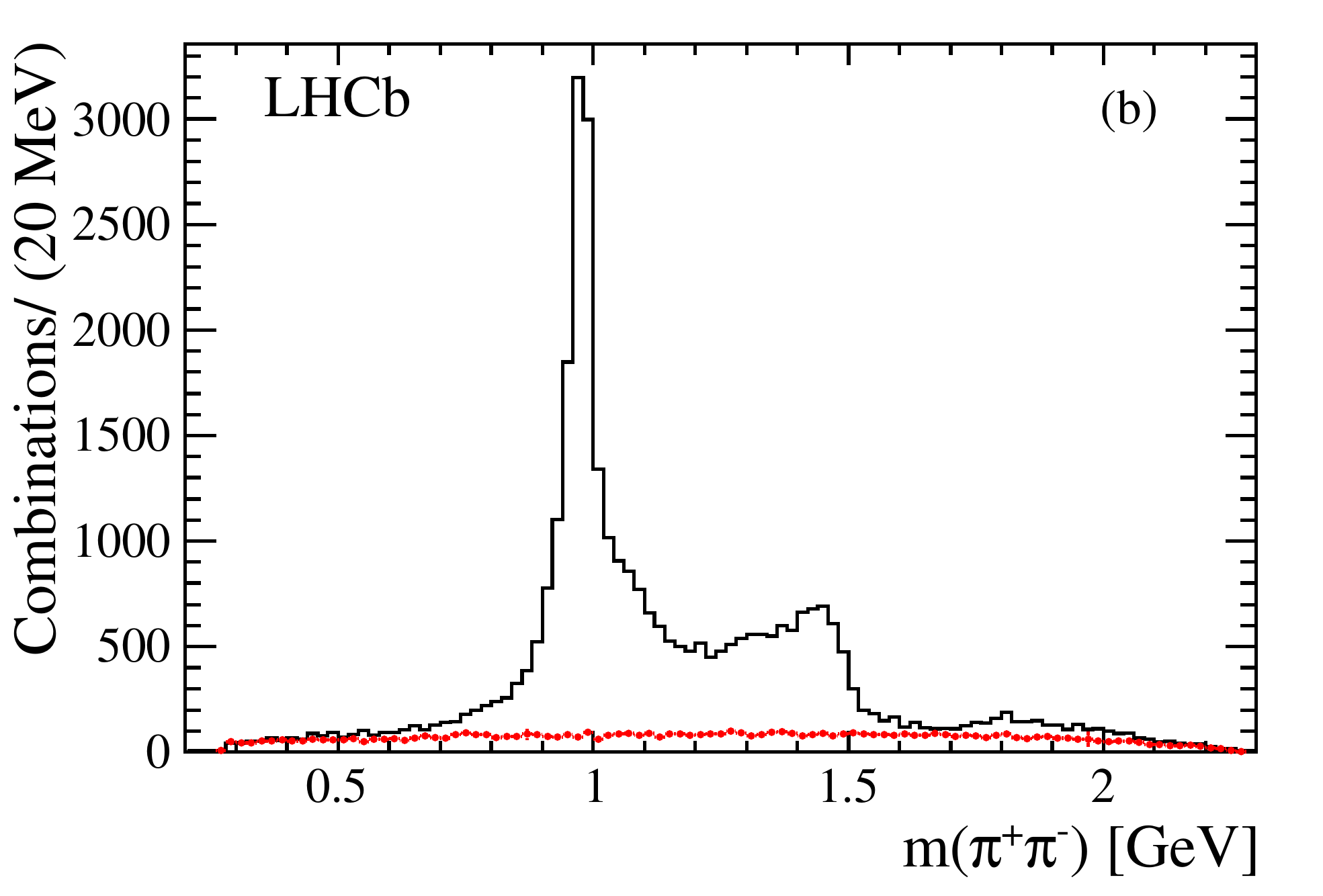}
\caption{Distributions of (a) $m(J/\psi \pi^+)$ and (b) $m(\pi^+\pi^-)$ for $\Bsb\to J/\psi \pi^+\pi^-$ candidate decays within $\pm20$ MeV of the $\Bsb$ mass. The (red) points with error bars show the background contribution determined from $m(J/\psi \pi^+\pi^-)$ fits performed in each bin of the plotted variables.}
\end{center}
\label{m-jpsipi}
\end{figure}

The event distribution for $m^2(\pi^+\pi^-)$ versus $m^2(J/\psi \pi^+)$ in Fig.~\ref{dalitz-1} shows
clear structures in $m^2(\pi^+\pi^-)$. The presence of possible exotic structures in  the $J/\psi\pi^+$ system, as claimed in similar decays \cite{Z4430,Ablikim:2013mio}, is investigated by examining the $J/\psi \pi^+$  mass distribution shown in Fig.~\ref{m-jpsipi} (a).  
No resonant effects are evident. Figure~\ref{m-jpsipi} (b) shows the $\pi^+\pi^-$ mass distribution. Apart from a large signal peak due to the  $f_0(980)$, there are visible structures at about 1450~\mev and 1800~\mev.

\subsection{Detection efficiency}
\label{sec:mc}
The detection efficiency is determined from a phase space simulation sample containing $4\times10^6$ $\Bsb\rightarrow J/\psi\pi^+\pi^-$ events with $J/\psi \rightarrow \mu^+\mu^-$. 
The efficiency can be parameterized in terms of analysis variables as
\begin{equation}
\varepsilon(m_{hh}, \angpi, \angmu, \chi)=\varepsilon_1(s_{12},s_{13})\times \varepsilon_2(m_{hh},\angmu)\times \varepsilon_3(m_{hh},\chi),\label{eq:eff}
\end{equation}
where $s_{12}\equiv m^2({\jpsi\pi^+})$ and $s_{13}\equiv m^2({\jpsi\pi^-})$ are functions of $(m_{hh}, \angpi)$; such parameter transformations in $\varepsilon_1$ are implemented in order to use the Dalitz-plot based efficiency model  developed in previous publications~\cite{LHCb:2012ae,Aaij:2013zpt}. The efficiency functions take into account correlations between $m_{hh}$ and each of the three angles as determined by the simulation.

\begin{figure}[b]
\centering
\includegraphics[width =0.55\textwidth]{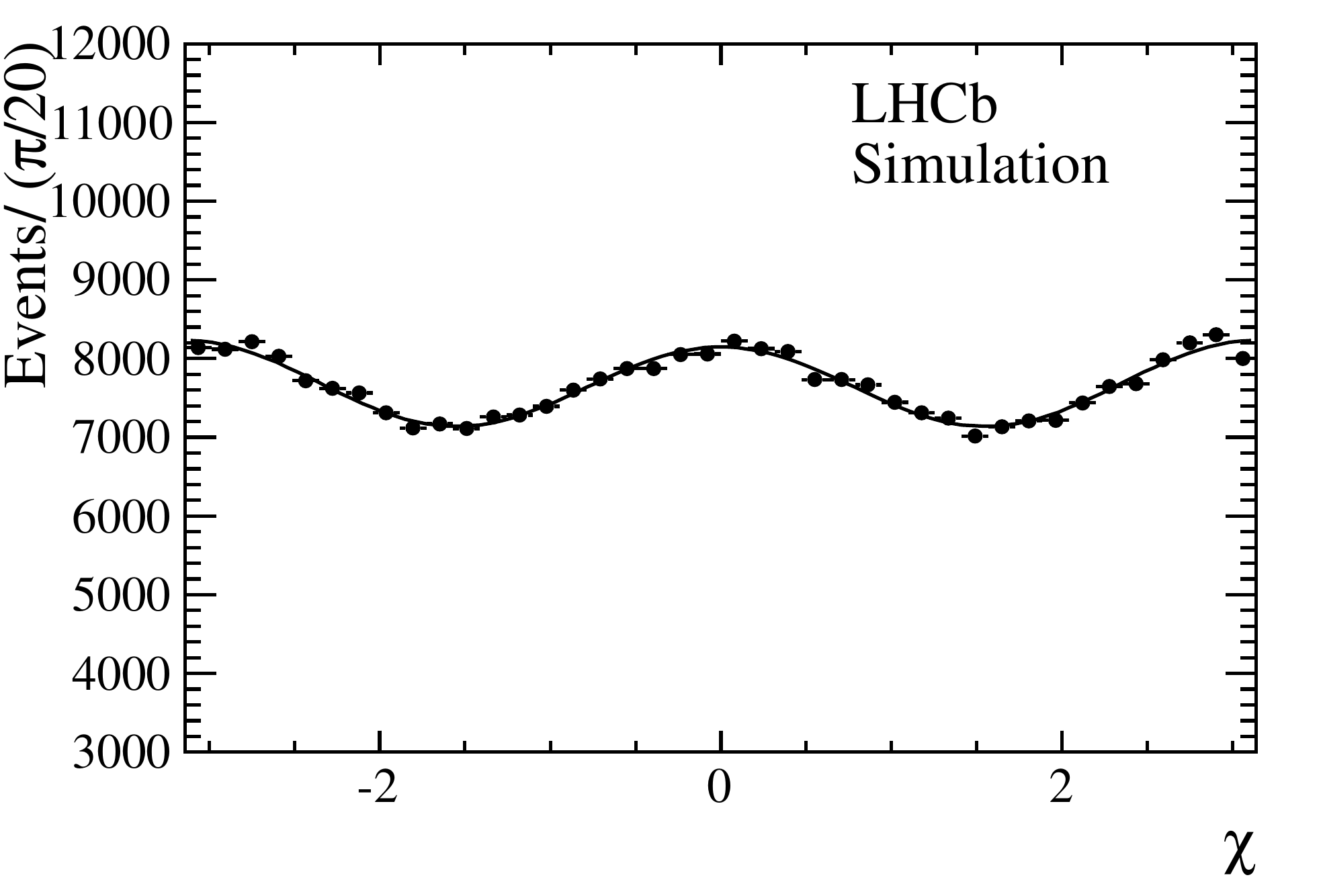}
\caption{Distribution of  the angle $\chi$  for the $J/\psi \pi^+\pi^-$ simulation sample fitted with Eq.~(\ref{Eq:chiacc}), used to determine the efficiency parameters.}
\label{fig:effchi}
\end{figure}

The efficiency as a function of the angle $\chi$ is shown in Fig.~\ref{fig:effchi}. To simplify the normalization of the PDF, the efficiency as a function of $\chi$ is parameterized in 26 bins of $m^2_{hh}$ as
\begin{equation}
\varepsilon_3(m_{hh},\chi)=\frac{1}{2\pi}(1+p_1\cos \chi+p_2\cos 2\chi),\label{Eq:chiacc}
\end{equation}
where $p_1=p_1^0+p_1^1 m_{hh}^2$ and $p_2=p_2^0+p_2^1m_{hh}^2+p_2^2m_{hh}^4$. A fit to the simulation determines $p_1^0=0.0087\pm0.0051$, $p_1^1=(-0.0062\pm0.0019)$\,GeV$^{-2}$, $p_2^0=0.0030\pm0.0077$, $p_2^1=(0.053\pm0.007)$\,GeV$^{-2}$, and $p_2^2=(-0.0077\pm0.0015)$\,GeV$^{-4}$.

\begin{figure}[!t]
\centering
\includegraphics[width =0.6\textwidth]{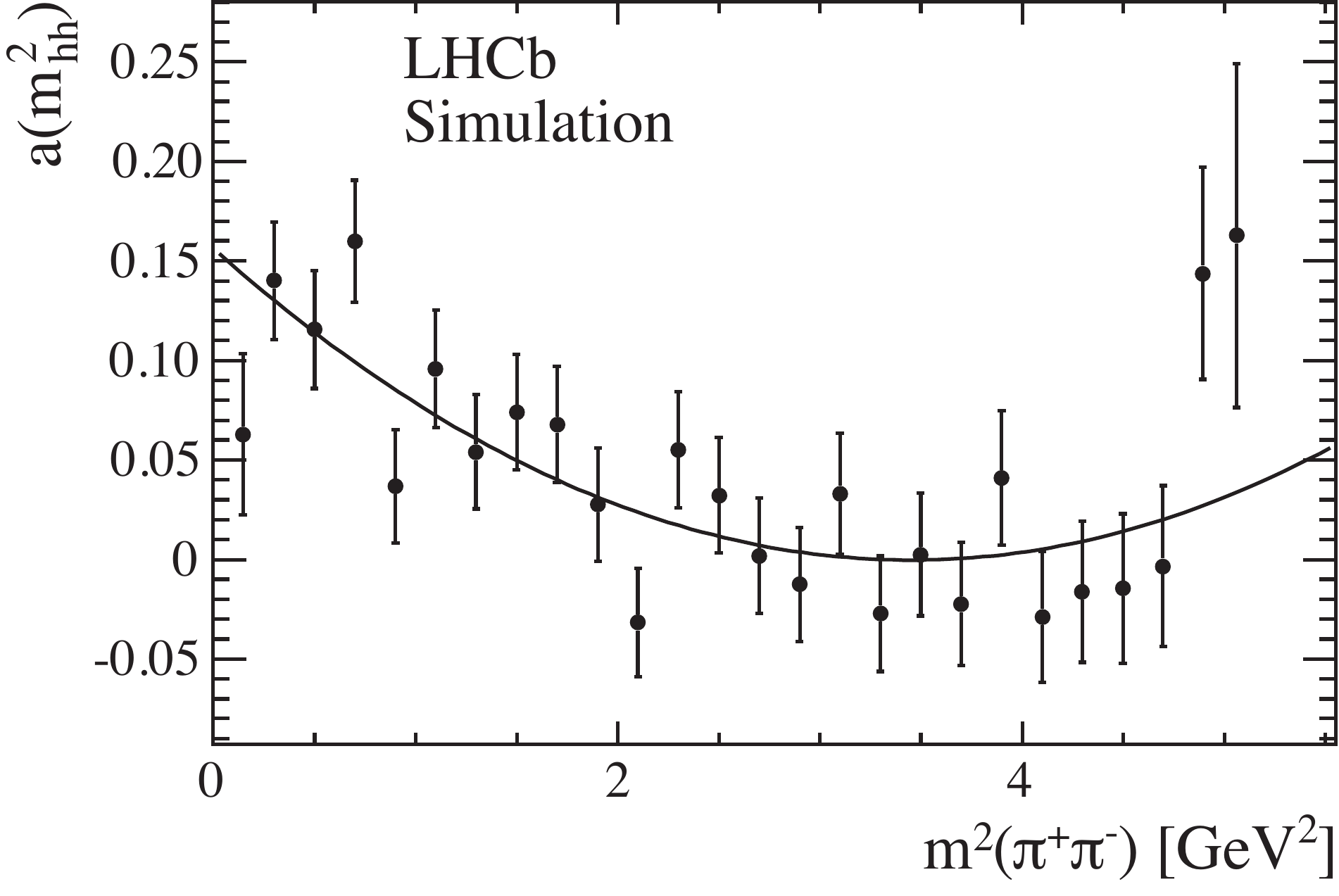}
\caption{Second order polynomial fit to the acceptance parameter $a(m^2_{hh})$ used in Eq.~\ref{eq:cosHacc}.}
\label{fig:cosHacc}
\end{figure}

The efficiency in $\cos \theta_{J/\psi}$ also depends on $m_{hh}$; we fit the $\cos \theta_{J/\psi}$ distributions of $J/\psi \pi^+\pi^-$ simulation sample with the function
\begin{equation}
\label{eq:cosHacc}
\varepsilon_2(m_{hh},\theta_{J/\psi})=\frac{1+ a(m^2_{hh})\cos^2\theta_{J/\psi}}{2+2a(m^2_{hh})/3},
\end{equation}
giving 26 values of $a$ as a function of $m^2_{hh}$. The resulting distribution in $a$ is shown in Fig.~\ref{fig:cosHacc} and is best described by a 2nd order polynomial function
\begin{equation}\label{eq:a}
a(m^2_{hh})= a_0+a_1m^2_{hh}+a_2m^4_{hh},
\end{equation}
with $a_0=0.156\pm0.020$, $a_1= (-0.091\pm0.018)$\,GeV$^{-2}$ and $a_2=(0.013\pm0.004)$\,GeV$^{-4}$.

\begin{figure}[hbtp]
\begin{center}
    \includegraphics[width=0.48\textwidth]{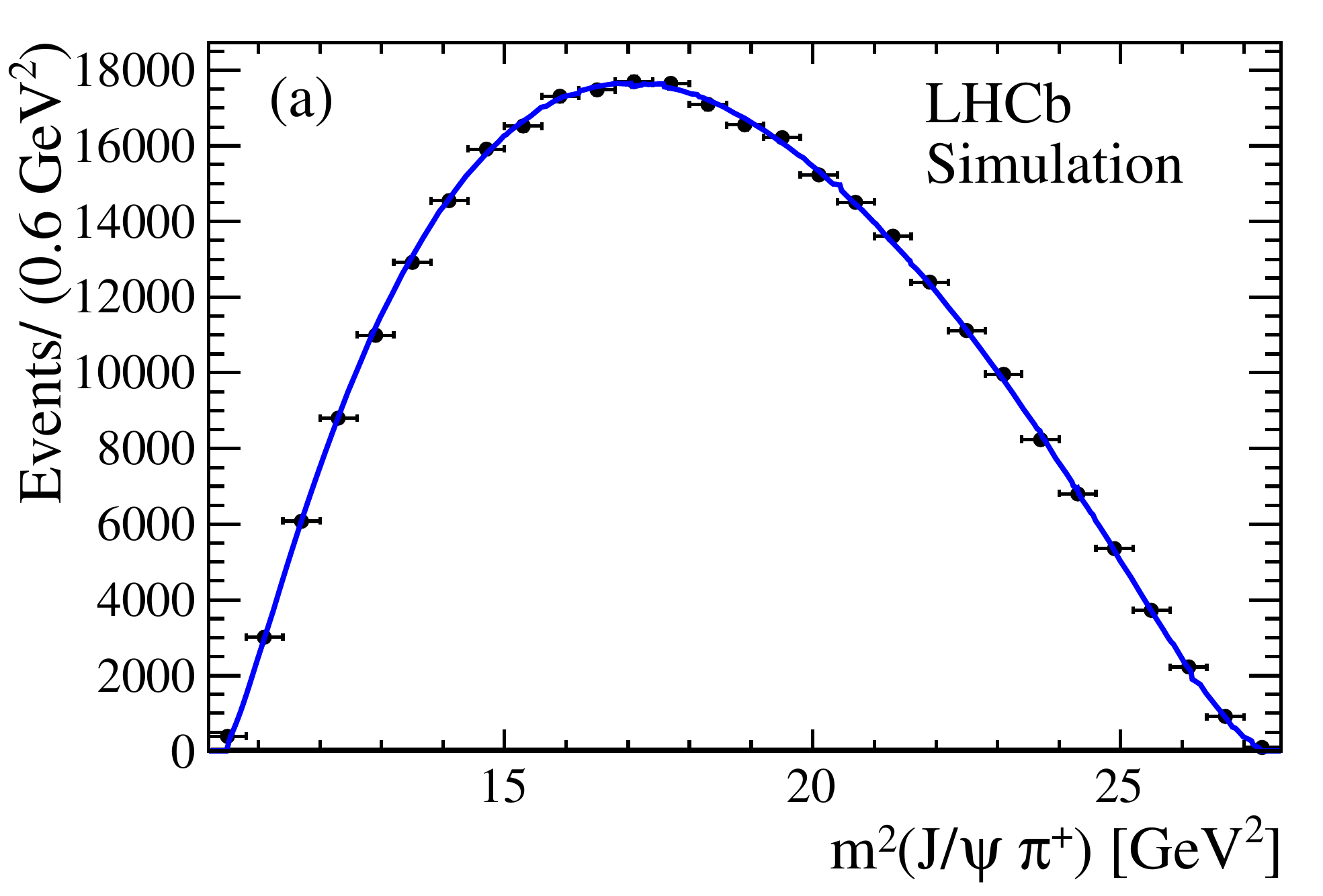}%
    \includegraphics[width =0.48\textwidth]{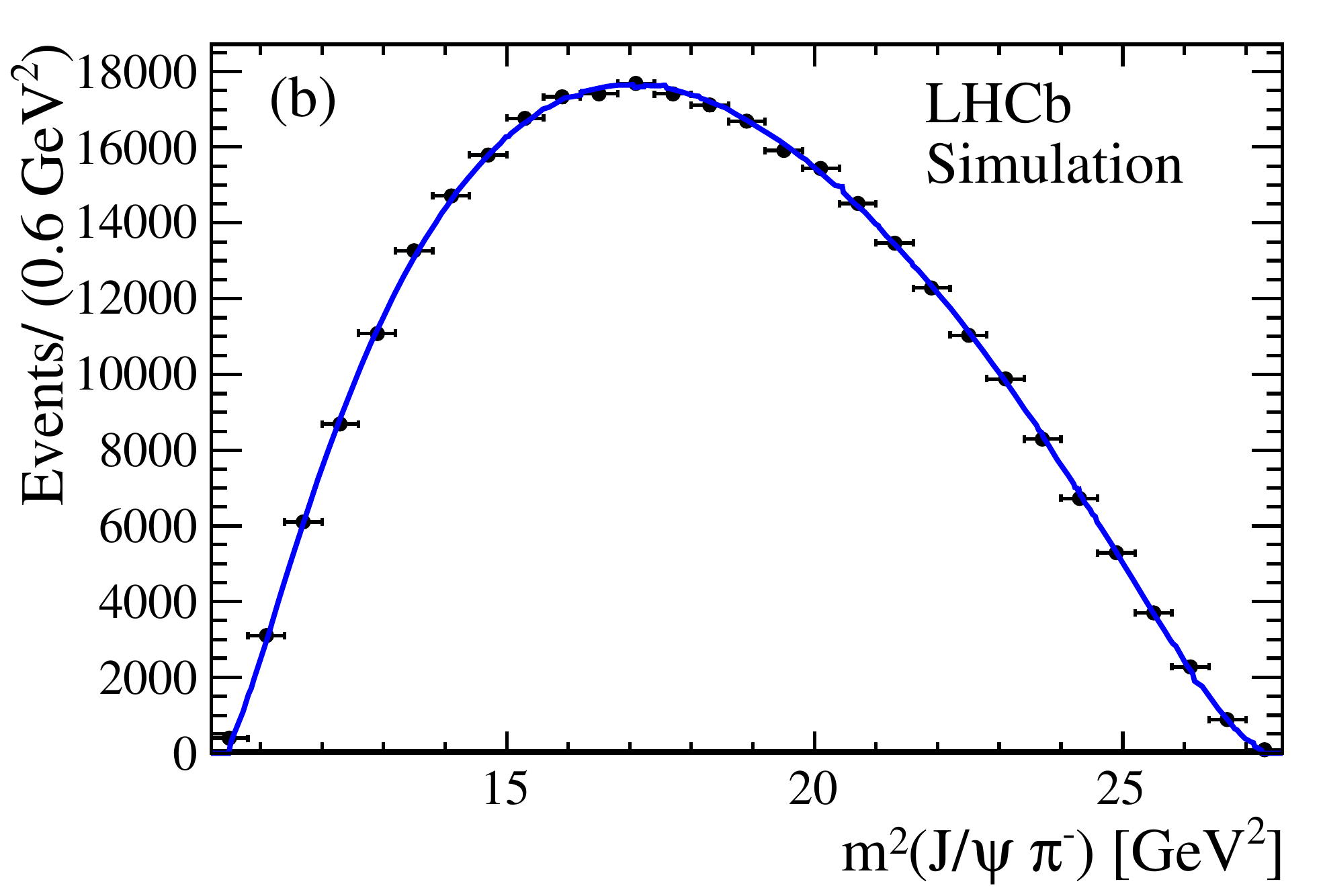}
\end{center}\label{eff2}
\vskip -0.5cm
\caption{Projections of invariant mass squared of (a) $m^2(J/\psi \pi^+)$ and (b) $m^2(J/\psi \pi^-)$ of the simulated Dalitz plot used to measure the efficiency parameters. The points represent the simulated event distributions and the curves the polynomial fit.}
\end{figure}

\begin{figure}[hbtp]
\begin{center}
     \includegraphics[scale=0.55]{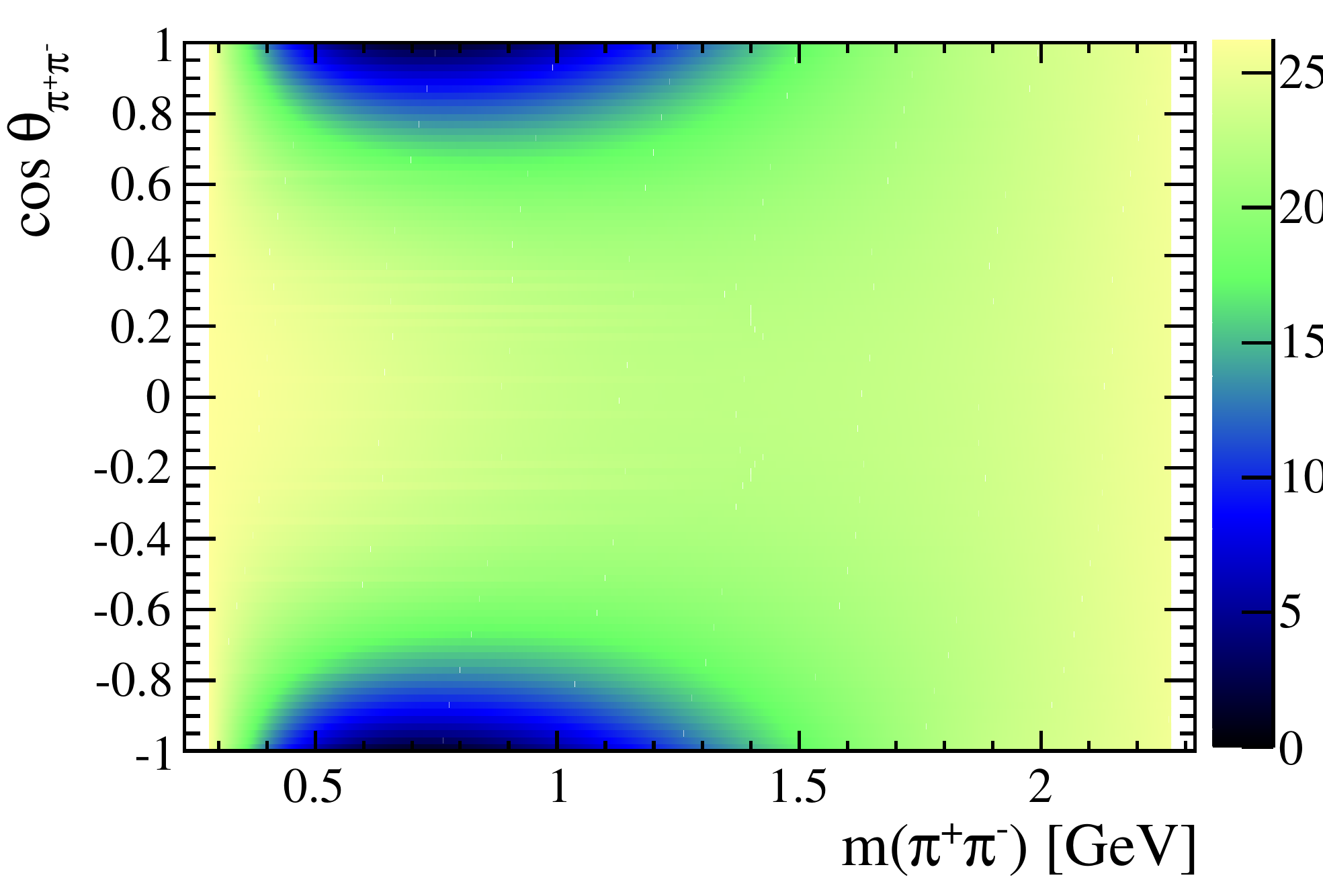}
\end{center}\label{eff1}
\vskip -0.5cm
\caption{Parameterization of the detection efficiency as a function
  of $\cos\theta_{\pi^+\pi^-}$ and $m(\pi^+\pi^-)$. The scale is arbitrary.}
\end{figure}

The function $\varepsilon_1(s_{12},s_{13})$ can be determined from the simulation after integrating over $\cos \theta_{J/\psi}$ and $\chi$, because the functions $\varepsilon_2$ and $\varepsilon_3$ are normalized in $\cos \theta_{J/\psi}$ and $\chi$, respectively. 
 It is parameterized as a symmetric 5th order polynomial function given by
\begin{eqnarray}
\varepsilon_1(s_{12},s_{13})&=& 1+\epsilon_1(x+y)+\epsilon_2(x+y)^2+\epsilon_3xy+\epsilon_4(x+y)^3
+\epsilon_5 xy(x+y)\nonumber \\
&&+\epsilon_6(x+y)^4+\epsilon_7 xy(x+y)^2+\epsilon_8 x^2y^2\nonumber \\&&+\epsilon_9(x+y)^5+\epsilon_{10} xy(x+y)^3+\epsilon_{11} x^2y^2(x+y),
\end{eqnarray}
where $x= s_{12}/{\rm GeV}^2-18.9$, and $y=s_{13}/{\rm GeV}^2-18.9$.
The phase space simulation is generated uniformly in the two-dimensional distribution of ($s_{12},s_{13})$, therefore the distribution of selected events reflects the efficiency and is fit to determine the efficiency parameters $\varepsilon_i$. The projections of the fit are shown in Fig.~\ref{eff2}, giving the efficiency as a function of $\cos\theta_{\pi^+\pi^-}$ versus $m(\pi^+\pi^-)$ in Fig.~\ref{eff1}. 

\subsection{Background composition}\label{sec:bkg}
The main background source is combinatorial and is taken from
the like-sign combinations within $\pm20$\,MeV of the $\Bsb$ mass peak. The like-sign combinations also contain the $\B^-$ background which is peaked at  $\cos\theta_{hh}=\pm1$.
The like-sign combinations cannot contain any $\rho^0$, which is measured to be 3.5\% of the total background. To obtain the $\rho^0$ contribution, the background $m(\pi^+\pi^-)$ distribution shown in Fig.~\ref{m-jpsipi} (b), found by fitting the $m(\jpsi\pi^+\pi^-)$ distribution in bins of $m(\pi^+\pi^-)$, is compared to $m(\pi^\pm\pi^\pm)$ distribution from the like-sign combinations. In this way simulated $\rho^0$ background is added into the like-sign candidates. 
The background PDF $B$ is the sum of functions for $B^-$ ($B_{B^-}$) and for the other ($B_{\rm other}$), given by
\begin{equation}\label{Eq:BkgF}
B(\m ,\angpi, \angmu, \chi)= \frac{1-f_{B^-}}{{\cal N_{\rm other}}}B_{\rm other}(\m ,\angpi, \angmu, \chi)+\frac{f_{B^-}}{{\cal N}_{B^-}}B_{B^-}(\m ,\angpi, \angmu, \chi),
\end{equation}
where ${\cal N}_{\rm other}$ and ${\cal N}_{B^-}$ are normalization factors, and $f_{B^-}$ is the fraction of the $\B^-$ background in the total background. The $J/\psi\pi^+\pi^-$ mass fit gives $f_{B^-}=(1.7\pm0.2)\%$.

The $\B^-$ background is separated because its invariant mass is very close to the highest allowed  limit, resulting in its $\cos \theta_{hh}$ distribution peaking at $\pm1$. The function for the $\B^-$ background is defined as
\begin{align}
B_{B^-}(m_{hh}, \angpi, \angmu, \chi) = &G(\m;m_0,\sigma_m)\times G(|\cos\angpi|;1,\sigma_\theta)\nonumber\\
 \times& \left(1-\cos^2\theta_{J/\psi}\right)\times (1+p_{b1}\cos\chi+p_{b2}\cos2\chi),
\end{align}
where $G$ is the Gaussian function, and the parameters $m_0$, $\sigma_m$, $\sigma_\theta$, $p_{b1}$, and $p_{b2}$ are determined by the fit. The last term is the same function for $\chi$.

The function for the other background is
\begin{equation}
B_{\rm other}(m_{hh}, \angpi, \angmu, \chi)=\m B_1(m_{hh}^2,\cos \theta_{hh})\times \left(1+\alpha\cos^2\theta_{J/\psi}\right)\times (1+p_{b1}\cos\chi+p_{b2}\cos2\chi),
\end{equation}
where the function
\begin{equation}
B_1(m_{hh}^2,\cos \theta_{hh})=B_2(\zeta)\frac{p_B}{m_B}\times\frac{1+c_1 q(\zeta)|\cos\theta_{hh}| +c_2p(\zeta)\cos^2 \theta_{hh}}{2[1+c_1q(\zeta)/2+c_2p(\zeta)/3]}.
\end{equation}
Here $\zeta\equiv 2(m_{hh}^2-m^2_{\rm min})/(m^2_{\rm max}-m^2_{\rm min})-1$, where $m_{\rm min}$ and $m_{\rm max}$ give the fit boundaries of $m_{hh}$, $B_2(\zeta)$ is a fifth-order Chebychev polynomial; $q(\zeta)$ and $p(\zeta)$ are both second-order Chebychev polynomials with the coefficients $c_1$ and $c_2$ being free parameters. In order to better approximate the real background in the $\Bsb$ signal region, the $\jpsi\pi^\pm\pi^\pm$ candidates are kinematically constrained to the $\Bsb$ mass, and $\mu^+\mu^-$ to the $\jpsi$ mass.

\begin{figure}[!b]
\begin{center}
\includegraphics[scale=0.5]{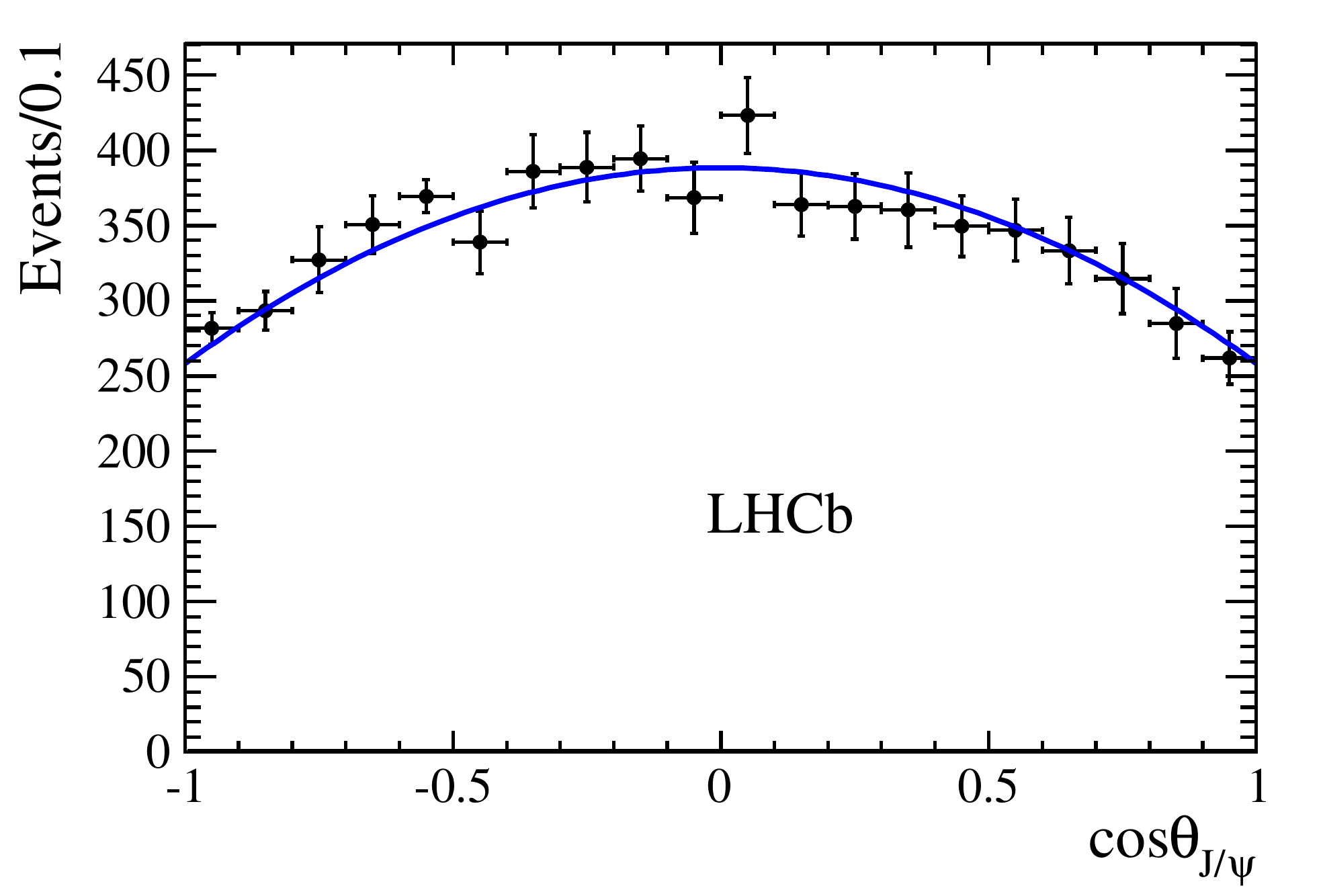}
\end{center}\label{bkg3}
\vskip -1cm
\caption{Distribution of $\cos\theta_{J/\psi}$ of the other background and the fitted function $1+\alpha\cos^2\theta_{J/\psi}$. The points with error bars show the background obtained from candidate mass fits in bins of $\cos\theta_{J/\psi}$.}
\end{figure}

\begin{figure}[htb]
\begin{center}
    \includegraphics[width=3 in]{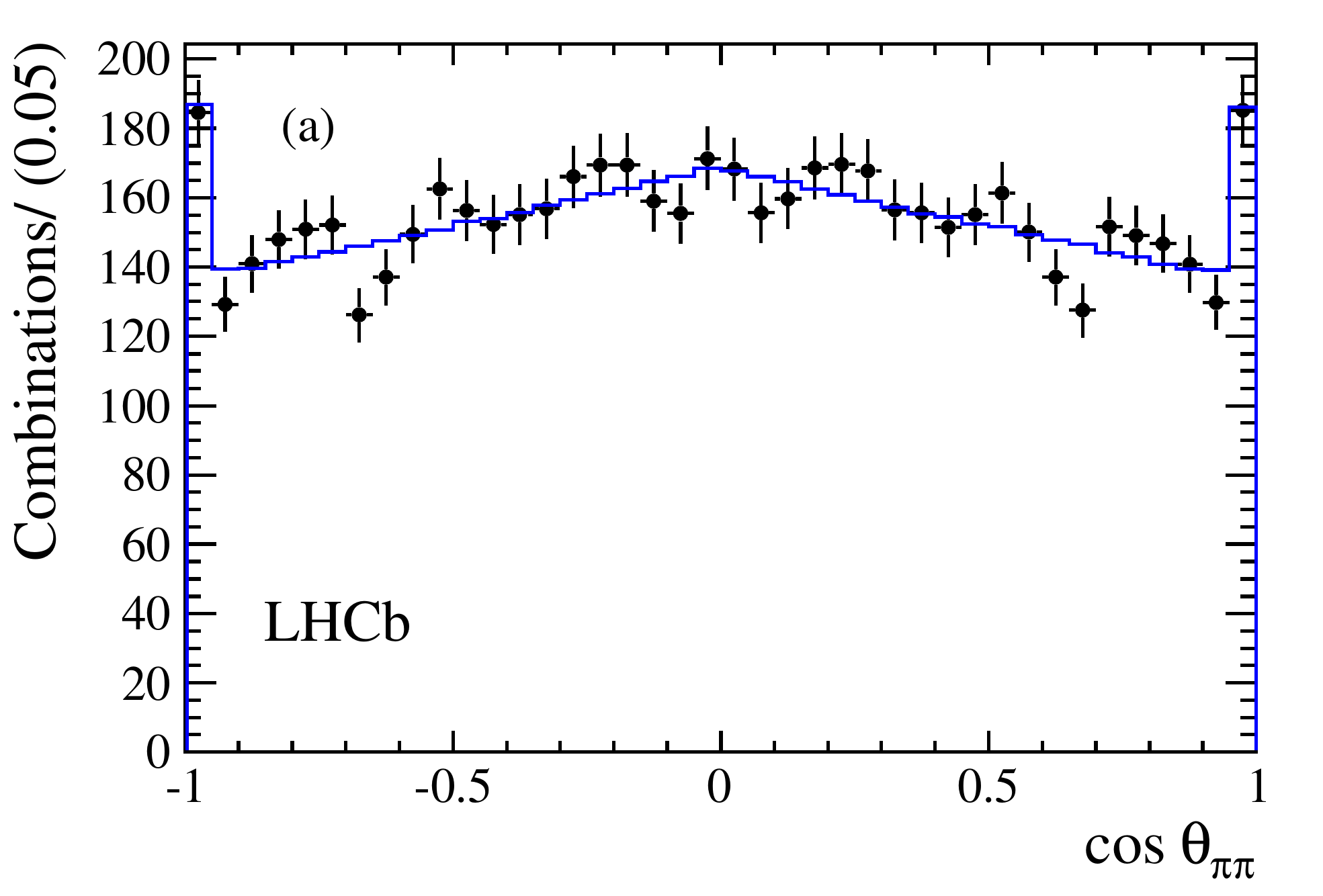}%
    \includegraphics[width=3 in]{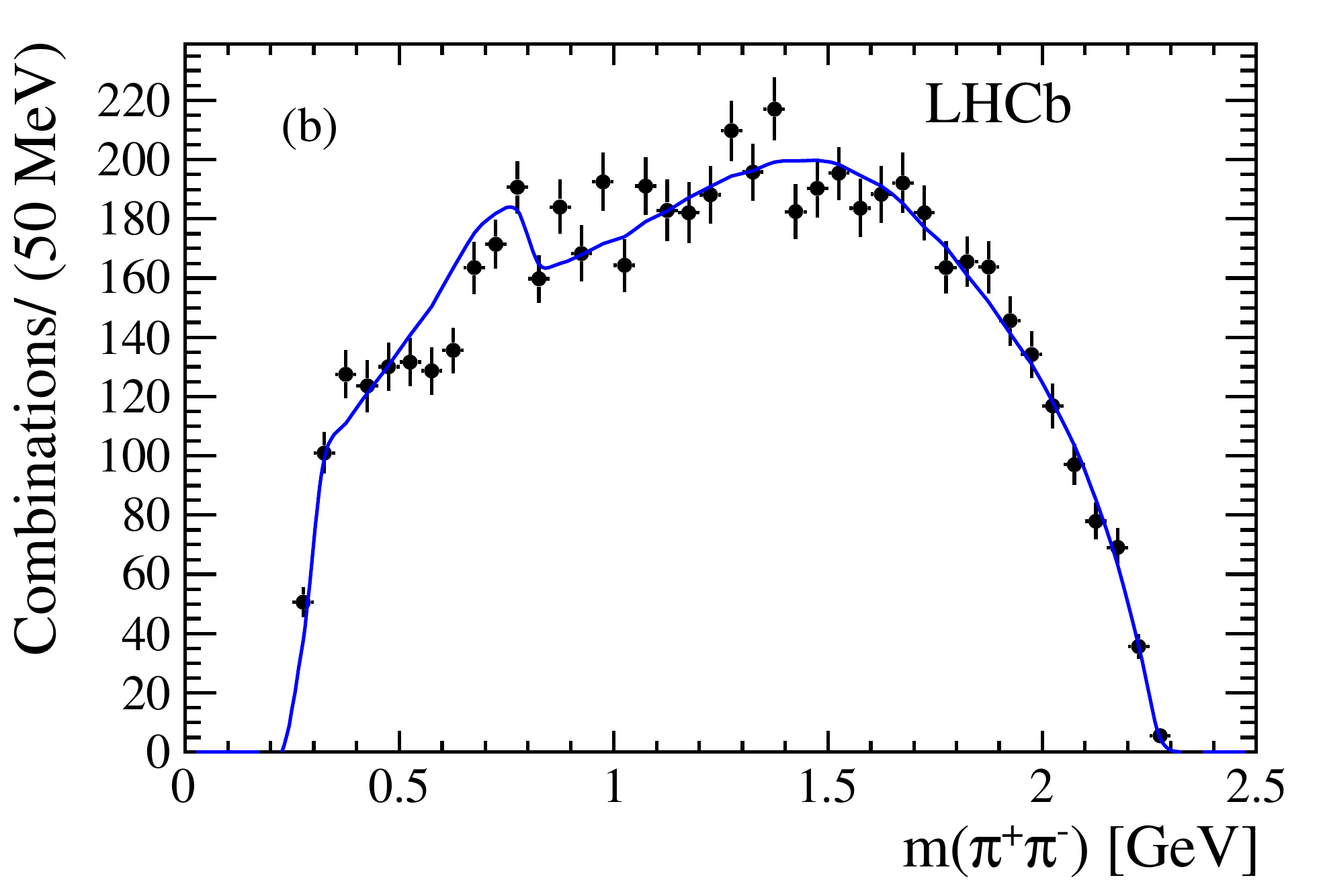}
\end{center}\label{bkg2}
\vskip -0.5cm
\caption{Projections of (a) $\cos\theta_{\pi\pi}$ and (b) $m(\pi^+\pi^-)$ of the total background. The (blue) histogram or curve is projection of the fit, and the points with error bars show the like-sign combinations added with additional $\rho^0$ background.}
\end{figure}

\begin{figure}[htb]
\begin{center}
\includegraphics[scale=0.5]{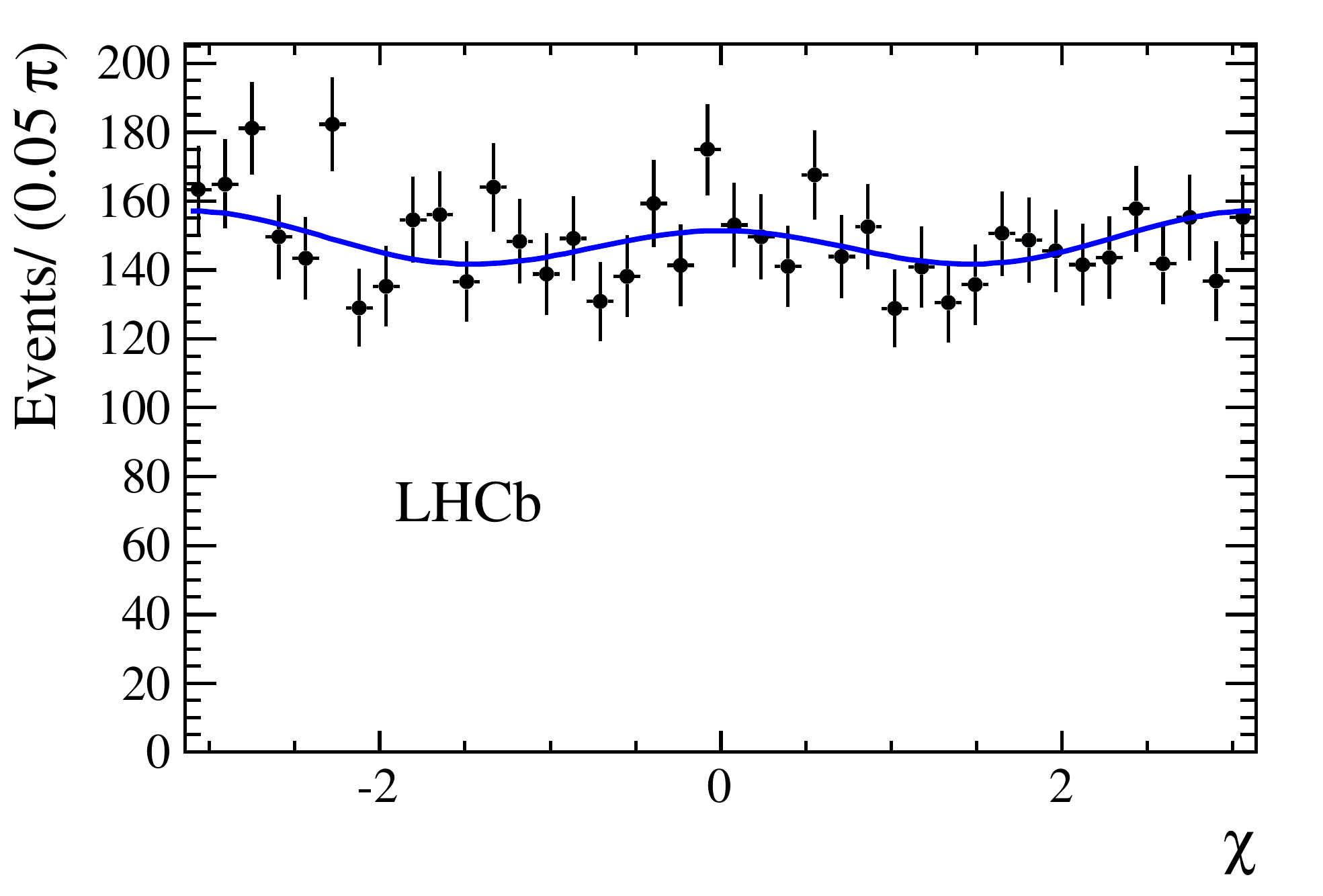}
\end{center}\label{bkg4}
\vskip -0.8cm
\caption{Distribution of $\chi$ of the total background and the fitted function. The points with error bars show the like-sign combinations added with additional $\rho^0$ background.}
\end{figure}

The second part $\left(1+\alpha \cos^{2}\theta_{J/\psi}\right)$ is a function of the $J/\psi$ helicity angle. The $\cos\theta_{J/\psi}$ distribution of background is shown in Fig.~\ref{bkg3}; fitting with the function determines the parameter $\alpha=-0.34\pm0.03$. A fit to the like-sign combinations added with additional $\rho^0$ background determines the parameters describing the $\m$, $\angpi$, and $\chi$ distributions. 
Figures ~\ref{bkg2} and ~\ref{bkg4} show the projections of $\cos\angpi$ and $\m$, and of $\chi$ of the total background, respectively. 

\section{Final state composition}\label{fit}
\label{sec:Results1}

\subsection{Resonance models}
To study the resonant structures of the decay $\Bsb\rightarrow J/\psi \pi^+\pi^-$ we use the 34\,471 candidates with invariant mass lying within $\pm20$ \mev of the $\Bsb$ mass peak which include 7075$\pm$101 background events.
The $\pi^+\pi^-$  resonance candidates that could contribute to $\Bsb\rightarrow J/\psi \pi^+\pi^-$ decay are listed in Table \ref{tab:reso1}.  The resonances  that decay into a $\pi^+\pi^-$ pair must be isoscalar ($I=0$), because the $s\bar{s}$ system forming the resonances in Fig.~\ref{feyn1} has $I=0$. To test the isoscalar argument, the isospin-1 $\rho(770)$ meson is also added to the baseline fit. The non-resonance (NR) is assumed to be S-wave, its shape is defined by Eq.~(\ref{eq:DP}) where the amplitude function $A_R(m_{hh})$ is set to be equal to one, and the Blatt-Weisskopf barrier factors $F_B^{(1)}$ and $F_R^{(0)}$ are both set to one.

In the previous analysis~\cite{LHCb:2012ad}, we observed a resonant state at $(1475\pm6)$\mev with a width of $(113\pm11)$\mev. We identified it with the $f_0(1370)$ though its mass and width values agreed neither with the $f_0(1500)$ or the $f_0(1370)$. W.~Ochs~\cite{Ochs:2013gi,*Ochs:2013vxa} argues that the better assignment is $f_0(1500)$; we follow his suggestion. In addition, a structure is clearly visible in the $1800$ MeV region (see Fig.~\ref{m-jpsipi} (b)), which was not the case in our previous analysis \cite{LHCb:2012ae}. This could be the $f_0(1790)$ resonance observed  by BES~\cite{Ablikim:2004wn} in $\jpsi\to \phi \pi^+\pi^-$ decays. 

From the measured ratios ${\cal B}\left(\Bsb\to\jpsi f_2^\prime(1525)\right)/{\cal B}\left(\Bsb\to\jpsi \phi\right)$~\cite{Aaij:2013orb} and ${\cal B}\left(\Bsb\to\jpsi \pi^+\pi^-\right)/{\cal B}\left(\Bsb\to\jpsi \phi\right)$~\cite{LHCb:2012ae}, using the measured $\pi^+\pi^-$ and $K^+K^-$ branching fractions \cite{PDG}, the expected $f_2^\prime(1525)$ fit fraction for the transversity $0$ component is $(0.45\pm0.13)\%$, and the ratio of helicity $\lambda=0$ to $|\lambda|=1$ components, which is equal to the ratio of transversity $0$ to the sum of $\perp$ and $\parallel$ components, is $1.9\pm0.8$, where the uncertainties are dominated by that on $ f_2^\prime(1525)$ fit fractions in $\Bsb\to\jpsi K^+K^-$ decays. This information is used as constraints in the fit.
\begin{table}[!b]
\begin{center}
\caption{Possible resonance candidates in the $\Bsb\rightarrow J/\psi \pi^+\pi^-$ decay mode and their parameters used in the fit.}
\begin{tabular}{ccccccc}
\hline
Resonance & Spin & Helicity & Resonance & Mass (\mev)& Width (\mev) & Source \\
&&& formalism \\
\hline
$f_0(500)$  & 0 & 0 & BW & $471\pm21$ & $534\pm53$ & LHCb~\cite{Aaij:2013zpt}\\
$f_0(980)$ & 0 & 0 & Flatt\'e &\multicolumn{3}{c}{see text} \\
$f_2(1270)$ & 2 &  $0,\pm 1$ & BW  & $1275.1\pm1.2$ & $185.1^{+2.9}_{-2.4}$&PDG \cite{PDG}\\
$f_0(1500)$ & 0 & 0 & BW &\multicolumn{3}{c}{see text} \\
$f_2^\prime(1525)$ & 2 &  $0,\pm 1$ & BW & $1522_{-3}^{+6}$ & $84_{-8}^{+12}$ &LHCb \cite{Aaij:2013orb}\\
$f_0(1710)$ & 0 & 0 & BW & $1720\pm 6$ & $135\pm 8$ &PDG \cite{PDG}\\
$f_0(1790)$ & 0 & 0 & BW &$1790_{-30}^{+40}$&$270_{-30}^{+60}$& BES \cite{Ablikim:2004wn} \\
$\rho(770)$ & 1 & $0,\pm 1$ & BW & $775.49\pm0.34$ &$149.1\pm0.8$&PDG \cite{PDG} \\
\hline
\end{tabular}\label{tab:reso1}
\end{center}
\end{table}

The masses and widths of the resonances are also listed in Table~\ref{tab:reso1}. When used in the fit they are fixed to these central values, except for the parameters of $f_0(980)$ and $f_0(1500)$ that are determined by the fit. In addition, the parameters of $f_0(1790)$ are constrained to those determined by the BES measurement~\cite{Ablikim:2004wn}.

As suggested by D.~V.~Bugg~\cite{Bugg:2008ig}, the Flatt\'e model \cite{Flatte:1976xv} for $f_0(980)$ is slightly modified, and is parameterized as
\begin{equation}
A_R(m_{\pi^+\pi^-})=\frac{1}{m_R^2-m^2_{\pi^+\pi^-}-im_R(g_{\pi\pi}\rho_{\pi\pi}+g_{KK}F_{KK}^2\rho_{KK})},
\end{equation}
where $m_R$ is the $f_0(980)$ pole mass, the parameters $g_{\pi\pi}$ and $g_{KK}$ are the $f_0(980)$ coupling constants to $\pi^+\pi^-$ and $K^+K^-$ final states, respectively, and the phase space $\rho$ factors are given by Lorentz-invariant phase spaces as
\begin{eqnarray}
\rho_{\pi\pi} &=& \frac{2}{3}\sqrt{1-\frac{4m^2_{\pi^{\pm}}}{m^2_{\pi^+\pi^-}}}+\frac{1}{3}\sqrt{1-\frac{4m^2_{\pi^{0}}}{m^2_{\pi^+\pi^-}}}\label{flatte1}, \\
\rho_{KK} &=& \frac{1}{2}\sqrt{1-\frac{4m^2_{K^{\pm}}}{m^2_{\pi^+\pi^-}}}+\frac{1}{2}\sqrt{1-\frac{4m^2_{K^{0}}}{m^2_{\pi^+\pi^-}}}.\label{flatte2}
\end{eqnarray}
Compared to the normal Flatt\'e function, a form factor $F_{KK} = \exp(-\alpha k^2)$ 
is introduced above the $KK$ threshold  and serves to reduce the $\rho_{KK}$ factor as $m^2_{\pi^+\pi^-}$ increases,
where $k$ is momentum of each kaon in the $KK$ rest frame, and $\alpha=(2.0\pm0.25)$ GeV$^{-2}$~\cite{Bugg:2008ig}. This parameterization slightly decreases the $f_0(980)$ width above the $KK$ threshold. The parameter $\alpha$ is fixed to $2.0$ GeV$^{-2}$ as it is not very sensitive to the fit.

To determine the complex amplitudes in a specific model, the data are fitted maximizing the unbinned likelihood given as
 \begin{equation}
\mathcal{L}=\prod_{i=1}^{N}F(\m^i,\angpi^i,\theta^i_{J/\psi},\chi^i),
\end{equation}
where $N$ is the total number of candidates, and $F$ is the total PDF defined in Eq.~(\ref{eq:pdf}). In order to converge properly in a maximum likelihood method, the PDFs of the signal and background need to be normalized. This is accomplished by first normalizing the $\chi$ and $\cos\angmu$ dependent parts analytically, 
and then normalizing the $\m$ and $\cos\angpi$ dependent parts using a numerical integration over 1000$\times$200 bins.

The fit determines amplitude magnitudes $a_i^{R_i}$ and phases $\phi_i^{R_i}$ defined in Eq.~(\ref{eq:amp}). The $a^{f_0(980)}_0$ amplitude is fixed to 1, since the overall normalization is related to the signal yield. As only relative phases are physically meaningful, $\phi_0^{f_0(980)}$ is fixed to 0. In addition, due to the averaging of $\Bs$ and $\Bsb$, the interference terms between opposite \CP states are cancelled out, making it not possible to measure the relative phase between \CP-even and odd states here, so one \CP-even phase, $\phi_{\perp}^{f_2(1270)}$, is also fixed to 0.

\subsection{Fit fraction}
Knowledge of the contribution of each component can be expressed by defining a fit fraction for each transversity $\tau$, ${\cal{F}}_\tau^R$, which is the squared amplitude of $R$ integrated over the phase space divided by the entire amplitude squared over the same area.  To determine ${\cal{F}}_\tau^R$ one needs to integrate over all the four fitted observables in the analysis. The interference terms between different helicity components vanish, after integrating over the two variables of $\cos\angmu$ and $\chi$.
Thus we define the transversity fit fraction as
\begin{equation}\label{eq:ff}
{\cal{F}}^R_{\tau}=\frac{\int\left| a^R_\tau e^{i\phi^R_\tau} {\cal A}_R(\m) d_{\lambda,0}^{J_R}(\angpi)\right|^2 {\rm d}\m\; {\rm d}\cos\angpi}{\int \left(|{\cal {H}}_0(m_{hh},\angpi)|^2+|{\cal {H}}_+(m_{hh},\angpi)|^2+|{\cal {H}}_-(m_{hh},\angpi)|^2\right) {\rm d}\m\; {\rm d}\cos\angpi},
\end{equation}
where $\lambda=0$ in the $d$-function for $\tau=0$, and $\lambda=1$ for $\tau=\perp$ or $\parallel$.

Note that the sum of the fit fractions is not necessarily unity due to the potential presence of interference between two resonances. Interference term fractions are given by
\begin{equation}
\label{eq:inter}
{\cal{F}}_{\tau}^{RR^\prime}=2\mathcal{R}e\left(\frac{\int a^R_\tau\; a^{R'}_\tau e^{i(\phi^R_\tau-\phi^{R'}_\tau)} {\cal A}_R(\m)  {\cal A}^{*}_{R'}(\m) d_{\lambda,0}^{J_R}(\angpi) d_{\lambda,0}^{J_{R'}}(\angpi)  {\rm d}\m\; {\rm d}\cos\angpi}{\int \left(|{\cal {H}}_0(m_{hh},\angpi)|^2+|{\cal {H}}_+(m_{hh},\angpi)|^2+|{\cal {H}}_-(m_{hh},\angpi)|^2\right) {\rm d}\m\; {\rm d}\cos\angpi}\right),
\end{equation}
and
\begin{equation}
\sum_{R,\tau} {\cal{F}}_\tau^R+\sum^{R > R'}_{RR',\tau} {\cal{F}}_\tau^{RR'} =1.
\end{equation}
Interference  between different spin-$J$ states vanishes, when integrated over angle, because the $d^J_{\lambda0}$ angular functions are orthogonal.

\subsection{Fit results}\label{sec:result}
In order to compare the different models quantitatively, an estimate of the goodness of fit is calculated from four-dimensional (4D) partitions of the four variables, $m(\pi^+\pi^-)$, $\cos\theta_{hh}$, $\cos \theta_{J/\psi}$ and $\chi$. We use the Poisson likelihood $\chi^2$ \cite{Baker:1983tu} defined as
\begin{equation}
\chi^2=2\sum_{i=1}^{N_{\rm bin}}\left[  x_i-n_i+n_i \text{ln}\left(\frac{n_i}{x_i}\right)\right],
\end{equation}
where $n_i$ is the number of events in the four-dimensional bin $i$ and $x_i$ is the expected number of events in that bin according to the fitted likelihood function. A total of 1845 bins are used to calculate the $\chi^2$, where $41(m_{hh})\times5(\cos\theta_{hh})\times3(\cos\theta_{J/\psi})\times3(\chi)$ equal size bins are used, and $m_{hh}$ is required to be between 0.25 and 2.30~GeV. The $\chi^2/\text{ndf}$, and the negative of the logarithm of the likelihood, $\rm -ln\mathcal{L}$, of the fits are given in Table~\ref{RMchi2}, where ndf is the number of degree of freedom given as 1845 subtracted by number of fitting parameters and 1. The nomenclature describing the models gives the base model first and then ``+" for any additions. The 5R model contains the resonances $f_0(980)$, $f_2(1270)$, $f_2^\prime(1525)$, $f_0(1500)$, and $f_0(1790)$. If adding NR to 5R model, two minima with similar likelihoods are found. One minimum is consistent with the 5R results and has NR fit fraction of $(0.3\pm0.3)\%$; we group any fit models that are consistent with this 5R fit into the ``Solution I" category. Another minimum has significant NR fit fraction of $(5.9\pm1.4)\%$, this model and other consistent models are classified in the ``Solution II" category.


\begin{table}[!t]
\begin{center}
\caption{Fit $\rm -ln\mathcal{L}$ and $\chi^2/\text{ndf}$ of different resonance models.}
\begin{tabular}{lccc}
\hline
Resonance model & $\rm -ln\mathcal{L}$& $\chi^2/\text{ndf}$  \\
\hline
5R (Solution I)&$-93738$ & 2005/1822 = 1.100 \\
5R+NR (Solution I) &$-93741$ &2003/1820 = 1.101\\
5R+$f_0(500)$ (Solution I) & $-93741$ &2004/1820 = 1.101\\
5R+$f_0(1710)$ (Solution I)&$-93744$ &1998/1820 = 1.098\\
5R+$\rho(770)$ (Solution I)&$-93742$ &2004/1816 = 1.104\\\hline
5R+NR (Solution II)&$-93739$ &2008/1820 = 1.103\\
5R+NR+$f_0(500)$ (Solution II)& $-93741$ &2004/1818 = 1.102\\
5R+NR+$f_0(1710)$ (Solution II)&$-93745$ &2004/1818 = 1.102\\
5R+NR+$\rho(770)$ (Solution II)&$-93746$ &1998/1814 = 1.101\\\hline
\hline
\end{tabular}
\label{RMchi2}
\end{center}
\end{table}


Among these resonance models, we select the baseline model by requiring each resonance in the model to have more than 3 standard deviation ($\sigma$) significance evaluated by the fit fraction divided by its uncertainty.
The baseline fits are 5R in Solution I and 5R+NR in Solution II. No additional components are significant when added to these baseline fits. Unfortunately, we cannot distinguish between these two solutions and will quote results for both of them. In both cases the dominant contribution is S-wave including $f_0(980)$, $f_0(1500)$ and $f_0(1790)$. The D-wave, $f_2(1270)$ and $f_2^\prime(1525)$, is only 2.3\% for both solutions.

\begin{table}[t]
\begin{center}
\caption{Fit fractions (\%) of contributing components for both solutions.}
\def\arraystretch{1.2}
\begin{tabular}{lcc}
\hline
 Component            & Solution I & Solution II  \\\hline
$f_0(980)$ & $70.3\pm1.5_{-5.1}^{+0.4}$ & $92.4\pm2.0_{-16.0}^{+~0.8}$\\
$f_0(1500)$ & $10.1\pm0.8_{-0.3}^{+1.1}$ & $9.1\pm0.9\pm0.3$\\
$f_0(1790)$ & $2.4\pm0.4_{-0.2}^{+5.0}$ & $0.9\pm0.3_{-0.1}^{+2.5}$\\
$f_2(1270)_0$ & $0.36\pm0.07\pm0.03$ & $0.42\pm0.07\pm0.04$\\
$f_2(1270)_{\|}$& $0.52\pm0.15_{-0.02}^{+0.05}$ & $0.42\pm0.13_{-0.02}^{+0.11}$\\
$f_2(1270)_{\perp}$& $0.63\pm0.34_{-0.08}^{+0.16}$ & $0.60\pm0.36_{-0.09}^{+0.12}$\\
$f_2^\prime(1525)_0$& $0.51\pm0.09_{-0.04}^{+0.05}$ & $0.52\pm0.09_{-0.04}^{+0.05}$\\
$f_2^\prime(1525)_{\|}$& $0.06_{-0.04}^{+0.13}\pm0.01$ &$0.11_{-0.07-0.04}^{+0.16+0.03}$\\
$f_2^\prime(1525)_{\perp}$& $0.26\pm0.18_{-0.04}^{+0.06}$  & $0.26\pm0.22_{-0.05}^{+0.06}$\\
NR&-&$5.9\pm1.4_{-4.6}^{+0.7}$\\ \hline
Sum & 85.2& 110.6\\ \hline
$\rm -ln\mathcal{L}$& $-93738$&$-93739$\\\hline

$\chi^2/\text{ndf}$ &2005/1822&$2008/1820$\\
\hline
\end{tabular}
\label{tab:FF}
\end{center}
\end{table}

\begin{figure}[t]
  \begin{center}
\includegraphics[width=1.0\textwidth]{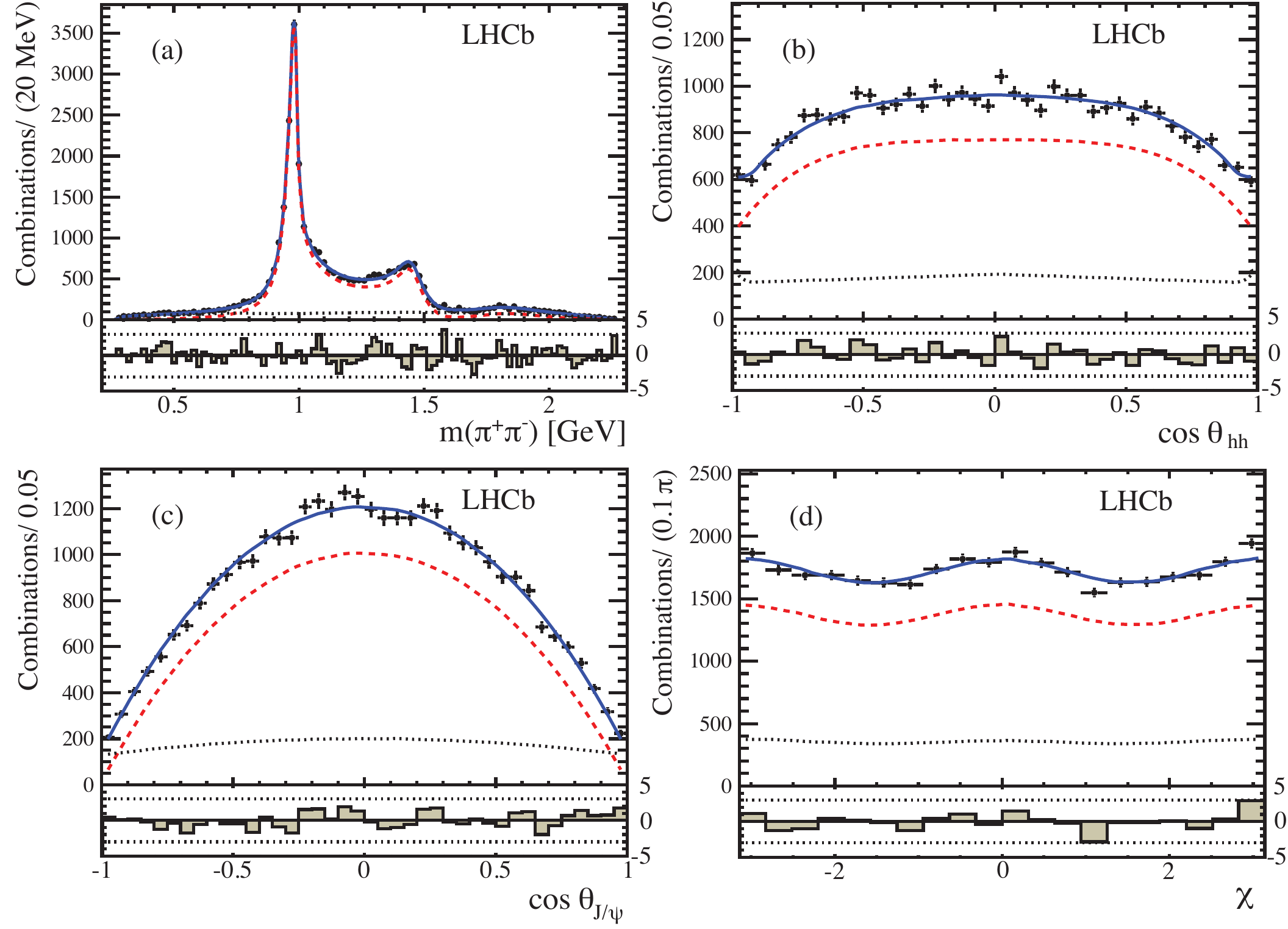}
     \vspace{-6mm}
    \caption{Projections of (a) $m(\pi^+\pi^-)$, (b) $\cos\theta_{\pi\pi}$, (c) $\cos \theta_{J/\psi}$ and (d) $\chi$ for 5R Solution~I. The points with error bars are data, the signal fit is shown with a (red) dashed line, the background with a (black) dotted line, and the (blue) solid line represents the total.}  \label{RM4}
  \end{center}
\end{figure}

\begin{figure}[t]
  \begin{center}
     \includegraphics[width=1.0\textwidth]{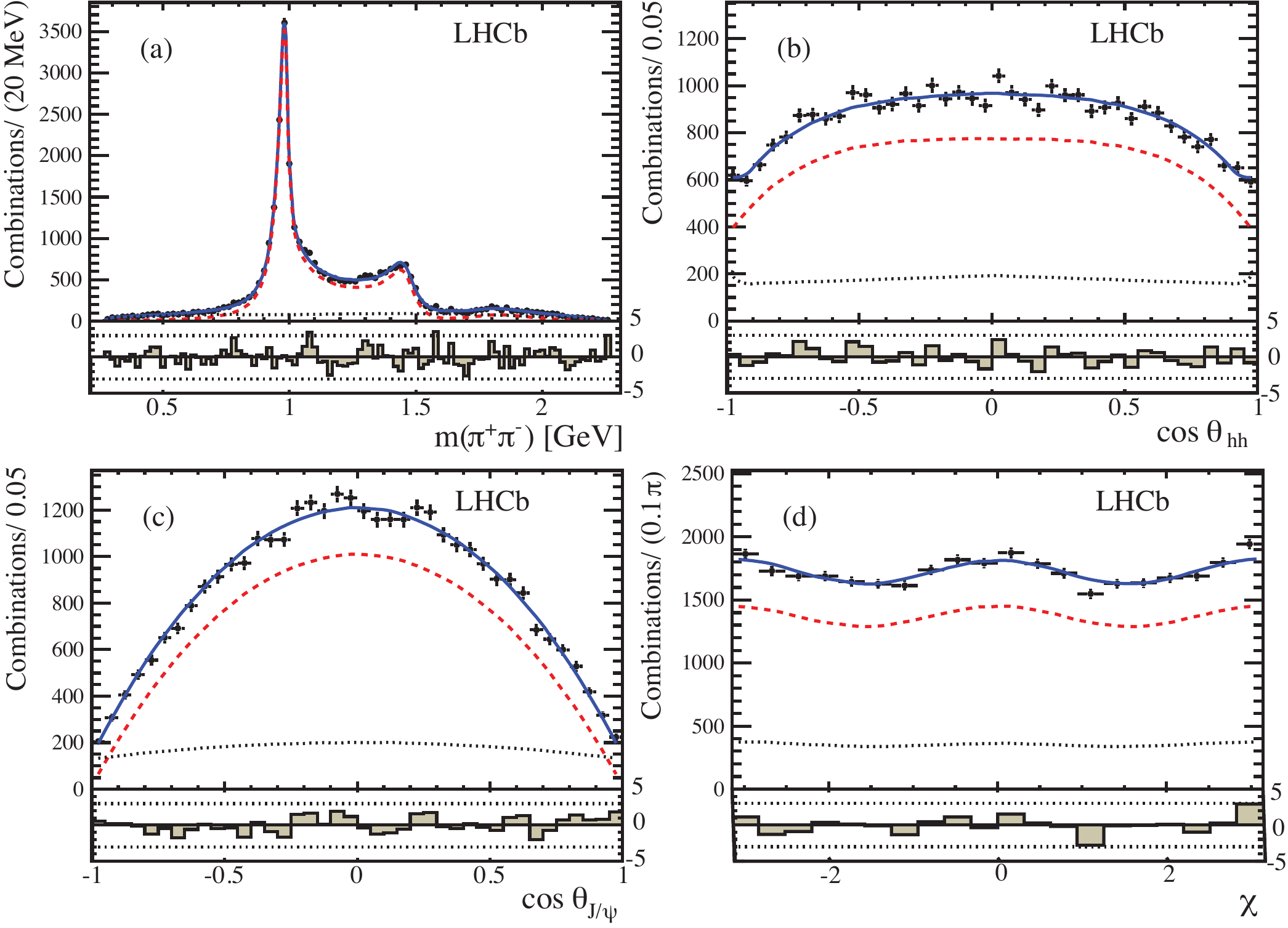}
     \vspace{-6mm}
    \caption{Projections of (a) $m(\pi^+\pi^-)$, (b) $\cos\theta_{\pi\pi}$, (c) $\cos \theta_{J/\psi}$ and (d) $\chi$ for 5R+NR Solution II. The points with error bars are data, the signal fit is shown with a (red) dashed line, the background with a (black) dotted line, and the (blue) solid line represents the total.}  \label{RM8}
  \end{center}
\end{figure}

\begin{figure}[!hbtp]
  \begin{center}
     \includegraphics[width=0.77\textwidth]{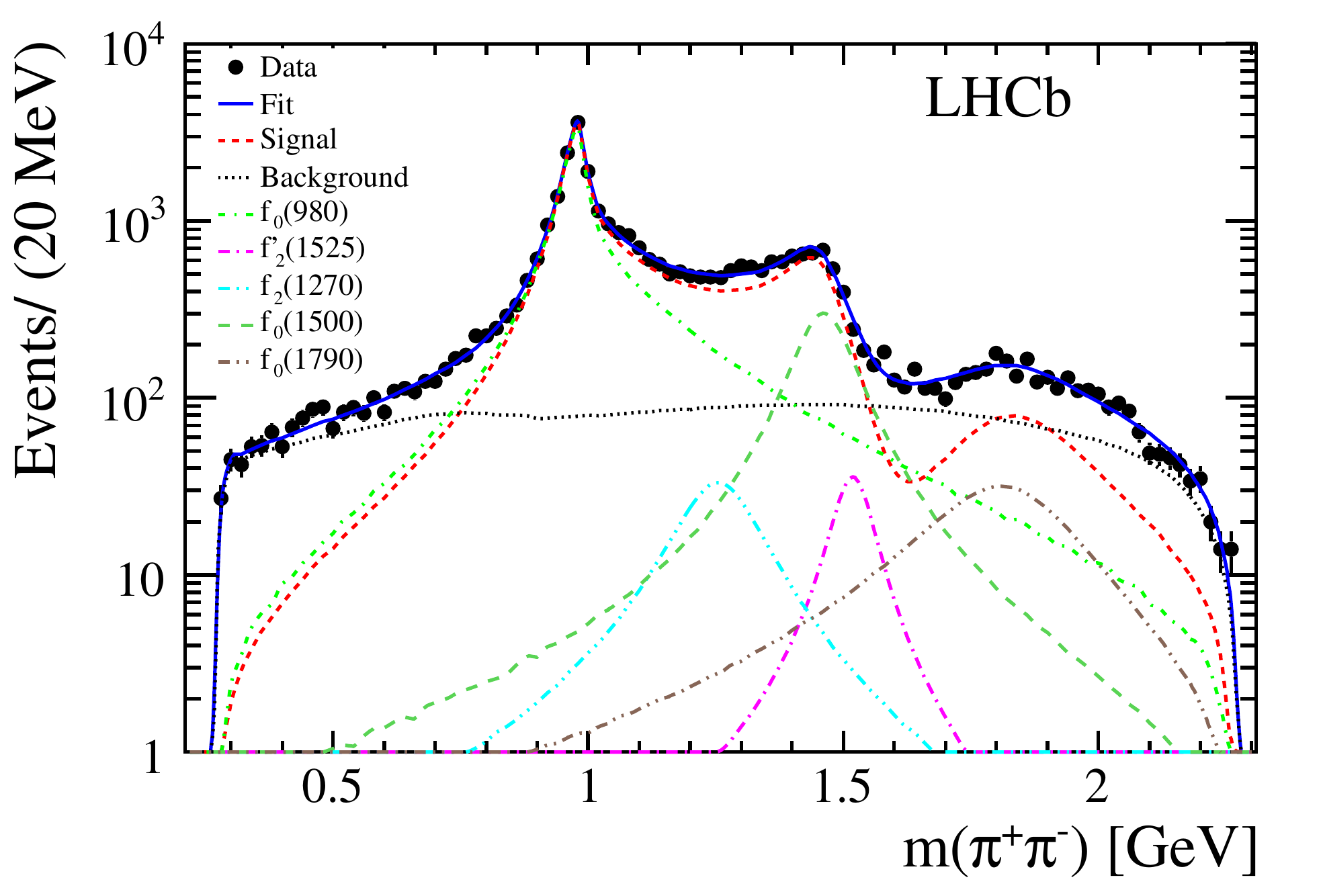}
    \caption{Distribution of $m(\pi^+\pi^-)$ with contributing components labeled from 5R Solution~I. }  \label{mpp-contr}
  \end{center}
\end{figure}

\begin{figure}[!hbtp]
  \begin{center}
     \includegraphics[width=0.77\textwidth]{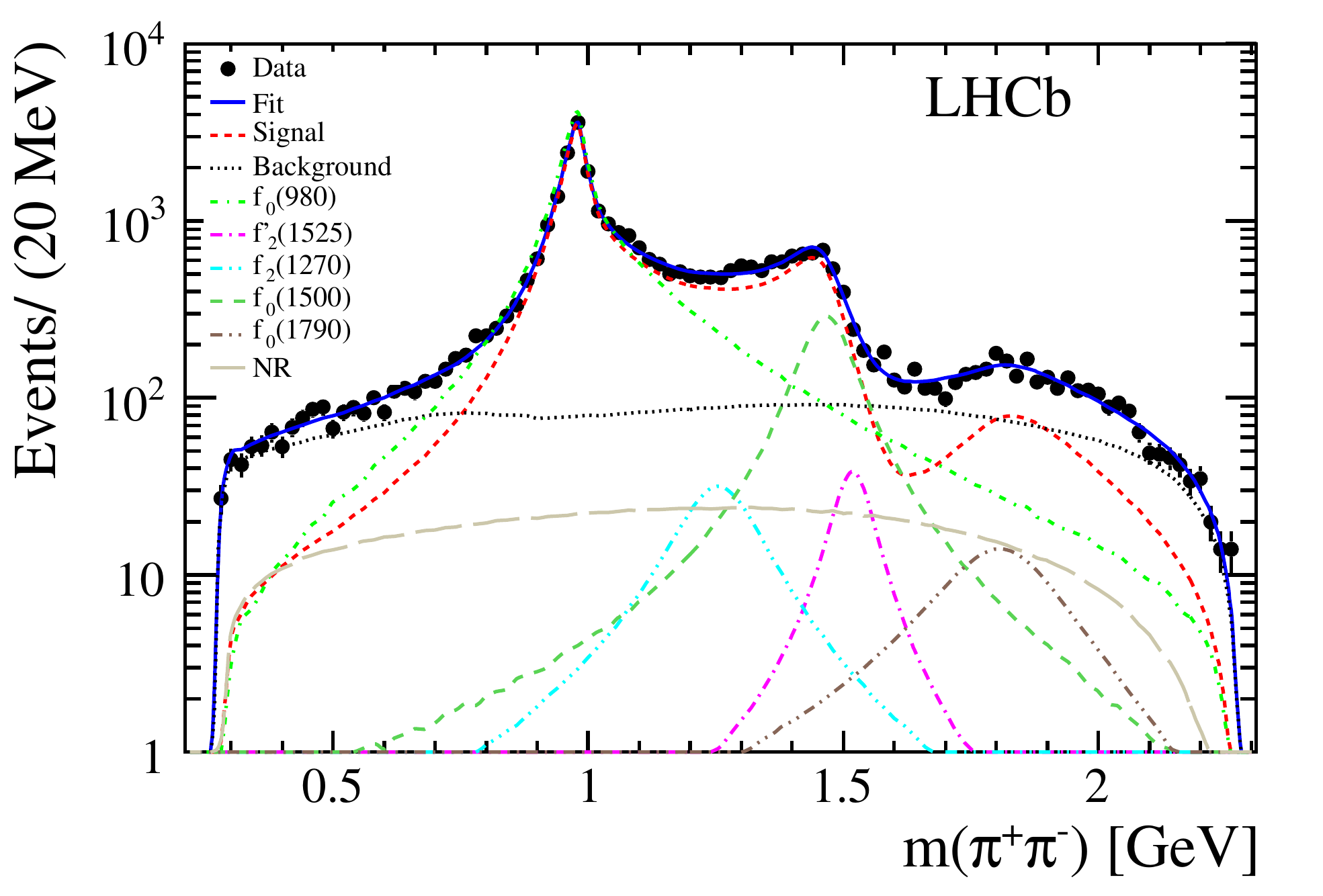}
    \caption{Distribution of $m(\pi^+\pi^-)$ with contributing components labeled from 5R+NR Solution II. }  \label{mpp-contr2}
  \end{center}
\end{figure}

Table~\ref{tab:FF} shows the fit fractions from the baseline fits of two solutions, where systematic uncertainties are included; they will be discussed in Sec.~\ref{Sec:sys}. Figures~\ref{RM4} and~\ref{RM8} show the fit projections of $m(\pi^+\pi^-)$, $\cos\theta_{\pi\pi}$, $\cos \theta_{J/\psi}$ and $\chi$ from 5R Solution I and 5R+NR Solution II, respectively.
Also shown in Figs.~\ref{mpp-contr} and ~\ref{mpp-contr2} are the contributions of each resonance as a function of $m(\pi^+\pi^-)$ from the baseline Solution I and II fits, respectively. Table~\ref{tab:inter} shows the fit fractions of the interference terms defined in Eq.~(\ref{eq:inter}). In addition, the phases are listed in Table~\ref{tab:phase}. The other fit results are listed in Table~\ref{tab:other} including the $f_0(980)$ mass, the Flatt\'e function parameters $g_{\pi\pi}$, $g_{KK}/g_{\pi\pi}$, and masses and widths of $f_0(1500)$ and $f_0(1790)$ resonances.

\begin{table}[tb]
\centering
\caption{Non-zero interference fraction (\%) for both solutions.}\label{tab:inter}
\begin{tabular}{ccc}\hline
Components& Solution I& Solution II\\\hline
$f_0(980)$+$f_0(1500)$ & ~~9.50& $-1.57$ \\
$f_0(980)$+$f_0(1790)$ & ~~7.93& ~~5.30\\
$f_0(1500)$+$f_0(1790)$&$-2.69$&$-2.26$\\
$f_2(1270)_0$+$f_2^\prime(1525)_0$ & ~~0.14& ~~0.09\\
$f_2(1270)_{\|}$+$f_2^\prime(1525)_{\|}$ & $-0.09$& $-0.16$\\
$f_2(1270)_{\perp}$+$f_2^\prime(1525)_{\perp}$ & ~~0.03& ~~0.05\\
$f_0(980)$+NR~~~~~ & - &$-16.41$\\
$f_0(1500)$+NR~~~~~~& -&~~5.26\\
$f_0(1790)$+NR~~~~~~&-&$-0.95$\\
\hline
\end{tabular}
\end{table}



\begin{table}[tb]
\centering
\caption{Fitted resonance phase differences ($^\circ$).}\label{tab:phase}
\begin{tabular}{lcc}\hline
Resonance & Solution I & Solution II \\\hline
$f_0(1500)-f_0(980)$ &$138\pm4$ & $177\pm6$\\
$f_0(1790)-f_0(980)$ & $78\pm9$& $95\pm16$\\
$f_2(1270)_0-f_0(980)$ & $96\pm7$& $123\pm8$\\
$f_2(1270)_{\|}-f_0(980)$& $-90\pm11$& $-84\pm13$\\
$f_2^\prime(1525)_0-f_0(980)$& $-132\pm6$& $-97\pm7$\\
$f_2^\prime(1525)_{\|}-f_0(980)$& $103\pm29$& $130\pm20$\\
NR $-f_0(980)$& - & $-104\pm5$\\\hline
$f_2^\prime(1525)_{\perp}-f_2(1270)_{\perp}$& $149\pm46$& $145\pm51$\\
\hline
\end{tabular}
\end{table}

\begin{table}[tb]
\centering
\caption{Other fit parameters. The uncertainties are only statistical.}\label{tab:other}
\begin{tabular}{lcc}\hline
Parameter & Solution I& Solution II\\\hline
$m_{f_0(980)}$ (\mev)&$945.4\pm2.2$&$949.9\pm2.1$\\
$g_{\pi\pi}$~(\mev)&$167\pm7$&$167\pm8$\\
$g_{KK}/g_{\pi\pi}$&$3.47\pm0.12$&$3.05\pm0.13$\\
$m_{f_0(1500)}$ (\mev)&$1460.9\pm2.9$&$1465.9\pm3.1$\\
$\Gamma_{f_0(1500)}$ (\mev)&$124\pm7$&$115\pm7$\\
$m_{f_0(1790)}$ (\mev)&$1814\pm18$&$1809\pm22$\\
$\Gamma_{f_0(1790)}$ (\mev)&$328\pm34$&$263\pm30$\\\hline
\end{tabular}
\end{table}

In both solutions the $f_0(500)$ state does not have a significant fit fraction. We set an upper limit for the fit fraction ratio between $f_0(500)$ and $f_0(980)$ of 0.3\% from Solution I and 3.4\% from Solution II, both at 90\% CL. A similar situation is found for the $\rho(770)$ state. When including it in the fit, the fit fraction of $\rho(770)$ is measured to be $(0.60\pm0.30^{+0.08}_{-0.14})\%$ in Solution I and $(1.02\pm0.36^{+0.09}_{-0.15})\%$ from Solution II.
The largest upper limit is obtained by Solution II,  where the $\rho(770)$ fit fraction is less than 1.7\% at 90\% CL.

Our previous study~\cite{LHCb:2012ae} did not consider the $f_0(1790)$ resonance, instead the NR component filled in the higher mass region near $1800$~MeV. It is found that including $f_0(1790)$ improves the fit significantly in both solutions. Inclusion of this state reduces $\rm -2ln\mathcal{L}$ by 276 (97) units and $\chi^2$ by 213 (91) units with 4 additional ndf, corresponding to 14 (9) $\sigma$ Gaussian significance, in Solution I(II), where the numbers are statistical only. When floating the parameters of $f_0(1790)$ resonance in the fits, we find its mass $m_{f_0(1790)}=1815\pm23$\mev and width $\Gamma_{f_0(1790)}=353\pm48$\mev in Solution I, and $m_{f_0(1790)}=1793\pm26$\mev and $\Gamma_{f_0(1790)}=180\pm83$\mev in Solution II, where the uncertainties are statistical only. The values in both solutions are consistent with the BES results $m_{f_0(1790)}=1790_{-30}^{+40}$\mev and $\Gamma_{f_0(1790)}=270_{-30}^{+60}$\mev~\cite{Ablikim:2004wn} at the level of $1\sigma$.

\begin{figure}[!hbt]
  \begin{center}\label{2model}
     \includegraphics[width=0.455\textwidth]{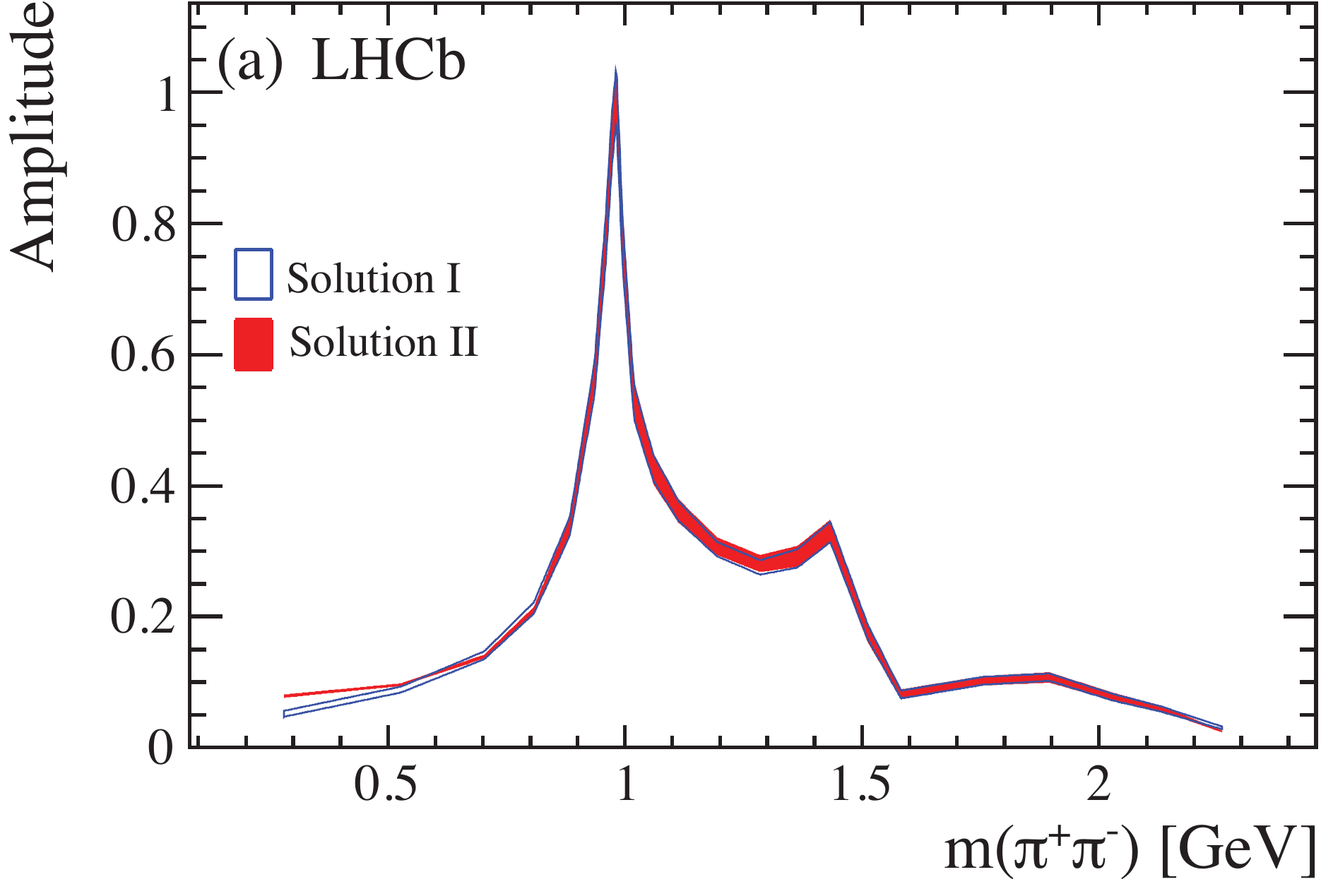}
      \includegraphics[width=0.48\textwidth]{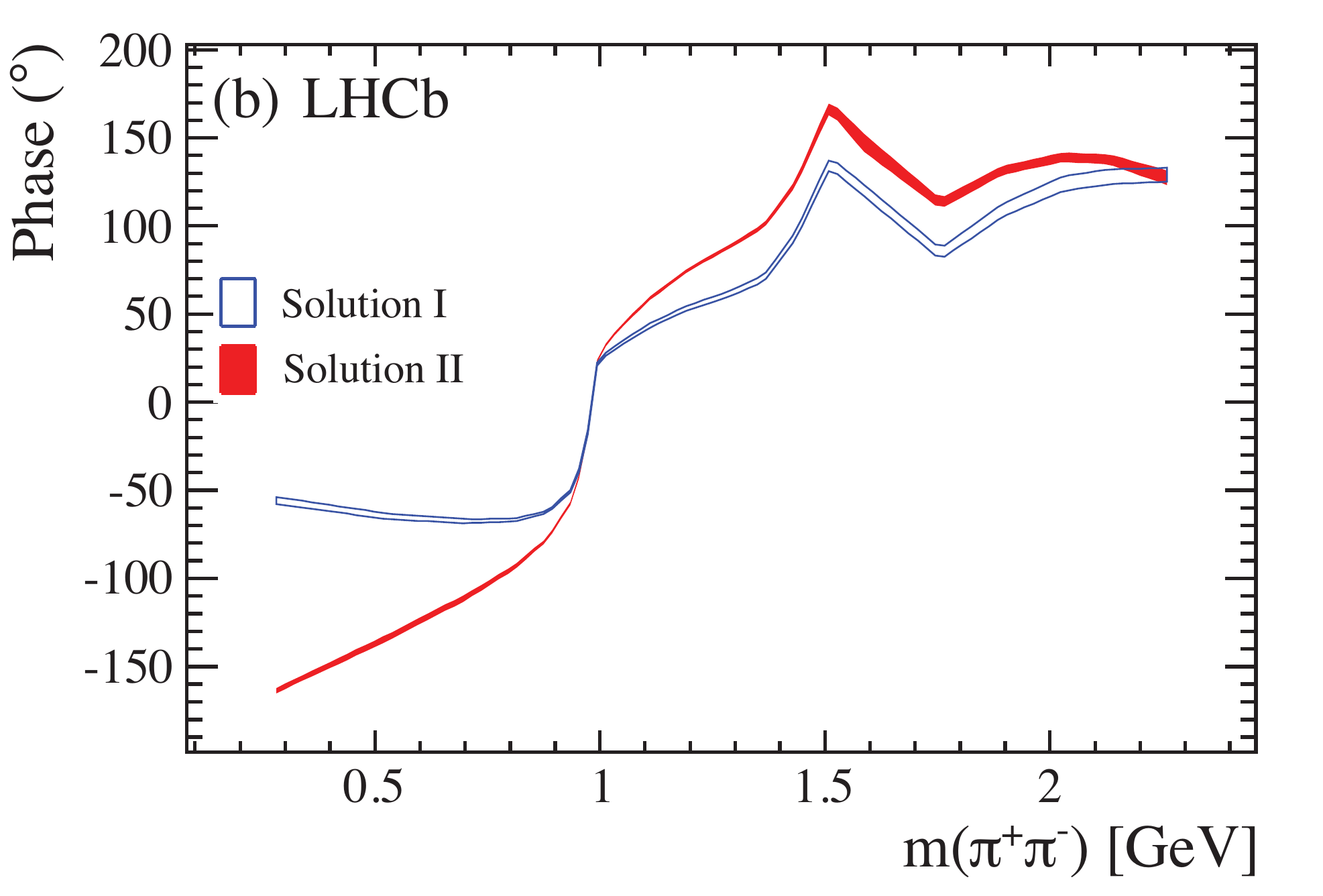}
 \caption{S-wave (a) amplitude strength and (b) phase as a function of $m(\pi^+\pi^-)$ from the 5R Solution I (open) and 5R+NR Solution II (solid), where the widths of the curves reflect $\pm1\sigma$ statistical uncertainties. The reference point is chosen at 980 MeV with amplitude strength equal to 1 and phase equal to 0. }
 \end{center}
\end{figure}

Figure~\ref{2model} compares the total S-wave amplitude strength and phase as a function of $m(\pi^+\pi^-)$ between the two solutions, showing consistent amplitude strength but distinct phase. The total S-wave amplitude is calculated as   Eq.~(\ref{eq:heart}) summing over all spin-0 component $R$ with $\lambda=0$, where the $d$-function is equal to 1. The amplitude strength can be well measured from the $m(\pi^+\pi^-)$ distribution, but this is not the case for the phase, which is determined from the interference with the small fraction of higher spin resonances.


\subsection{Angular moments}

\begin{figure}[!htbp]
\begin{center}
   \hspace*{1mm} \includegraphics[width=0.4075\textwidth]{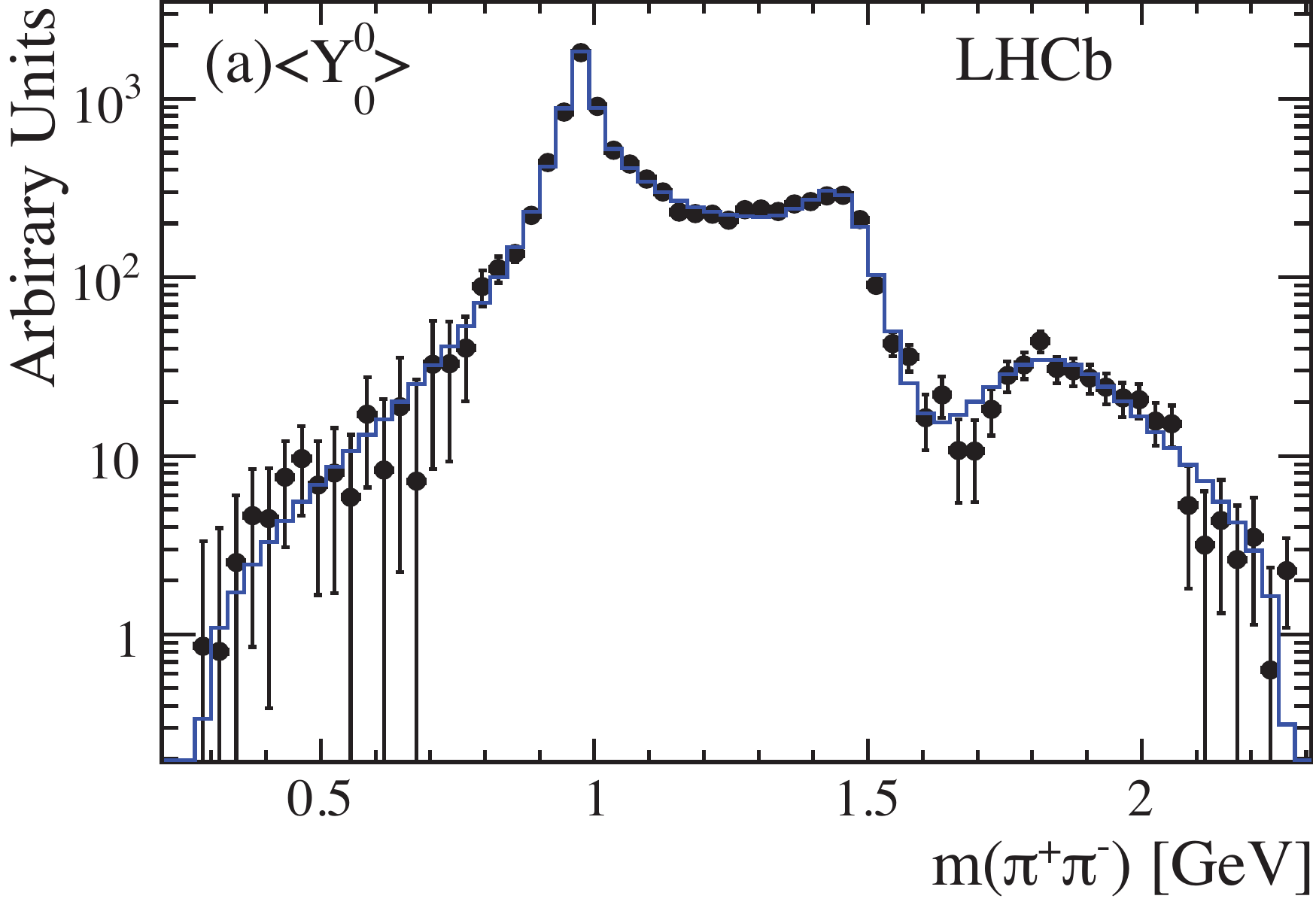}\hspace*{2mm}
    \includegraphics[width=0.44\textwidth]{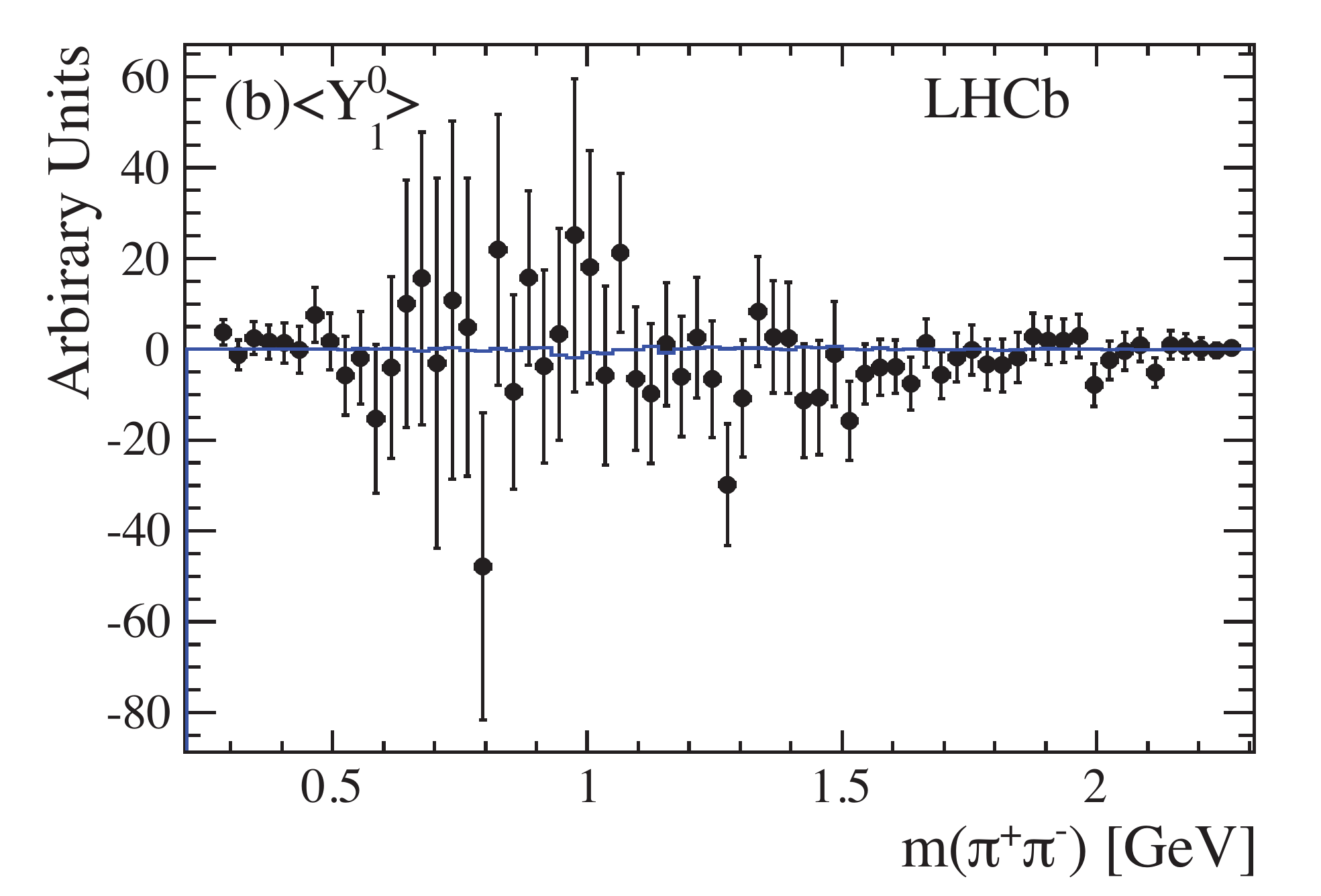}
    \includegraphics[width=0.44\textwidth]{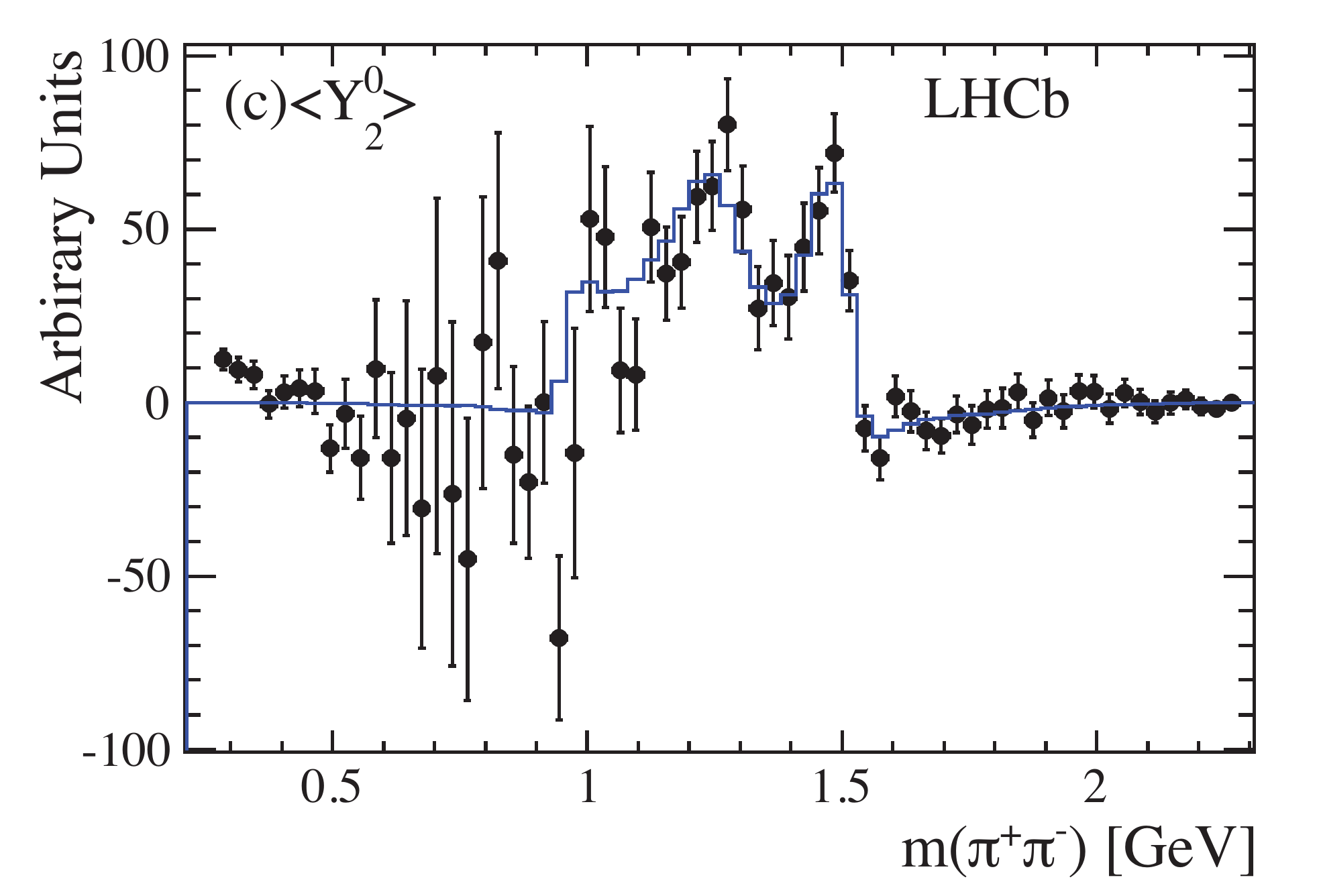}%
    \includegraphics[width=0.44\textwidth]{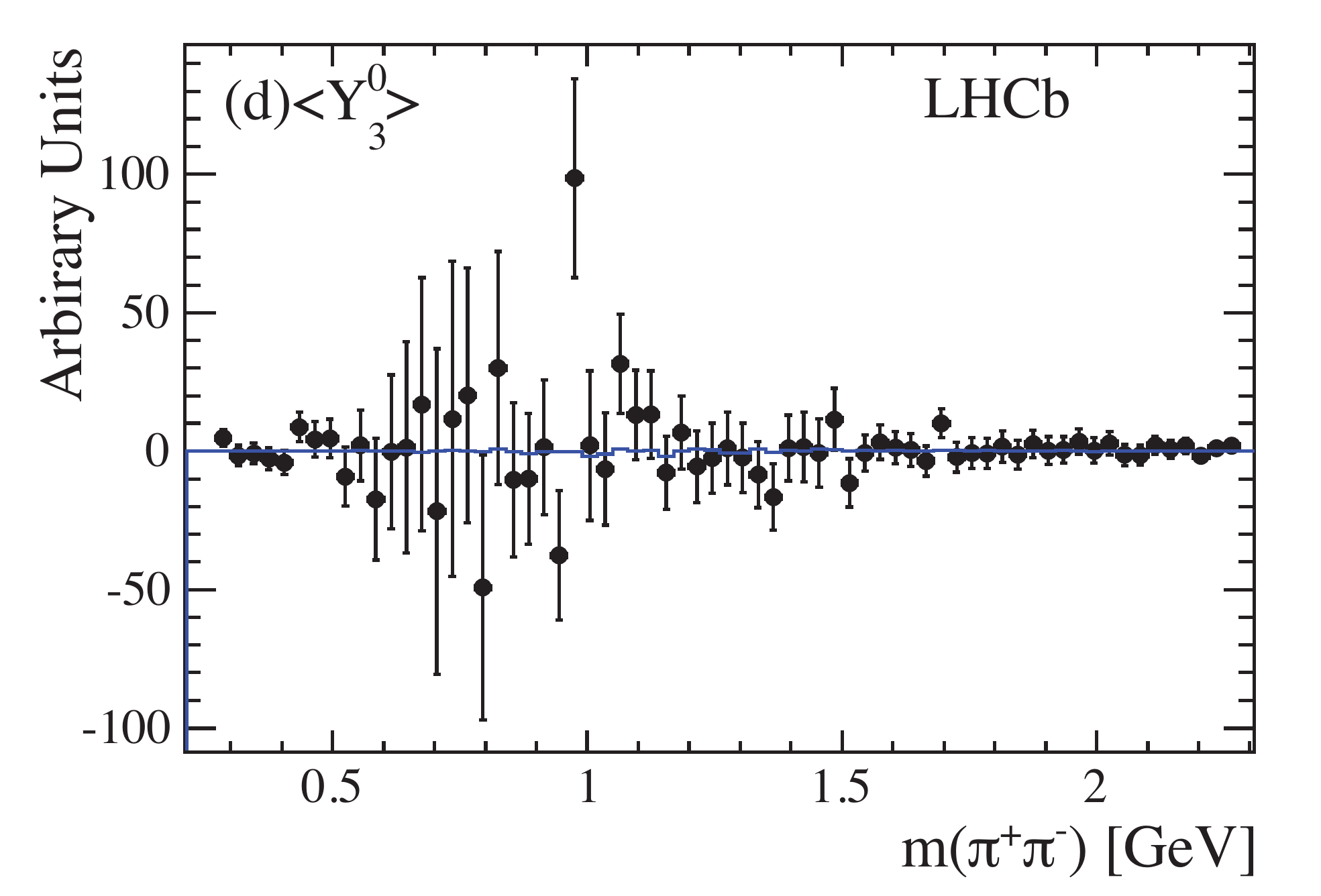}
    \includegraphics[width=0.44\textwidth]{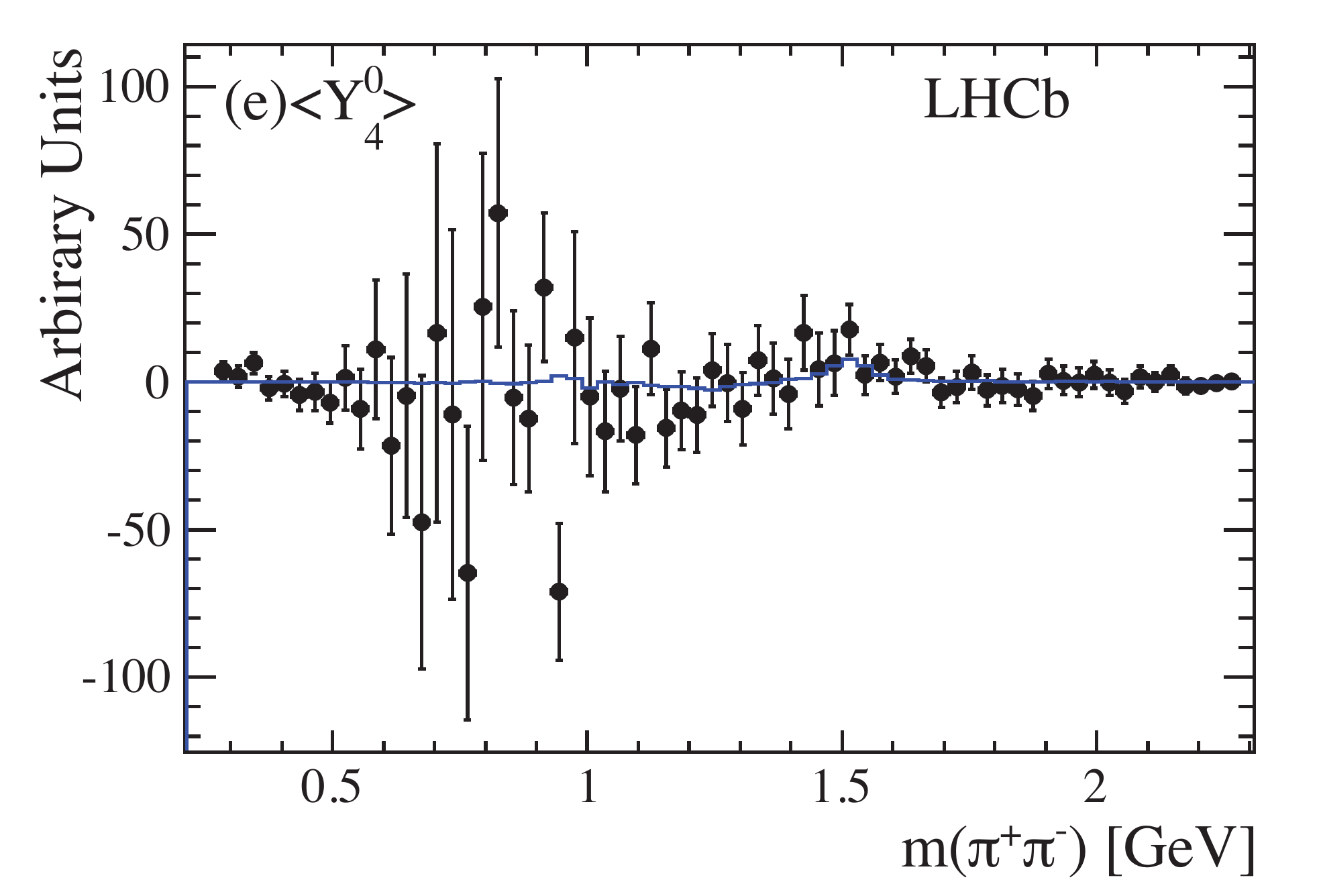}%
    \includegraphics[width=0.44\textwidth]{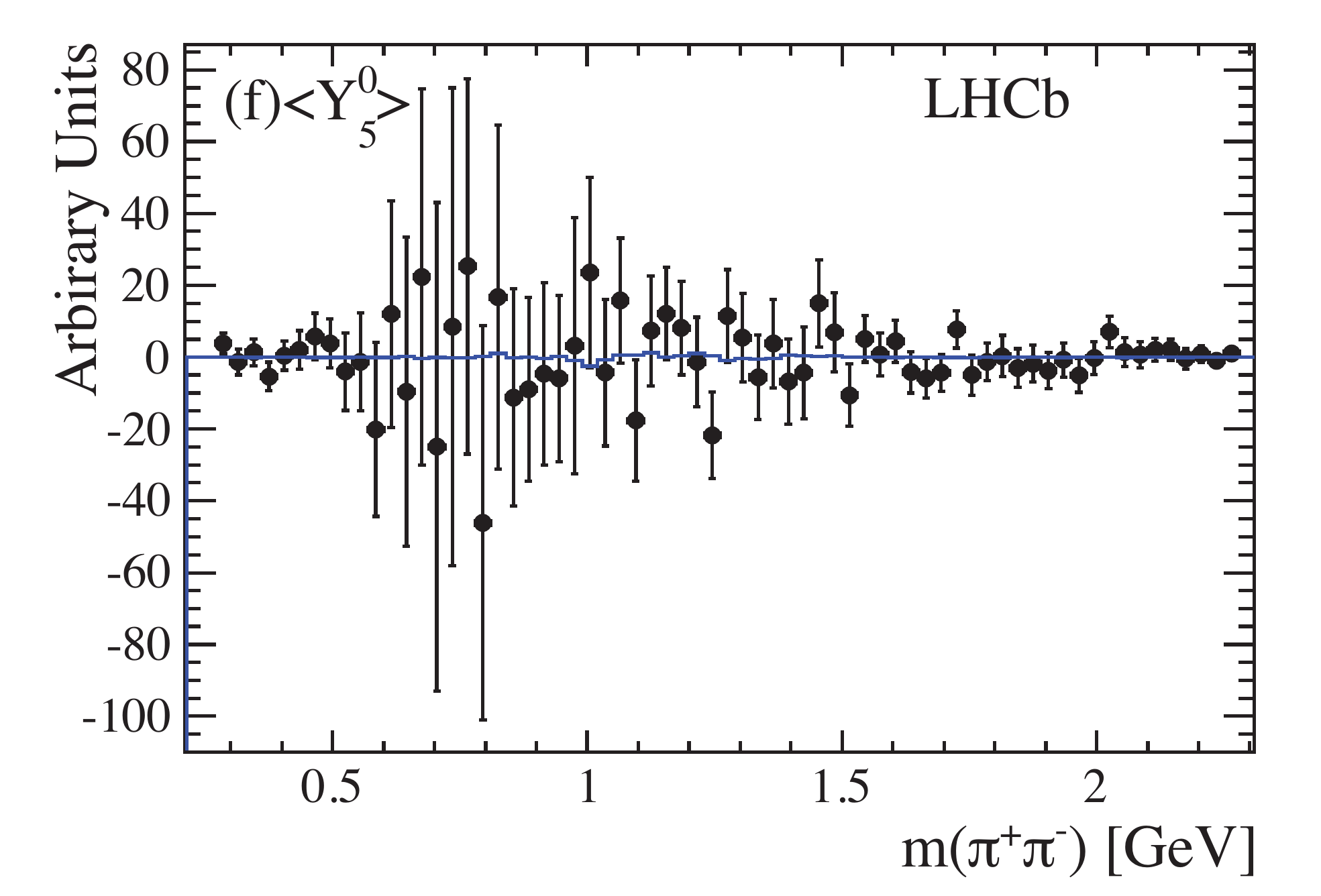}
\end{center}
\vskip -0.5cm
\caption{The $\pi^+\pi^-$ mass dependence of the spherical harmonic moments of $\cos \theta_{\pi\pi}$ after efficiency corrections and background subtraction:
(a) $\langle Y^0_0\rangle$ ($\chi^2$/ndf =78/70),  (b) $\langle Y^0_1\rangle$ ($\chi^2$/ndf =37/70), (c) $\langle Y^0_2\rangle$ ($\chi^2$/ndf =79/70), (d) $\langle Y^0_3\rangle$ ($\chi^2$/ndf =42/70), (e) $\langle Y^0_4\rangle$ ($\chi^2$/ndf =43/70), (f) $\langle Y^0_5\rangle$ ($\chi^2$/ndf =35/70).
The points with error bars are the data points and the solid curves are derived from the model 5R Solution I.}
\label{SPH2}
\end{figure}

\begin{figure}[!htbp]
\begin{center}
    \includegraphics[width=0.44\textwidth]{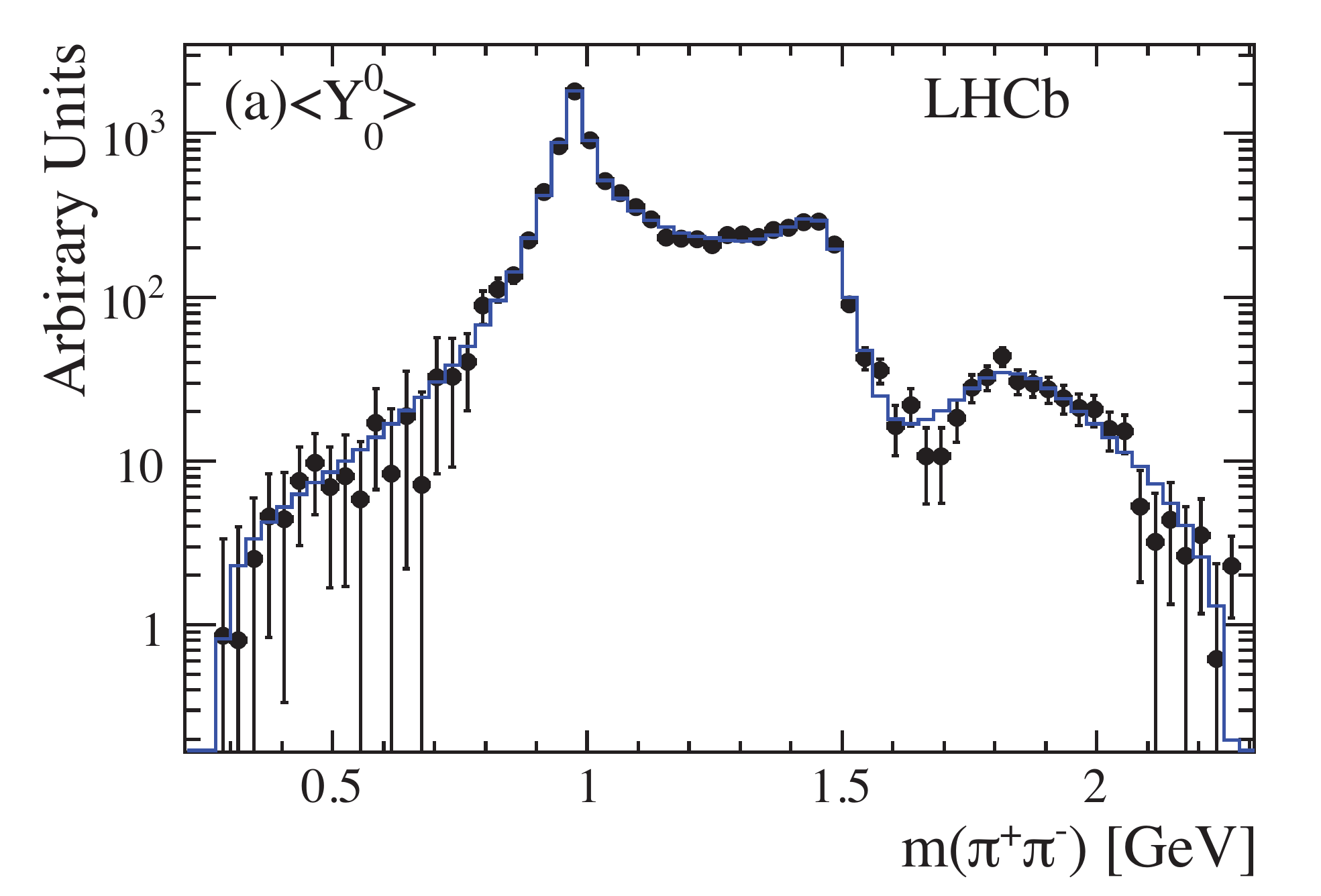}%
    \includegraphics[width=0.44\textwidth]{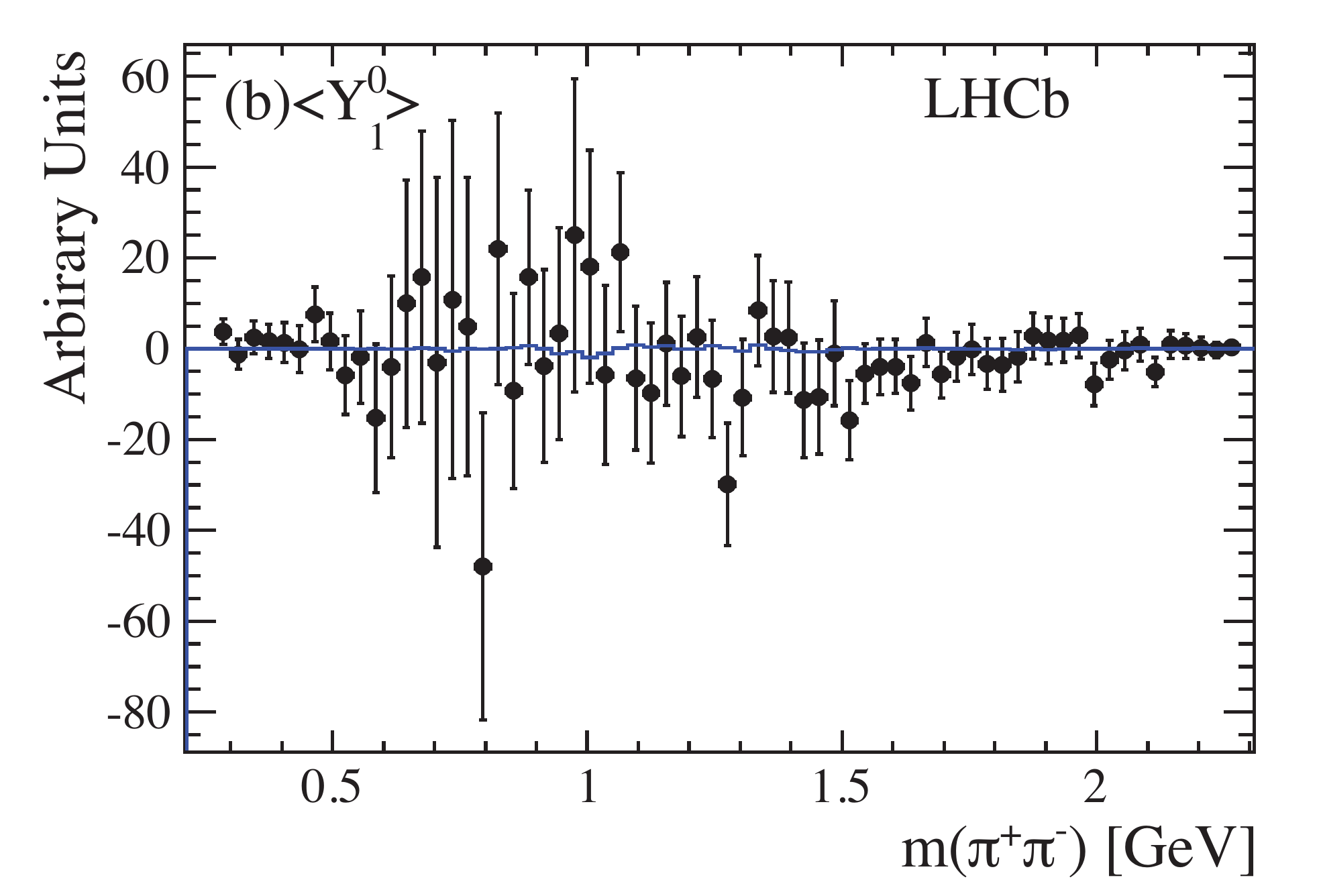}
    \includegraphics[width=0.44\textwidth]{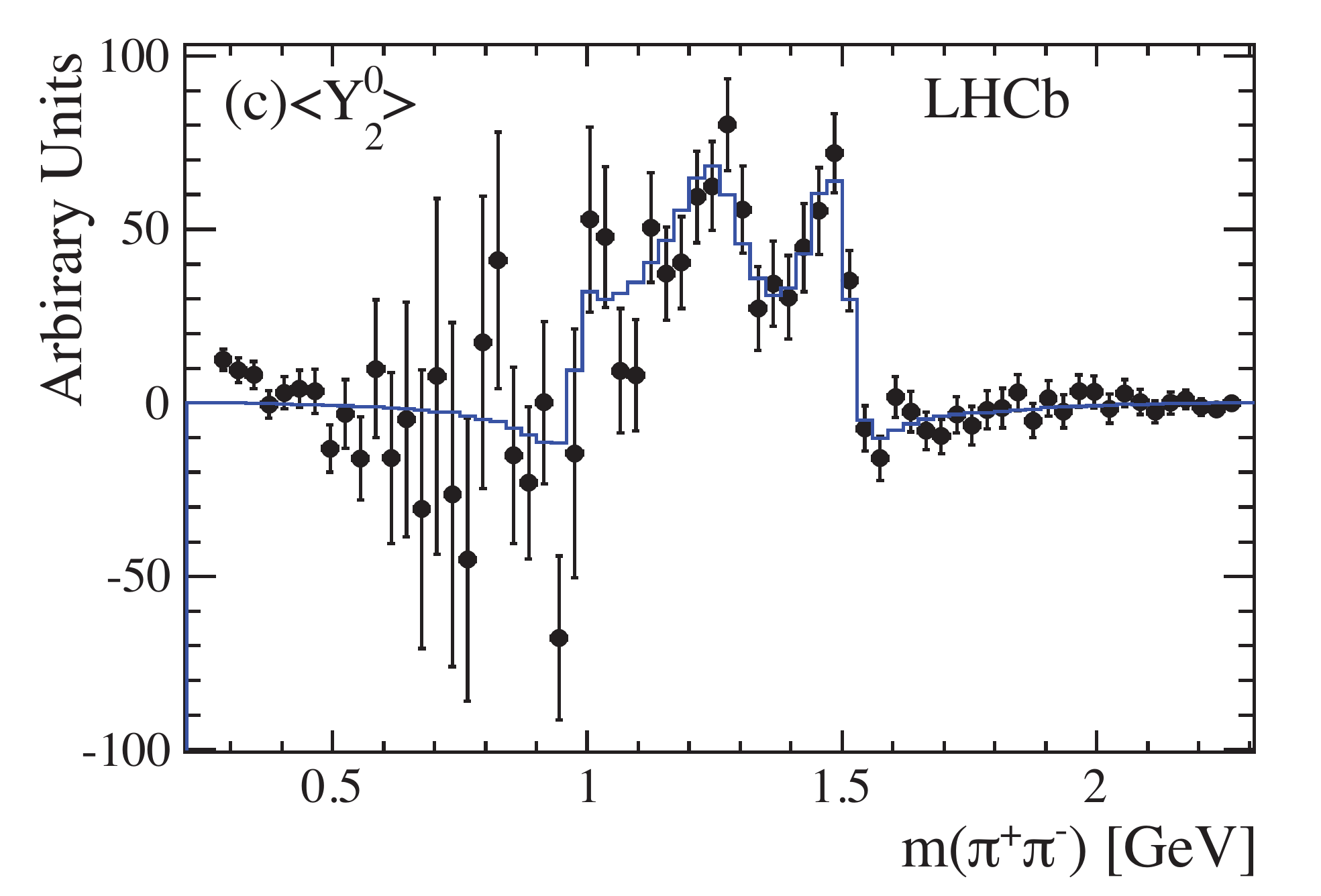}%
    \includegraphics[width=0.44\textwidth]{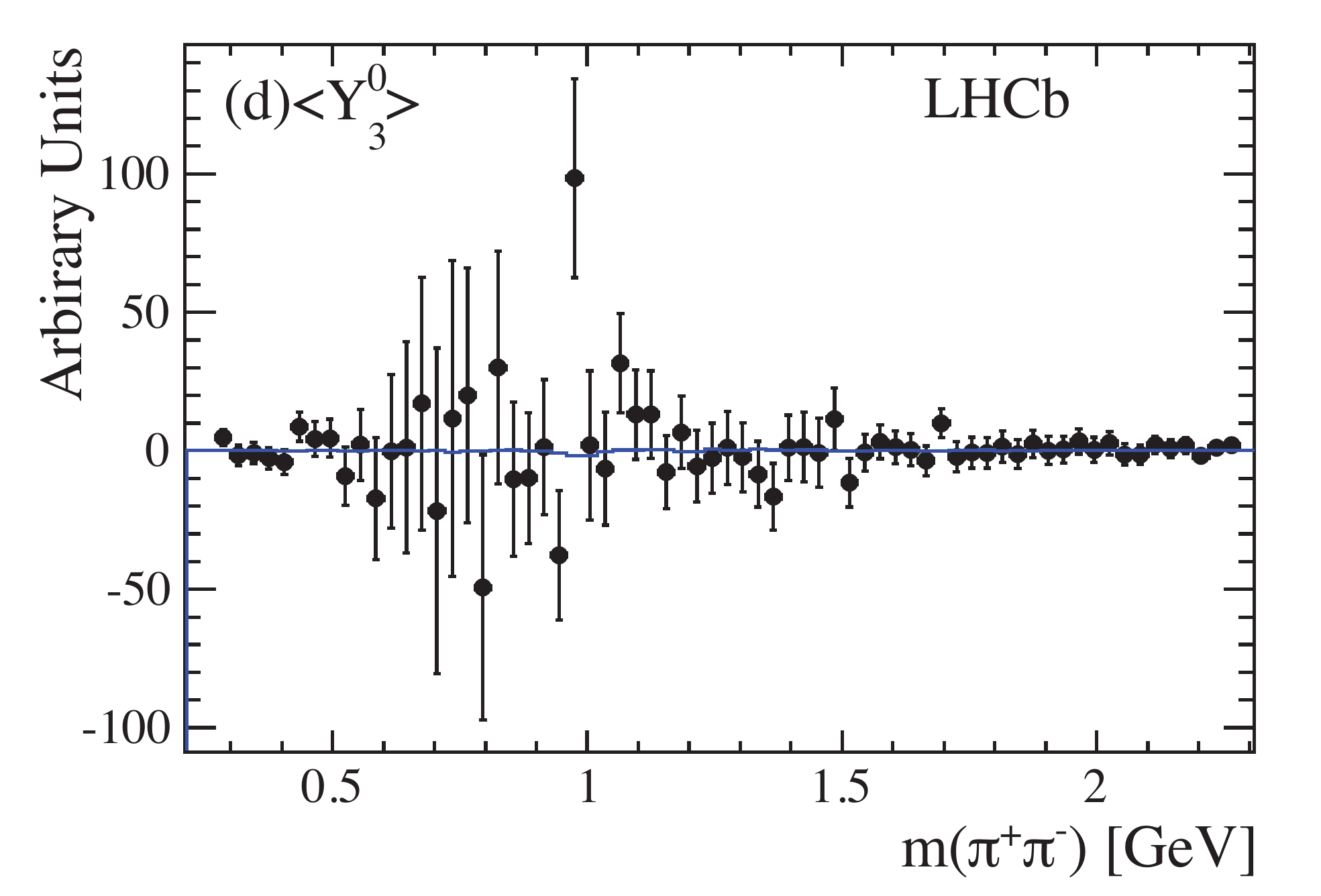}
    \includegraphics[width=0.44\textwidth]{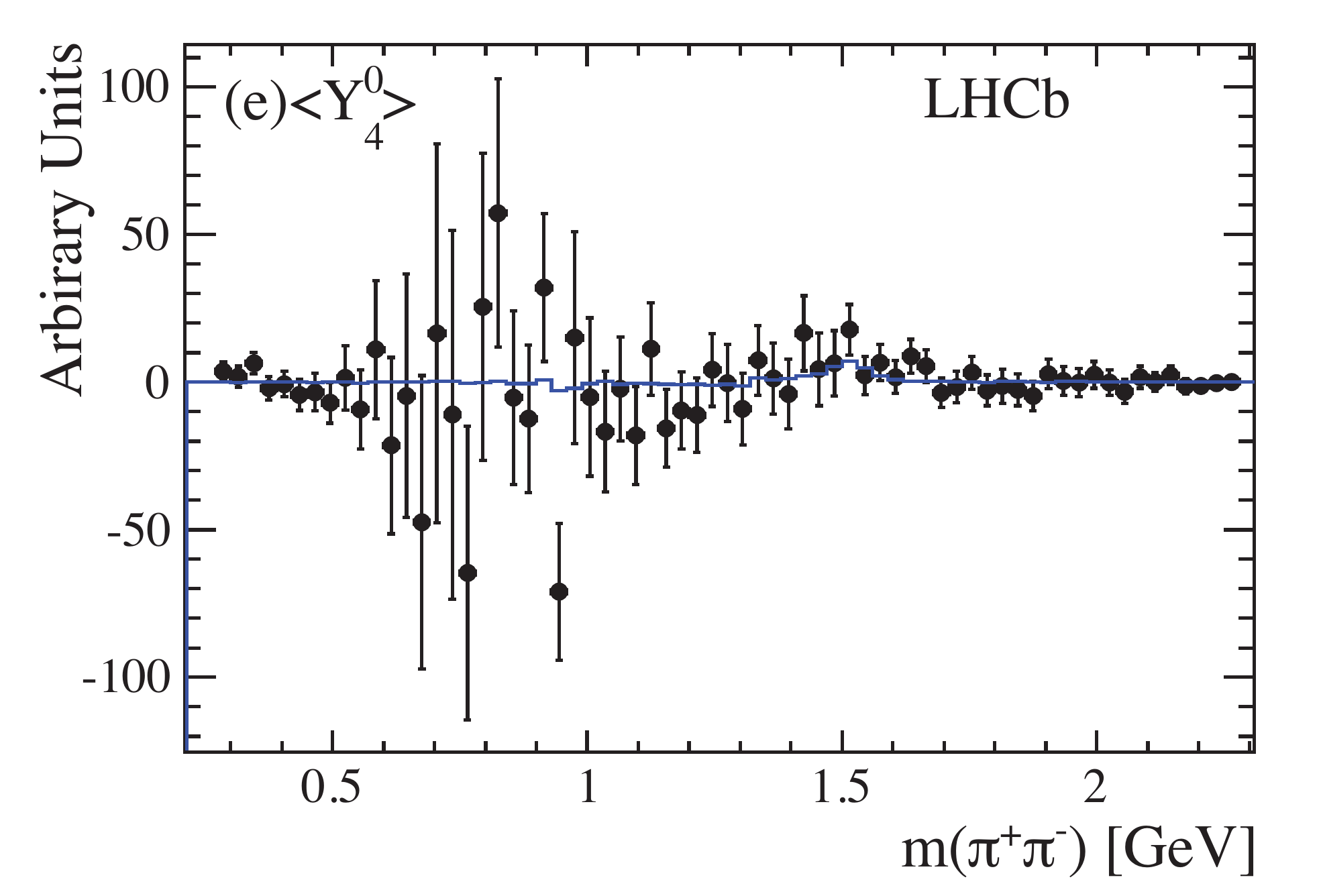}%
    \includegraphics[width=0.44\textwidth]{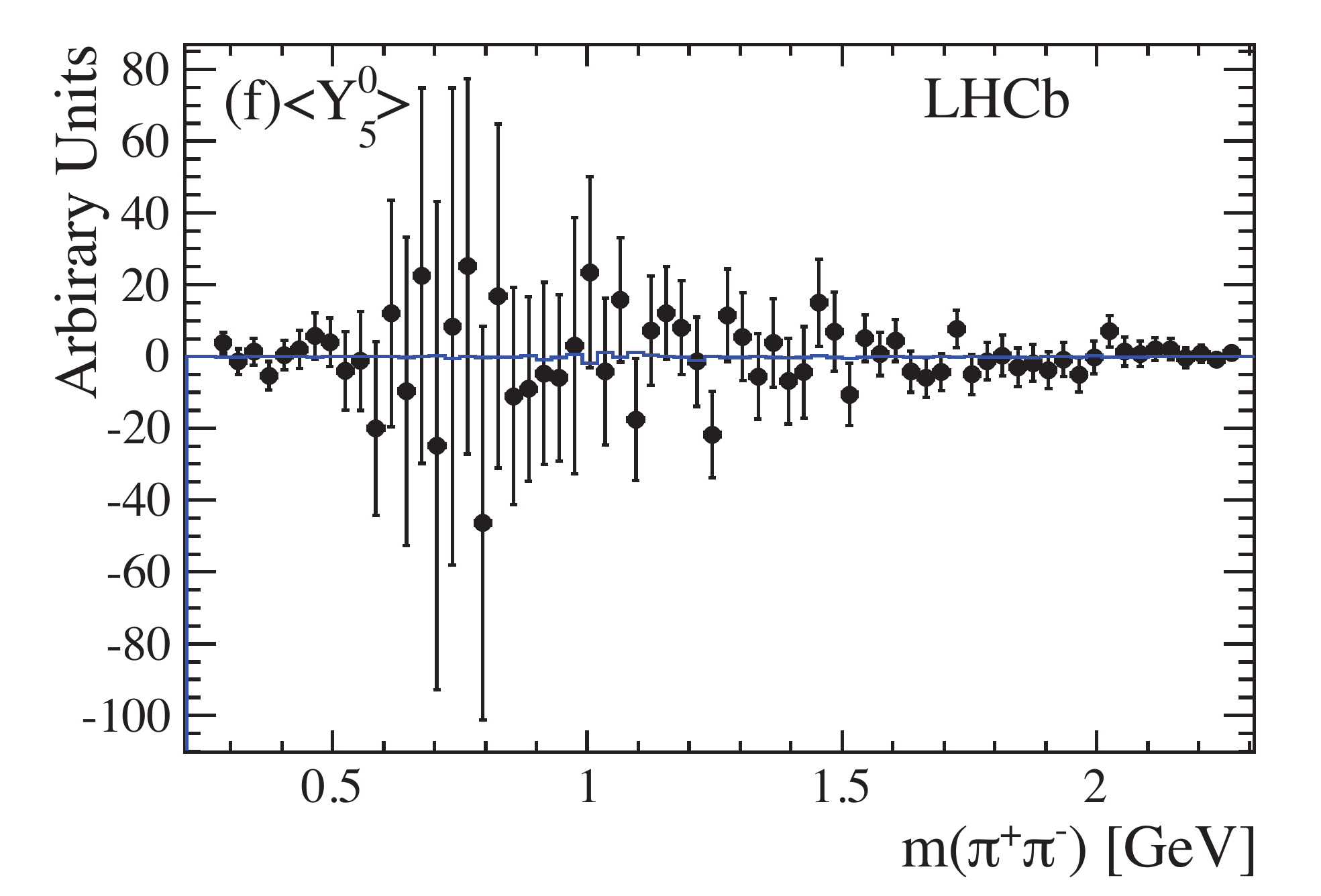}
\end{center}
\vskip -0.5cm
\caption{The $\pi^+\pi^-$ mass dependence of the spherical harmonic moments of $\cos \theta_{\pi\pi}$ after efficiency corrections and background subtraction:
(a) $\langle Y^0_0\rangle$ ($\chi^2$/ndf =73/70),  (b) $\langle Y^0_1\rangle$ ($\chi^2$/ndf =36/70), (c) $\langle Y^0_2\rangle$ ($\chi^2$/ndf =72/70), (d) $\langle Y^0_3\rangle$ ($\chi^2$/ndf =43/70), (e) $\langle Y^0_4\rangle$ ($\chi^2$/ndf =41/70), (f) $\langle Y^0_5\rangle$ ($\chi^2$/ndf =34/70).
The points with error bars are the data points and the solid curves are derived from the model 5R+NR Solution II.}
\label{SPH4}
\end{figure}

 We define the moments of the cosine of the helicity angle $\theta_{\pi\pi}$, $\langle Y^0_{l}(\cos \theta_{\pi\pi})\rangle$ as the efficiency corrected and background subtracted $\pi^+\pi^-$ invariant mass distributions, weighted by spherical harmonic functions.
The moment distributions provide an additional way of visualizing the presence of different resonances and their interferences, similar to a partial wave analysis.
Figures ~\ref{SPH2} and ~\ref{SPH4} show the distributions of the angular moments for 5R Solution I and 5R+NR Solution II, respectively. In general the interpretation of these moments~\cite{LHCb:2012ae} is that $\langle Y^0_0\rangle$ is the efficiency corrected and background subtracted event distribution, $\langle Y^0_1\rangle$  the interference of the sum of S-wave and P-wave and P-wave and D-wave amplitudes, $\langle Y^0_2\rangle$  the sum of the  P-wave, D-wave and the interference of S-wave and D-wave amplitudes, $\langle Y^0_3\rangle$  the interference between P-wave and D-wave,  $\langle Y^0_4\rangle$ the D-wave, and $\langle Y^0_5\rangle$ the F-wave. The values of $\langle Y^0_1\rangle$ and $\langle Y^0_3\rangle$ are almost zero because the opposite contributions from $\Bs$ and $\Bsb$ decays are summed.  Note, in this analysis the P-wave contributions are zero so the above description simplifies somewhat. The $f_2(1270)$ and $f_2^\prime(1525)$ interference with S-waves are clearly shown in the $\langle Y^0_2\rangle$ plot (see Figs.~\ref{SPH2} (c) and ~\ref{SPH4} (c)).


\section{Systematic uncertainties}\label{Sec:sys}
The sources of the systematic uncertainties on the results of the amplitude analysis are summarized in Table~\ref{tab:sys-dlz} for Solution I and Table~\ref{tab:sys-dlz2} for Solution II.
The contributions to the systematic error  due to $\phi_s$, the function $\varepsilon(t)$, $\Gs$ and $\DGs$\cite{PDG} uncertainties, and $L_B$ choices for transversity 0 and $\|$ of spin $\ge1$ resonances, are negligible. The systematic errors associated to the acceptance or background modeling are estimated by repeating the fit to the data 100 times. In each fit the parameters in the acceptance or background function are randomly generated according to the corresponding error matrix.
The uncertainties due to the fit model include possible contributions from each resonance listed in Table \ref{tab:reso1} but not used in the baseline fit models, varying the hadron scale $r$ parameters in the Blatt-Weisskopf barrier factors for the $B$ meson and $R$ resonance from 5.0 GeV$^{-1}$ and 1.5 GeV$^{-1}$, respectively, to both 3.0 GeV$^{-1}$, and using $F_{KK}=1$ in the Flatt\'e function. Compared to the nominal Flatt\'e function, the new one improves the likelihood fit $\rm -2ln\mathcal{L}$ by 6.8 and 14.0 units for Solution I and Solution II, respectively.  The largest variation among those changes is assigned as the systematic uncertainties for modeling.

Finally, we repeat the data fit by varying the mass and width of resonances within their errors one at a time, and add the changes in quadrature. To assign a systematic uncertainty from the possible presence of the $f_0(500)$ or $\rho(770)$, we repeat the above procedures using the model that has the baseline resonances plus $f_0(500)$ or $\rho(770)$. 
\begin{table}[t]
\centering
\caption{Absolute systematic uncertainties for Solution I.}
\def\arraystretch{1.2}
\begin{tabular}{lccccc}
\hline
Item& Acceptance& Background& Fit model & Resonance parameters& Total\\\hline
\multicolumn{6}{c}{Fit fractions (\%)}\\\hline
$f_0(980)$ & $\pm0.17$ &$\pm0.36$& $_{-5.04}^{+0.00}$&$\pm0.03$&$_{-5.1}^{+0.4}$\\
$f_0(1500)$ & $\pm0.06$ &$\pm0.14$& $_{-0.29}^{+1.11}$&$\pm0.02$&$_{-0.3}^{+1.1}$\\
$f_0(1790)$ & $\pm0.02$ &$\pm0.11$& $_{-0.11}^{+4.98}$&$\pm0.01$&$_{-0.2}^{+5.0}$\\
$f_2(1270)_0$ & $\pm0.03$ &$\pm0.01$& $\pm0.01$&$\pm0.01$&$\pm0.03$\\
$f_2(1270)_{\|}$& $\pm0.007$ &$\pm0.009$& $_{-0.020}^{+0.050}$&$\pm0.004$&$_{-0.02}^{+0.05}$\\
$f_2(1270)_{\perp}$& $\pm0.04$ &$\pm0.05$& $_{-0.04}^{+0.14}$&$\pm0.03$&$_{-0.08}^{+0.16}$\\
$f_2^\prime(1525)_0$& $\pm0.007$ &$\pm0.012$& $_{-0.000}^{+0.030}$&$\pm0.03$&$_{-0.04}^{+0.05}$\\
$f_2^\prime(1525)_{\|}$& $\pm0.003$ &$\pm0.004$& $_{-0.020}^{+0.000}$&$\pm0.004$&$_{-0.02}^{+0.05}$\\
$f_2^\prime(1525)_{\perp}$& $\pm0.007$ &$\pm0.016$& $_{-0.01}^{+0.04}$&$\pm0.04$&$^{+0.06}_{-0.04}$\\\hline
\multicolumn{6}{c}{Other fraction (\%)}\\\hline
$f_0(500)/f_0(980)$ & $\pm0.005$& $\pm$0.051& $^{+0.150}_{-0.020}$& $\pm$0.017 & $_{-0.06}^{+0.16}$\\\hline
$\rho(770)$ & $\pm0.013$ & $\pm0.065$& $^{+0.040}_{-0.120}$ &$\pm0.013$ & $_{-0.14}^{+0.08}$\\\hline
\CP-even& $\pm0.04$& $\pm0.06$&$^{+0.59}_{-0.05}$&$\pm0.05$&$^{+0.59}_{-0.10}$\\\hline
\end{tabular}\label{tab:sys-dlz}
\end{table}

\begin{table}[t]
\centering
\caption{Absolute systematic uncertainties for Solution II.}
\def\arraystretch{1.2}
\begin{tabular}{lccccc}
\hline
Item& Acceptance& Background& Fit model & Resonance parameters& Total\\\hline
\multicolumn{6}{c}{Fit fractions (\%)}\\\hline
$f_0(980)$ & $\pm0.12$ &$\pm0.79$& $_{-15.97}^{+~0.00}$&$\pm0.00$&$_{-16.0}^{+~0.8}$\\
$f_0(1500)$ & $\pm0.05$ &$\pm0.15$& $\pm0.27$&$\pm0.07$&$\pm0.3$\\
$f_0(1790)$ & $\pm0.02$ &$\pm0.09$& $_{-0.10}^{+2.46}$&$\pm0.01$&$_{-0.1}^{+2.5}$\\
$f_2(1270)_0$ & $\pm0.02$ &$\pm0.01$& $_{-0.03}^{+0.02}$&$\pm0.02$&$\pm0.04$\\
$f_2(1270)_{\|}$& $\pm0.005$ &$\pm0.009$& $_{-0.010}^{+0.110}$&$\pm0.020$&$_{-0.02}^{+0.11}$\\
$f_2(1270)_{\perp}$& $\pm0.04$ &$\pm0.05$& $_{-0.05}^{+0.10}$&$\pm0.03$&$_{-0.09}^{+0.12}$\\
$f_2^\prime(1525)_0$& $\pm0.006$ &$\pm0.012$& $_{-0.010}^{+0.03}$&$\pm0.031$&$_{-0.04}^{+0.05}$\\
$f_2^\prime(1525)_{\|}$& $\pm0.004$ &$\pm0.008$& $_{-0.040}^{+0.030}$&$\pm0.008$&$_{-0.04}^{+0.03}$\\
$f_2^\prime(1525)_{\perp}$& $\pm0.01$ &$\pm0.02$& $_{-0.00}^{+0.03}$&$\pm0.05$&$^{+0.06}_{-0.05}$\\
NR&$\pm0.07$ &$\pm0.63$& $_{-4.52}^{+0.34}$&$\pm0.04$&$_{-4.6}^{+0.7}$\\\hline
\multicolumn{6}{c}{Other fraction (\%)}\\\hline
$f_0(500)/f_0(980)$ & $\pm$0.005& $\pm$0.051& $^{+0.300}_{-0.120}$& $\pm$0.017 & $_{-0.14}^{+0.31}$\\\hline
$\rho(770)$& $\pm0.015$ & $\pm0.080$& $^{+0.040}_{-0.120}$ &$\pm0.016$ & $_{-0.15}^{+0.09}$\\\hline
\CP-even& $\pm0.04$& $\pm0.06$&$^{+0.66}_{-0.03}$&$\pm0.06$&$^{+0.66}_{-0.10}$\\\hline
\end{tabular}\label{tab:sys-dlz2}
\end{table}

\section{Further results}
\subsection{Fit fraction intervals}
The fit fractions shown in Table~\ref{tab:FF} differ considerably for some of the states between the two solutions. Table~\ref{tab:FF68} lists the  $1\sigma$ regions for the fit fractions taking into account the differences between the solutions and  including systematic uncertainties. The regions covers both $1\sigma$ intervals of the two solutions.
\begin{table}[b]
\begin{center}
\caption{Fit fraction ranges taking $1\sigma$ regions for both solutions including systematic uncertainties.}
\def\arraystretch{1.2}
\begin{tabular}{lcc}
\hline
 Component            & Fit fraction (\%)  \\\hline
$f_0(980)$ & $65.0-94.5$\\
$f_0(1500)$ & $8.2-11.5$\\
$f_0(1790)$ & $0.6-7.4$\\
$f_2(1270)_0$ & $0.28-0.50$\\
$f_2(1270)_{\|}$& $0.29-0.68$\\
$f_2(1270)_{\perp}$& $0.23-1.00$\\
$f_2^\prime(1525)_0$& $0.41-0.62$\\
$f_2^\prime(1525)_{\|}$& $0.02-0.27$\\
$f_2^\prime(1525)_{\perp}$& $0.03-0.49$\\
NR& $0-7.5$\\\hline
\end{tabular}
\label{tab:FF68}
\end{center}
\end{table}

\subsection{\CP content}
The only \CP-even content arises from the $\perp$ projections of the $f_2(1270)$ and $f_2^\prime(1525)$ resonances, in addition to the 0 and $\|$ of any possible $\rho(770)$ resonance. The \CP-even measured values are $(0.89\pm0.38_{-0.10}^{+0.59})\%$ and $(0.86\pm0.42_{-0.10}^{+0.66})\%$  for Solutions I and II,  respectively (see Table~\ref{tab:FF}),  where the systematic uncertainty is dominated by the forbidden $\rho(770)$ transversity $0$ and $\|$ components added in quadrature. To obtain the corresponding upper limit, the covariance matrix and parameter values from the fit are used to generate 2000 sample parameter sets. For each
set, the \CP-even fraction is calculated and is then smeared by the systematic uncertainty. The integral of 95\% of the area of the distribution yields an upper limit on the \CP-even component of 2.3\% at 95\% CL, where the larger value given by Solution II is used. The upper limit is the same as our previous measurement~\cite{LHCb:2012ae}, while the current measurement also adds in a possible $f_2^\prime(1525)$ contribution.

\subsection{Mixing angle and interpretation of light scalars}

The $I=0$ resonanances, $f_0(500)$ and $f_0(980)$, are thought to be mixtures of underlying states whose mixing angle has been estimated previously (see references cited in Ref.~\cite{Fleischer:2011au}). The mixing is parameterized by a normal 2$\times$2 rotation matrix characterized by the angle $\varphi_m$, giving in our case
\begin{eqnarray}
  \label{eq:fmix}
 \ket{f_0(980)}&=&\;\;\;\cos\varphi_m\ket{s\overline{s}}+\sin\varphi_m\ket{n\overline{n}}\nonumber\\
  \ket{f_0(500)}&=&-\sin\varphi_m\ket{s\overline{s}}+\cos\varphi_m\ket{n\overline{n}},\nonumber\\
  {\rm where~} \ket{n\overline{n}}&\equiv&\frac{1}{\sqrt{2}}\left(\ket{u\overline{u}}+\ket{d\overline{d}}\right).
\end{eqnarray}

In this case only the $\ket{s\overline{s}}$ wave function contributes. Thus we have \cite{Stone:2013eaa}
\begin{equation}
\tan^2\varphi_m=\frac{{\cal{B}}\left(\Bsb\to\jpsi f_0(500)\right)}{{\cal{B}}\left(\Bsb\to\jpsi f_0(980)\right)}\frac{\Phi(980)}{\Phi(500)},
\end{equation}
where the $\Phi$'s are phase space factors. The phase space in this pseudoscalar to vector-pseudoscalar decay is proportional to the cube of the $f_0$ momenta. Taking the average of the momentum dependent phase space over the resonant line shapes results in the ratio of phase space factors $\frac{\Phi(500)}{\Phi(980)}=1.25$.

Our measured upper limit is
\begin{equation}
\frac{{\cal{B}}\left(\Bsb\to\jpsi f_0(500),~f_0(500)\to\pi^+\pi^-\right)}{{\cal{B}}\left(\Bsb\to\jpsi f_0(980),~f_0(980)\to\pi^+\pi^-\right)}<3.4\%~{\rm at~90\%~CL,}
\end{equation}
where the larger value of the two solutions (II) is used.
This value must be corrected for the individual branching fractions of the $f_0$ resonances into $\pi^+\pi^-$. BaBar measures the relative branching ratios of $f_0(980) \to K^+ K^-$ to $\pi^+\pi^-$ of $0.69\pm0.32$ using $B\to KKK$ and $B\to K\pi\pi$ decays \cite{Aubert:2006nu}. BES has extracted relative branching ratios using $\psi(2S)\to\gamma \chi_{c0}$ decays where the $\chi_{c0}\to f_0(980)f_0(980)$, and either both $f_0(980)$'s decay into $\pi^+\pi^-$ or one into $\pi^+\pi^-$ and the other into $K^+K^-$ \cite{Ablikim:2004cg,*Ablikim:2005kp}. Averaging the two measurements gives
\begin{equation}
\frac{{\cal{B}}\left(f_0(980)\to K^+ K^-\right)}{{\cal{B}}\left(f_0(980)\to\pi^+\pi^-\right)}=0.35_{-0.14}^{+0.15}
\end{equation}

Assuming that the $\pi\pi$ and $KK$ decays are dominant we can also extract
\begin{equation}
{\cal{B}}\left(f_0(980)\to\pi^+\pi^-\right)
=\left(46\pm6\right)\%
\end{equation}
where we have assumed that the only other decays are to $\pi^0\pi^0$, $\frac{1}{2}$ of the $\pi^+\pi^-$ rate, and to neutral kaons, equal to charged kaons.
We use ${\cal{B}}\left(f_0(500)\to\pi^+\pi^-\right)
=\frac{2}{3}$, which results from isopsin Clebsch-Gordon coefficients, and assuming that the only decays are into two pions. Since we have only an upper limit on the $\jpsi f_0(500)$, we will only find an upper limit on the mixing angle, so if any other decay modes of the $f_0(500)$ exist, they would make the limit more stringent. Including uncertainty of ${\cal{B}}\left(f_0(980)\to\pi^+\pi^-\right)$, our limit is
\begin{equation}
\tan^2\varphi_m=\frac{{\cal{B}}\left(\Bzb\to\jpsi f_0(500)\right)}{{\cal{B}}\left(\Bzb\to\jpsi f_0(980)\right)}\frac{\Phi(980)}{\Phi(500)} <1.8\%~{\rm at~90\%~CL},
\end{equation}
which translates into a limit
\begin{equation}
|\varphi_m| < 7.7^{\circ}~{\rm at~90\%~CL}.
\end{equation}

This limit is the most constraining ever placed on this mixing angle~\cite{Aaij:2013zpt}. The value of $\tan^2\varphi_m$ is consistent with the tetraquark model, which predicts zero within a few degrees\cite{Stone:2013eaa,Fleischer:2011au}.

\section{Conclusions}
The $\Bsb\to\jpsi\pi^+\pi^-$ decay can be described by the interfering sum of five resonant components: $f_0(980), f_0(1500), f_0(1790), f_2(1270)$ and $f_2^{\prime}(1525)$. In addition we find that a second model including these states plus non-resonant $\jpsi \pi^+\pi^-$ also provides a good description of the data. In both models the largest component of the decay is the
$f_0(980)$ with the $f_0(1500)$ being almost an order of magnitude smaller. We also find including the $f_0(1790)$ resonance improves the data fit significantly. The $\pi^+\pi^-$ system is mostly S-wave, with the D-wave components totaling only 2.3\% in either model. No significant $\Bsb\to\jpsi\rho(770)$ decay is observed; a 90\% CL upper limit on the fit fraction is set to be 1.7\%.

The most important result of this analysis is that the \CP content is consistent with being purely odd, with the \CP-even component limited to 2.3\% at 95\% CL. Also of importance is the limit on the absolute value of the mixing angle between the $f_0(500)$ and $f_0(980)$ resonances of $7.7^{\circ}$ at 90\% CL, the most stringent limit ever reported. This is also consistent with these states being tetraquarks.

\section*{Acknowledgements}
\noindent We express our gratitude to our colleagues in the CERN
accelerator departments for the excellent performance of the LHC. We
thank the technical and administrative staff at the LHCb
institutes. We acknowledge support from CERN and from the national
agencies: CAPES, CNPq, FAPERJ and FINEP (Brazil); NSFC (China);
CNRS/IN2P3 and Region Auvergne (France); BMBF, DFG, HGF and MPG
(Germany); SFI (Ireland); INFN (Italy); FOM and NWO (The Netherlands);
SCSR (Poland); MEN/IFA (Romania); MinES, Rosatom, RFBR and NRC
``Kurchatov Institute'' (Russia); MinECo, XuntaGal and GENCAT (Spain);
SNSF and SER (Switzerland); NAS Ukraine (Ukraine); STFC (United
Kingdom); NSF (USA). We also acknowledge the support received from the
ERC under FP7. The Tier1 computing centers are supported by IN2P3
(France), KIT and BMBF (Germany), INFN (Italy), NWO and SURF (The
Netherlands), PIC (Spain), GridPP (United Kingdom).
We are indebted to the communities behind the multiple open source software packages we depend on.
We are also thankful for the computing resources and the access to software R\&D tools provided by Yandex LLC (Russia).

\newpage
\clearpage
\ifx\mcitethebibliography\mciteundefinedmacro
\PackageError{LHCb.bst}{mciteplus.sty has not been loaded}
{This bibstyle requires the use of the mciteplus package.}\fi
\providecommand{\href}[2]{#2}

\end{document}